\documentclass[a4paper,12pt,twoside,openright]{iitbthesis}
\usepackage{fancyhdr}

\usepackage{epsfig}
\usepackage[T1]{fontenc}
\usepackage[utf8]{inputenc}
\usepackage{mathptmx}
\usepackage{amssymb}
\usepackage{amsmath,epsfig}
\usepackage{graphicx,graphics}
\usepackage{url}
\usepackage{hyperref}
\usepackage[capitalise]{cleveref}
\usepackage{textcomp}
\usepackage{enumitem}
\usepackage{pdflscape}
\renewcommand\bibname{References}
\usepackage{titlesec}
\usepackage{bm,bbm}
\usepackage{float,lscape}
\usepackage{subfig}

\usepackage{epstopdf}
\usepackage{tabu}
\usepackage{multirow}
\usepackage{array}
\usepackage{booktabs}
\usepackage{graphicx}
\usepackage{caption}
\usepackage{lettrine}
\usepackage{enumitem}
\usepackage{lscape}
\usepackage{rotating}
\usepackage{booktabs}
\usepackage{longtable}
\usepackage{mathptmx}
\usepackage{anyfontsize}
\usepackage{t1enc}
\usepackage{relsize}
\usepackage{placeins}
\usepackage{nomencl}
\usepackage{rotating}
\usepackage{glossaries}
\usepackage{scrlayer}
\usepackage{enumitem}

\usepackage[square,comma,numbers,sort&compress]{natbib}

\newcommand{\qbinom}[2]{\genfrac{[}{]}{0pt}{}{#1}{#2}_{\mathbbm{q}}}
\def\q{{\mathbbm{q}}}
\def\ie{{\textit{i.e.}}}
\usepackage{xcolor}
\newcommand{\Z}{{\mathbb Z}}
\newcommand{\textcol}[1]{\textcolor{black}{#1}}
\newtheorem{proposition}{Proposition}
\def\Tr{{\text{Tr}}}
\def\CL{{\mathcal L}}
\def\lk{{\ell k}}
\newcommand{\Q}{{\mathbb Q}}
\def\CW{{\mathcal W}}
\usepackage{comment}
\newcommand{\C}{{\mathbb C}}
\def\CI{{\mathcal I}}
\usepackage{quiver}

\usepackage[titletoc]{appendix}
\usepackage[nottoc]{tocbibind}
\newtheorem{theorem}{Theorem}[chapter]

\newtheorem{conjecture}[theorem]{Conjecture}

\newtheorem{definition}[theorem]{Definition}

\usepackage[...]{youngtab}
\usepackage{pgfplots}
\newcommand{\mdot}{\raise1.5pt \hbox{.}}
\usepackage{braids}
\newcommand{\be}{\begin{equation}}
\newcommand{\ee}{\end{equation}}

\begin{document}

%=====================================================================
% Include the prelude for Title page, abstract, table of contents, etc
% You need to modify it to contain your details
\abovedisplayskip=2mm
\abovedisplayshortskip=2mm
\belowdisplayskip=2mm
\belowdisplayshortskip=2mm

\clubpenalty=10000 \widowpenalty=10000
% prelude.tex
%   - titlepage
%   - dedication (optional)
%   - approval sheet
%   - course certificate
%   - table of contents, list of tables and list of figures
%   - nomenclature
%   - abstract
%============================================================================

%\clearpage\pagenumbering{roman}  % This makes the page numbers Roman (i, ii, etc)

\pagenumbering{gobble}

% TITLE PAGE
%   - define \title{} \author{} \date{}
\title{$q$-Series Invariants of Three-Manifolds \\
	and \\\vspace{4mm}
	 Knots-Quivers Correspondence}
\author{Sachin Chauhan}
\date{2024}
%  - Roll number, required for title page, approval sheet, and
%    certificate of course work 
\rollnum{194129001} 

%   - The default degree is ``Doctor of Philosophy''
%     (unless the document style msthesis is specified
%      and then the default degree is ``Master of Science'')
%     Degree can be changed using the command \iitbdegree{}
\iitbdegree{Doctor of Philosophy}

%   - The default report type is preliminary report.
%      * for a PhD thesis, specify \thesis
\thesis
%      * for a M.Tech./M.Phil./M.Des./M.S. dissertation, specify \dissertation
%\dissertation
%      * for a DIIT/B.Tech./M.Sc.project report, specify \project
%\project
%      * for any other type, use  \reporttype{}
%\reporttype{}

%   - The default department is ``Unknown Department''
%     The department can be changed using the command \department{}
\department{Department of Physics}

%    - Set the guide's name
\setguide{Prof. P. Ramadevi}
%    - Set the coguide's name (if you have one)
%\setcoguide{Prof. Gulab Singh}
%    - Set external guide (if you have one)
%\setexguide{Prof External Guide}

%   - once the above are defined, use \maketitle to generate the titlepage
\maketitle
%
%\newpage\null\thispagestyle{empty}\newpage % % For blank pages

%--------------------------------------------------------------------%
% DEDICATION
%   Dedications, if any, must be first page after title page.
\begin{dedication}
\large{Dedicated to my parents.}
\end{dedication}

%\newpage\null\thispagestyle{empty}\newpage % % For blank pages

%--------------------------------------------------------------------%
% APPROVAL SHEET
%   - for final thesis, you need Approval Sheet. So, uncomment the
%     \makeapproval command.
\chapter*{}
\thispagestyle{empty}
\vspace{-1in}
\begin{center}
{\Large  {\bf Thesis Approval}}
\end{center}
\vspace*{0.1in} \noindent This thesis entitled {\large \bf $q$-Series Invariants of Three-Manifolds and Knots-Quivers Correspondence} by {\bf \large
Sachin Chauhan} is approved for the degree of  {\large \bf Doctor of Philosophy}.\\\\
\hspace*{4 in} Examiners:\\\\
\hspace*{3.5 in} \ldots\ldots \ldots \ldots \ldots \ldots\ldots
\ldots \ldots \ldots\ldots\\\\
\hspace*{3.5 in} \ldots\ldots \ldots \ldots \ldots \ldots\ldots
\ldots \ldots \ldots\ldots\\\\
\hspace*{3.5 in} \ldots\ldots \ldots \ldots \ldots \ldots\ldots
\ldots \ldots \ldots\ldots\\\\
  \hspace*{3.5in} \ldots\ldots \ldots \ldots \ldots \ldots\ldots
\ldots \ldots \ldots\ldots\\\\
 Adviser:\hspace*{3.5 in}  Chairperson:\\\\\\
\ldots\ldots \ldots \ldots \ldots \ldots\ldots
\ldots \ldots \ldots\ldots
\hspace*{1.0 in} \ldots\ldots \ldots \ldots \ldots \ldots\ldots
\ldots \ldots \ldots\ldots\\\\\\
%  \hspace*{4 in} Chairman:\\\\
% \hspace*{3.8 in} \ldots\ldots \ldots \ldots \ldots \ldots\ldots
% \ldots \ldots \ldots\ldots

\noindent
Date: \ldots\ldots \ldots \ldots\\\\
Place: \ldots\ldots \ldots \ldots
%\pagebreak \clearpage{\pagestyle{empty}\cleardoublepage}
% \clearpage{\pagestyle{empty}\cleardoublepage}

%\newpage\null\thispagestyle{empty}\newpage % % For blank pages

%     it should come after dedication, if dedication is
%     present. Otherwise it is the first page after title page.
%\makeapproval
%\input{Acceptancecertificate}
%\addcontentsline{toc}{chapter}{Acceptance Certificate}
\clearpage
\thispagestyle{empty}

\begin{center}
\Large  {\bf Declaration }
\end{center}
\vspace{-6in}
I declare that this written submission represents my ideas in my own words and where others ideas or words have been included, I have adequately cited and referenced the original sources. I also declare that I have adhered to all principles of academic honesty and integrity and have not misrepresented or fabricated or falsified any idea/data/fact/source in my submission. I understand that any violation of the above will be cause for disciplinary action by the Institute and can also evoke penal action from the sources which have thus not been properly cited or from whom proper permission has not been taken when needed.
\vspace{0.5in}

%\begin{flushleft}
%Date:
%\end{flushleft}
%\begin{flushright}
%Surendar M\\
%(Roll. No. 114310007)
%\end{flushright}

%
 \begin{table}[h]
 \begin{flushleft}

\vspace{-3.2in} 
 \begin{tabular}{ccccc}
 % %\hline 	\rule[5ex]{0pt}{-10ex} &&(Signature) (Date) && \\ 
 \rule[5ex]{0pt}{-10ex}&& Date: && \\ 
 \end{tabular}
\end{flushleft}

\vspace{-0.5in} 
\begin{flushright}
 \begin{tabular}{ccccc}
 % %\hline 	\rule[5ex]{0pt}{-10ex} &&(Signature) (Date) && \\ 
 
 \hline 	\rule[5ex]{0pt}{-10ex}&& Sachin Chauhan&& \\ 
 \rule[5ex]{0pt}{-10ex}&& Roll No. 194129001&& \\ \\
 \end{tabular}
\end{flushright}
\end{table}

\pagebreak
%\addcontentsline{toc}{chapter}{Declaration}

%--------------------------------------------------------------------%
% CERTIFICATE OF COURSE WORK
%   - for final thesis, a course certificate is required.
%   - specify the  PhD joining date for the certificate.
%     Contact you department office or academic office if you do not
%     know it.
%\joiningdate{16 July 2015}
%\begin{coursecertificate}
%\addcourse{GNR 647}{Microwave Remote Sensing}{6}
%\addcourse{GNR 618}{Remote Sensing and GIS Applications to Cryosphere }{6}
%\addcourse{GNRS 01}{Seminar}{4}
%\addppcourse{HS 699}{Communication and Presentation Skills}{PP}
%\end{coursecertificate}

%for quotations
%\quotationpage

%--------------------------------------------------------------------%
% COPYRIGHT PAGE
%   - To include a copyright page use \copyrightpage
%\copyrightpage
%\newpage\null\thispagestyle{empty}\newpage % % For blank pages
%--------------------------------------------------------------------%

\pagenumbering{roman}
\chapter*{Acknowledgments}
\label{ch:Acknowledgments}
\addcontentsline{toc}{chapter}{\nameref{ch:Acknowledgments}}
\vspace{10mm}
I would like to express my deepest gratitude to my adviser, Pichai Ramadevi, for her invaluable guidance and unwavering support throughout all stages of my PhD journey. Her mentorship has been instrumental in my academic and research endeavors, and I feel incredibly fortunate to have had such a remarkable adviser.

I extend my heartfelt thanks to S. Shankaranarayanan, Urjit Yajnik, and Uma Sankar for serving on my research progress committee. Their insightful feedback has significantly contributed to the advancement of my research.

My sincere appreciation goes to my collaborators, Vivek, Aditya, Siddharth, and B.P. Mandal, for their valuable suggestions and contributions.

I am profoundly grateful to Piotr Sulkowski and Piotr Kucharski for welcoming me into their groups and fostering enriching interactions. Additionally, I thank Pavel Putrov, Sunghyuk Park, and Dmitry Noshchenko for their time and assistance in clarifying my doubts. I am especially thankful to Sergei Gukov for his thoughtful feedback on my work. I would also like to express my gratitude to Paul Wedrich for his warm hospitality and the fruitful discussions on categorification.

I extend my heartfelt thanks to the organizers of the conferences String Math 2022, String Math 2023, Indian Strings Meeting 2023, "Simons Semester on Knots, Homologies, and Physics," and the Learning Workshop on BPS States and 3-Manifolds for creating such an enriching learning atmosphere and providing me with the opportunity to present my work.

A special thank you to my officemates—Akhil, Amol, Archana, Ayaz, Himanshu, Lekhika, Ravi, Sudeep, and Zafri—for their engaging questions and discussions, which have greatly enhanced my understanding of the subject. I would like to thank Himanshu for his help with computational issues.

I am immensely grateful to my friends, Ashu and Naba, for their steadfast support throughout my PhD journey.

Finally, I dedicate this thesis to my parents and brother. Their unwavering support and encouragement have been the cornerstone of my achievements. None of this would have been possible without them.

% ABSTRACT
\begin{abstract}
  
The Gukov-Pei-Putrov-Vafa (GPPV) conjecture is a relationship between two three-manifold invariants: the Witten-Reshetikhin-Turaev (WRT) invariant and the \(\widehat{Z}\) (``Z-hat'') invariant. In fact, WRT invariant is defined at roots of unity, $\q\left(\exp\left(\frac{2\pi i}{k+2}\right),~k\in\mathbb{Z}_+,~\text{for}~SU(2)\right)$, and is generally a complex number, whereas $\widehat{Z}$-invariant is a $q$-series with integer coefficients such that $|q|<1$. Therefore, $\widehat{Z}$-invariant can be obtained from WRT-invariant by performing a particular analytic continuation, $\q\rightarrow q$. In this thesis, we first examine this conjecture for $SO(3)$ and the ortho-symplectic supergroup $OSp(1|2)$. This is done by setting up the WRT invariant for the respective groups and then performing the particular analytic continuation to extract $\widehat{Z}$. As a result of this exercise, we found that $\widehat{Z}^{SU(2)}=\widehat{Z}^{SO(3)}$ and identified a relation between $\widehat{Z}^{SU(2)}$ and $\widehat{Z}^{OSp(1|2)}$. Motivated by the equality of $\widehat{Z}$ for $SU(2)$ and $SO(3)$ groups, we study this conjecture for $SU(N)/\mathbb{Z}_m$ groups, where $\mathbb{Z}_m$ is a subgroup of $\mathbb{Z}_N$, in our second paper. We subsequently found that $\widehat{Z}^{SU(N)/\mathbb{Z}_m}=\widehat{Z}^{SU(N)}$.

Another theme of the thesis is to study a conjecture between knot theory and quiver representation theory. More precisely, this conjecture relates the generating function of the symmetric $r$-colored HOMFLY-PT polynomial with the motivic generating series associated with a symmetric quiver. In particular, we obtain a quiver representation for a family of knots called double twist knots $K(p,-m)$. Primarily, we exploit the reverse engineering of Melvin-Morton-Rozansky (MMR) formalism to deduce the pattern of the matrix for these quivers.

\end{abstract}

\begin{listofpub}
  \section*{International Journals}

\begin{enumerate}

\item[1.] Sachin Chauhan and P. Ramadevi ; \textit{$\hat Z$-invariant for $SO(3)$ and $OSp(1|2)$ Groups}, \href{https://doi.org/10.1007/s00023-023-01332-y}{Annales Henri Poincaré, Volume 24, pages 3347–3371, (2023)}

\item[2.] Sachin Chauhan and P. Ramadevi; \textit{Gukov-Pei-Putrov-Vafa conjecture for $SU(N)/\mathbb{Z}_m$}, \href{https://doi.org/10.1007/s11005-024-01791-2}{Lett Math Phys 114, 42 (2024)}.

\item[3.] Vivek Kumar Singh, Sachin Chauhan, P. Ramadevi, Aditya Dwivedi, B.P. Mandal, and Siddharth Dwivedi; \textit{Knot-quiver correspondence for double twist knots}, \href{https://doi.org/10.1103/PhysRevD.108.106023}{Phys. Rev. D 108, 106023}

\end{enumerate}  
\end{listofpub}

%--------------------------------------------------------------------%
% CONTENTS, TABLES, FIGURES

\tableofcontents
%\listoftables
%\listoffigures

% Here is the file for Abbrivations
%\input{abbreviations}

% To automate abbriavtions using Nomencluture  package.
% Comment the  \input{abbreviations}
% Then include NOMEN...... package 
% Refer for this q31_nom_tex 

%\addcontentsline{toc}{chapter}{Abbreviations and Nomenclature}
%--------------------------------------------------------------------%
% NOMENCLATURE
%\begin{nomenclature}
%\begin{description}
%\item{\makebox[0.75in][l]{$C_1$}} Constant 1
%
%\item{\makebox[0.75in][l]{$V$}}    Voltage 
%
%\item{\makebox[0.75in][l]{\$}}     US Dollars
%\end{description}
%\end{nomenclature}

%\cleardoublepage\pagenumbering{arabic} % Make the page numbers Arabic (1, 2, etc)

%\include{abbreviations}
%\include{symbols}

\setlength{\parskip}{1.5mm}
\titlespacing{\chapter}{0cm}{8mm}{5mm}
\titleformat{\chapter}[display]
{\normalfont\huge\bfseries\raggedright}
{\chaptertitlename\ \thechapter}{20pt}{\Huge}

\titlespacing*{\section}
{0pt}{6mm}{6mm}
\titlespacing*{\subsection}
{0pt}{6mm}{6mm}
\pagebreak
%\newpage  
%\cleardoublepage\pagenumbering{arabic}
\pagenumbering{arabic}

\makeatletter
\def\cleardoublepage{\clearpage\if@twoside \ifodd\c@page\else
	\hbox{}
	\vspace*{\fill}
	\begin{center}
		This page was intentionally left blank.
	\end{center}
	\vspace{\fill}
	\thispagestyle{empty}
	\newpage
	\if@twocolumn\hbox{}\newpage\fi\fi\fi}
\makeatother
%=====================================================================
% Include the technical part of the report
%%\include{chap_intro}             % Chapter 1: Introduction

\newpage
%\pagebreak
\cleardoublepage
\chapter{Introduction}
\label{chap:1}
Over the past few decades, the interplay between mathematics and string theory has proven to be highly fruitful, leading to significant advancements in both fields. String dualities have facilitated the prediction of new conjectures across diverse areas of mathematics, while mathematical conjectures have, in turn, informed the discovery of novel dualities in string theory \cite{aganagic2015string,bah2022panorama}.

One area that has particularly benefited from this interdisciplinary approach is low-dimensional topology, which has seen remarkable progress driven by insights from quantum field theory. This collaboration has introduced a wealth of new topological invariants for knots and 3-manifolds, collectively referred to as quantum invariants. These developments have given rise to the field known as quantum topology.

\section{Knot Theory}
\begin{figure}[ht]
	\centering
\includegraphics[scale=0.38]{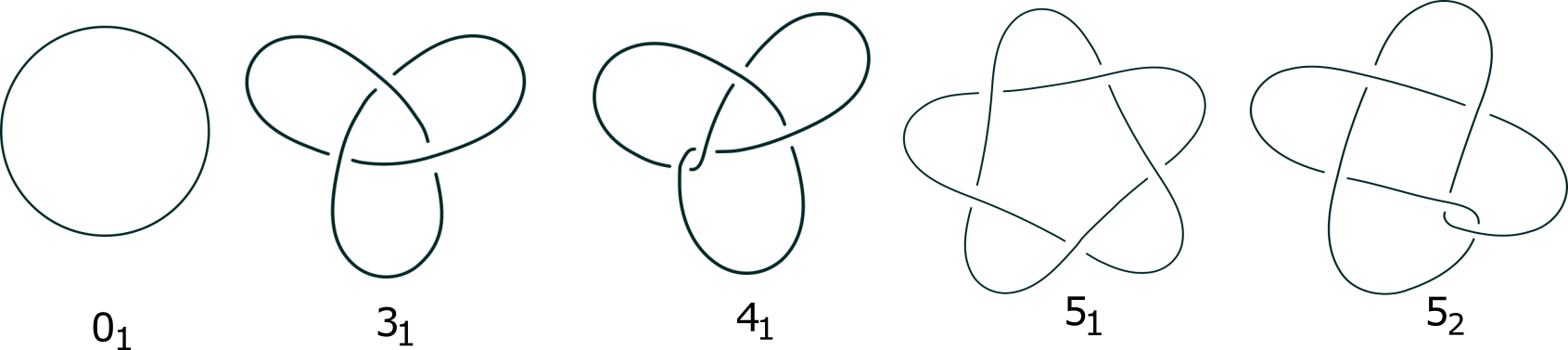}
	\caption{Diagrams of knots shown via their projection onto a plane.}
	\label{knotspic}
\end{figure}

Knot theory, a key component of this field, focuses on the study and classification of knots as mathematical objects. A knot is defined as the image of a smooth (or piecewise smooth) embedding \(S^1 \rightarrow \mathbb{R}^3\). Further, topologically equivalent knots $K$ and $K'$ can be converted into each other via Reidemeister moves RI, RII, and RIII \cite{Rmoves}. Figure \ref{knotspic} illustrates some examples of knots in which we have used the Alexander-Briggs notation, where each number represents the crossing number, and each subscript indicates the number of inequivalent knots with the same crossing.
A collection of knots that do not intersect but may be linked or knotted together is called a link. A link consisting of a single component is simply a knot. The framing of a knot \( K \) is defined as a continuous choice of a vector field normal to the knot \( K \). This framing is often visualized by specifying a ribbon-like band around the knot, representing the normal vector field. A framed link 
$[\mathcal L,\mathbf f]$ is therefore a link with a specified framing number $\mathbf f= (f_1,f_2,\ldots )$ for each of its components.

Among the significant classes of knot invariants are polynomial knot invariants, such as the Alexander polynomial $\Delta(K;x)$, and the Jones polynomial, $J(K;\q)$ \cite{Alexanderpoly,Jonespoly}. The Jones polynomial was subsequently extended to the two-parameter HOMFLY-PT polynomial, $P(K;a,\q)$ \cite{bams/1183552531,przytycki2016invariants}. There is another two-variable polynomial invariant for knots/links called the Kauffman polynomial \cite{louis115kauffman}. Furthermore, the HOMFLY-PT polynomial was generalized to the colored HOMFLY-PT polynomial by incorporating additional data from the representation theory of quantum groups \cite{Turaev:1988eb}. We denote the colored HOMFLY-PT polynomial of knot $K$ by \(P_R(K;a,\q)\), where \(R\) represents the color (or representation) of the gauge group \(SU(N)\) and \(a=\q^N\). In particular, when $N=2$, we obtain colored Jones polynomial for knot $K$ and we denote it by $J_R(K;\q)$. Moreover, when $R=$ \textit{fundamental representation}, we reproduce the Jones polynomial $J(K;\q)$ and the HOMFLY-PT polynomial $P(K;a,\q)$. In this thesis, we will be mostly using the colored HOMFLY-PT polynomials for symmetric representations and will denote it by $P_r(K;a,\q)$ where $r$ represents the number of boxes \ie, $R=\underbrace{\yng(4)}_r$. 

For a general $m$-component link $\mathcal{L}_m$ made up of component knots 
$K_i$'s carrying different representations $r_i$'s, we can define multicolored HOMFLY-PT invariant $P_{r_1,r_2\ldots r_m}(\mathcal{L}_m;a,\q)$ and the corresponding multicolored Jones invariant as $J_{r_1,r_2,\ldots r_m}(\mathcal{L}_m;\q)$.

The seminal work on relating Chern-Simons field theory to the Jones polynomial \cite{Witten:1988hf}, conducted by Witten, provided the first three-dimensional interpretation of the Jones polynomial. In fact, according to \cite{Witten:1988hf}, colored HOMFLY-PT polynomials of any knot $K$ can be viewed as expectation values of Wilson loop in representation $R$ within Chern-Simons theory on $S^3$:
\begin{equation}
	\overline{P}_R(K;a,\q)=\langle \text{Tr}_RU_K\rangle=\frac{\int[DA]\exp\left(iS_{CS}\right)\text{Tr}_RU_K}{\int[DA]\exp\left(iS_{CS}\right)}\label{expectval},
\end{equation}
where $S_{CS}=\frac{k}{4\pi}\int_{S^3}\text{Tr}\left(A\wedge dA+\frac{2}{3}A\wedge A\wedge A\right)$, $A$ is a one form valued in the Lie algebra of SU(N) gauge group, $k$ is the coupling constant which takes integer values, and $U_K=P \text{exp}\oint_KA$ is the holonomy of the $SU(N)$ Chern-Simons gauge field along a knot $K$. Further, we denote the denominator of eqn.(\ref{expectval}) by $Z^{SU(N)}_k[S^3;
\q]$:
\begin{equation}
Z^{SU(N)}_k[S^3;
\q]={\int[DA]\exp\left(iS_{CS}\right)}~.\label{chern-simons-partition-function-sun}
\end{equation}
Note that the polynomial variable  
\begin{equation}
\q= \exp\left({2 \pi i\over k+N}\right)
\end{equation} 
is a root of unity which depends on the Chern-Simons coupling $k$ and rank $N$ of $SU(N)$ group. For any $m$-component link, the link invariant is 
\begin{equation}
\overline{P}_{R_1,R_2 \ldots R_m}(\mathcal{L}_m;a,\q)=\langle \prod_{i=1}^m\text{Tr}_{R_i} U_{K_i}\rangle~.\label{linkinvariant}
\end{equation}
$SU(N)$ Chern-Simons knot or link invariants are referred to as unreduced colored HOMFLY-PT invariant. There is a normalisation involved to relate them to colored HOMFLY-PT (also known as reduced colored HOMFLY-PT). For any knot, the relation between unreduced $\overline{P}_R(K;a,\q)=$ and reduced HOMFLY-PT ${P}_R(K;a,\q)$ is
\begin{equation}
    \overline{P}_R(K;a,\q)=\overline{P}_R(0_1;a,\q)P_R(K;a,\q).
\end{equation}
Observe that the reduced colored HOMFLY-PT polynomial for the unknot is 1. Similarly, the Chern-Simons knot invariant for $SU(2)$ group will give the unreduced colored Jones polynomial $\overline{J}_R(K;\q)$. For any $r$-symmetric representation, the unreduced colored HOMFLY-PT polynomial $\overline{P}_r(0_1;a,\q)$ for the unknot is
\begin{equation}
    \overline{P}_r(0_1;a,\q)=a^{-\frac{r}{2}}\q^{\frac{r}{2}}\frac{(a;\q)_r}{(\q;\q)_r}.
\end{equation}
We have tabulated the reduced $r$-colored HOMFLY-PT polynomials for some knots in Table\ref{table1}.
\renewcommand{\arraystretch}{2.3}
\begin{table}[htp]
	\begin{center}
		\begin{tabular}{|l|l|}
			\hline
			\textbf{Knots}& \textbf{Colored HOMFLY-PT polynomials} \\
			\hline
			Trefoil knot, $3_1$& $\frac{a^{r}}{\q^{r}}\sum_{k=0}^{r}\qbinom{r}{k}\q^{k(r+1)}\prod_{i=1}^{k}(1-a\q^{(i-2)})$\\
			\hline
			Figure eight knot, $4_1$& $\sum_{0\leq k\leq r}\qbinom{r}{k}a^{k}\q^{(k^2-k)}(a^{-1}\q;q^{-1})_k(a^{-1}\q^{-r};\q^{-1})_k$ \\
			\hline
			Cinquefoil knot, $5_1$ & $\frac{a^{2r}}{\q^{2r}}\sum_{0\leq k_2\leq k_1\leq r}\qbinom{r}{k_1}\qbinom{k_1}{k_2}\q^{(2r+1)(k_1+k_2)-rk_1-k_1k_2}(a\q^{-1};\q)_{k_1}$ \\
			\hline
		\end{tabular}
	\end{center}
\caption{$P_r(K;a,\q)$ for some knots} \label{table1}
\end{table}
In all these examples, we use the following convention for $\q$-Pochhammer and $\q$-binomial symbols:
\begin{equation}
	(x;\q)_r=\prod_{k=0}^{r-1}(1-x\q^k),~~~~~~~~\qbinom{r}{k}=\frac{(\q;\q)_r}{(\q;\q)_k(\q;\q)_{r-k}}.	
\end{equation}
In fact, if one computes these polynomials explicitly, one would realise that these polynomials have integer coefficients. Towards the end of 20th century, attempts to give a topological interpretation for these integer coefficients for Jones polynomial (HOMFLY-PT)  as well as the corresponding colored invariants  for any knot  $K$
\begin{equation}
	J_{r}(K;\q)  = \sum_s  a_s^K \q^s~, ~ \{ a_s^K \} \in \mathbb Z \label{laurent}
\end{equation}
has resulted in developments on homology theories as well as physics explanation.
We will discuss these `homological invariants' and their appearance in string/M-theory in the following section.

\subsection{Knot homologies}
\label{subsectionknothomologies}
We will first review the developments on homological invariants of knots accounting for these integers $a_s^K$ (\ref{laurent}) as  dimension of the vector space $\mathcal H^{K}$ of a homological  theory. Then, we will present the topological string/M-theory approach where these integers count number of BPS states. 
\paragraph{Homological Invariants of Knots:}
The pioneering work of Khovanov\cite{khovanov2000categorification} on bi-graded homology theory led to categorification of  the Jones polynomial. This was extended to colored $\mathfrak{sl2}$  knot homology $\mathcal H^{K;\mathfrak {sl}_2;r}_{i,j}$ \cite{webster2017knot,cooper2012categorification,frenkel2012categorifying} leading to new homological invariants $J_r^{\mathfrak{sl}_2}[K;\q,t]$ which categorifies the colored  Jones polynomial: 
\begin{equation}
	J_r^{\mathfrak{sl}_2}(K;\q,t)=\sum_{i,j} t^j   \q^i {\rm dim} \mathcal {H}^{K;\mathfrak {sl}_2;r}_{i,j}~. \label{sl2homo}
\end{equation}
The subscripts $i$ and $j$ on the colored $\mathfrak {sl}_2$ homology $\mathcal{H}^{K;\mathfrak {sl}_2;r}_{i,j}$ are called the polynomial grading and the homological grading respectively. In fact, the $\q$-graded Euler characteristic of the colored $\mathfrak {sl}_2$ knot homology gives the colored Jones invariant:
\begin{equation} 
	J_{r}(K;\q) = \sum_{i,j} (-1)^j   \q^i {\rm dim} \mathcal H^{K;\mathfrak {sl}_2;r}_{i,j}~,\label{sl2colorhomo}
\end{equation} 
explaining the reasons behind the integers $a_s^K$(\ref {laurent}).
Khovanov and Rozansky \cite{khovanov2004matrix} constructed  $\mathfrak {sl}_N$ homology using matrix factorizations. This  led to the categorification of HOMFLY-PT polynomials of knots.
There has been interesting insight on these homological invariants within topological strings context and  $M$-theory. We will now discuss the essential features from physics approach.\\ \\
{\bf Topological Strings and M-theory:}
The  parallel  developments from topological strings and intersecting branes in $M$-theory \cite{Gopakumar:1998jq,Gopakumar:1998ii,Ooguri:1999bv,Witten:2011zz} interpreted the integers of unreduced HOMFLY-PT  (\ref{laurent})   as counting of BPS states. Invoking topological string duality in the presence of any knot $K$, Ooguri-Vafa conjectured \cite{Ooguri:1999bv} the form for the reformulated invariant as
\begin{equation} 
	f_{R}[K;\q,\lambda=\q^N]= {1 \over (\q^{1/2} - \q^{-1/2})} \sum_{Q,s} N_{R, Q,s}^K \lambda^Q \q^s~,\label{ovconj}
\end{equation}
where $N_{R, Q,s}^K$ are integers and widely known as Labastida-Marino-Ooguri-Vafa (LMOV) invariant. For the defining representation,
\begin{equation}
f_{{\tiny {\yng(1)}}}[K; \q,\lambda] = \overline{P}_{\tiny{\yng (1)}}(K;\lambda,\q)~,\label{ovconjfundamental}
\end{equation}
justifying the series(\ref {laurent}) for HOMLY-PT polynomial.
The integrality structure of $f_{R\neq {\tiny{\yng(1)}}}(K;\q,\lambda)$ (\ref{ovconj}), which can be rewritten in terms of unreduced colored HOMFLY-PT polynomials, was verified for unknot in Ref.\cite{Ooguri:1999bv} and other knots in Refs.\cite{Labastida:2000zp,Ramadevi:2000gq,Labastida:2000yw}. 
In fact, the LMOV integers count the number of D2 branes intersecting D4 branes in the type-IIA string theory on \textit{resolved conifold}$\times \mathbb{R}^{3,1}$ \cite{Ooguri:1999bv}. Note that the $D4$-branes are wrapped on the knot conormal $L_K$ corresponding to Lagrangian cycle on the deformed conifold ($T^*S^3$) which intersects $S^3$ along the knot $K$. 

Further, the relation between the  BPS spectrum and $sl_N$/Khovanov-Rozansky\cite{khovanov2004matrix} knot homology was conjectured within the topological string context by Gukov et al.\cite{Gukov:2004hz} :
\begin{equation}
	N_{\tiny\yng(1), Q,s}^K= \sum_j (-1) ^j  D_{Q,s,j}^K~,\label{gukovkhovrojtopstring}
\end{equation} 
where the integers $D^K_{Q,s,j}$ are referred to as refined BPS invariants. The extra charge/ homological grading $j$ are explainable by the appearance of extra $U(1)$ symmetry in $M$-theory compactified on $\mathbb{R}\times X\times TN$ with the M5 branes on $\mathbb{R}\times L_K\times D$. Here, $X$ refers to Calabi-Yau 3-fold ($CY_3$), $L_K\subset X$ denotes knot conormal, and the disc $D$ is inside Taub-Nut (TN) geometry. The topological string duality and the   dualities of  physical string theories compactified on $CY_3$ implies that the vector space of knot homologies are the Hilbert space of BPS states(see review \cite{Nawata:2015wya} and references therein):
$$\mathcal H^{K} \equiv \mathcal H^{BPS}~.$$

Associating quivers to knots 
\cite{Kucharski:2017ogk,PhysRevD.96.121902} as a conjecture through the relation between $r$-colored HOMFLY-PT polynomial and motivic generating series attracted lots of attention in recent years. This conjecture relates LMOV invariants to certain linear combinations of Donaldson-Thomas (DT) invariants of symmetric quivers, using integer coefficients. Additionally, it is very well known in mathematics literature that DT-invariants for symmetric quivers are integers \cite{efimov2012cohomological}. Hence, the knot-quiver correspondence conjecture also provides an indirect proof for the integrality of LMOV invariants(\ref{ovconj} for $R=$ symmetric representations. We will briefly present this conjecture in the following section.
\subsection{Knots-quivers correspondence}
Knots-quivers correspondence (KQC) conjectured by Kucharski-Reineke-Stosic-Sulkowski \cite{Kucharski:2017ogk} provides a new encoding of HOMFLY-PT invariants of knots  in terms of the representation theory of quivers. Such a correspondence was motivated by studying the supersymmetric quiver quantum mechanics description of BPS states in brane systems describing knots \cite{PhysRevD.96.121902}.

Quivers are denoted as directed graphs with a finite number of vertices connected by oriented edges. For a quiver with $n$ number of vertices, the directed graph is encoded in a $n\times n$ quiver matrix $C$.  The diagonal elements $C_{ii}$  refer to the number of loops at the `$i$'-th vertex, and the off-diagonal elements $C_{ij}$  give the number of oriented edges from the vertex `$i$' to the vertex `$j$'. Hence, the elements in the matrix $C$ are non-negative integers. An example of three nodes symmetric quiver and the corresponding quiver matrix $C$ is,

\renewcommand{\arraystretch}{1} 
\begin{equation}
 \begin{tikzcd}
	&& \bullet \\
	\\
	\\
	\bullet &&&& \bullet
	\arrow[curve={height=12pt}, from=1-3, to=4-1]
	\arrow[curve={height=-12pt}, from=1-3, to=4-5]
	\arrow[curve={height=12pt}, from=4-1, to=1-3]
	\arrow[from=4-1, to=4-1, loop, in=190, out=260, distance=10mm]
	\arrow[from=4-1, to=4-1, loop, in=195, out=255, distance=5mm]
	\arrow[curve={height=12pt}, from=4-1, to=4-5]
	\arrow[curve={height=18pt}, from=4-1, to=4-5]
	\arrow[curve={height=-12pt}, from=4-5, to=1-3]
	\arrow[curve={height=12pt}, from=4-5, to=4-1]
	\arrow[curve={height=18pt}, from=4-5, to=4-1]
	\arrow[from=4-5, to=4-5, loop, in=275, out=355, distance=15mm]
	\arrow[from=4-5, to=4-5, loop, in=280, out=350, distance=10mm]
	\arrow[from=4-5, to=4-5, loop, in=285, out=345, distance=5mm]
\end{tikzcd},
~~~~~~~~~~~~~C=\begin{pmatrix}
    0&1&1\\
    1&2&2\\
    1&2&3
\end{pmatrix}.
\end{equation}

According to the conjecture, at least one quiver graph $Q_K$ is associated with the knot $K$ as elaborated in the Refs.\cite{Kucharski:2017ogk,Stosic:2024egq,Ekholm:2021gyu,Ekholm:2021irc}. 
In the context of KQC, the quivers are {\it symmetric} quivers. That is., $C_{ij}^{(K)}= C_{ji}^{(K)}$. Particularly, for the knot $K$ which obey exponential growth property,
\begin{equation}
    P_r(K;a,\q=1)=\left(P(K;a,\q=1)\right)^r,
\end{equation}
we can write the  generating series for $\Tilde{P}_r(K;a,\q)$, which is related to colored HOMFLY-PT\footnote{$\Tilde{P}_r(K;a,\q)=\frac{P_r(K;a^2,\q^2)}{(\q^2;\q^2)_r}$}, in two equivalent forms:
\begin{eqnarray} {\label{KQ1}}
	P_{Q_K} (x,a,\q) &=& \sum_r \Tilde{P}_r(K;a,\q) x^r \nonumber\\
	~&=&\sum_{\mathbf{d}}(-\q)^{\sum_{i,j}^n d_i C_{ij}^{(K)}d_j} \mathbf{x}^{\mathbf{d}}\prod_{i=1}^n\prod_{j=1}^{d_i}\frac{1}{1-\q^{2j}}\nonumber\\
	~&=&\prod_{\mathbf{d}\neq 0}\prod_{j\in \mathbb{Z}}\prod_{k\geq 0}\left(1-(-1)^j\mathbf{x^d}\q^{j+2k+1}\right)^{-\Omega_{\mathbf{d},j}},\nonumber\\
	\label{DTinv}
\end{eqnarray}
to extract the quiver matrix $C^{(K)}$ as well as the motivic Donaldson-Thomas (DT) invariants
$\Omega_{\mathbf{d},j}$. Here ${\mathbf{d}}\equiv (d_1,d_2, \ldots d_n)\geq 0$ and $\mathbf{x^d}= \prod_{i=1}^n x_i^{d_i}$ with $x_i=x a^{\beta_i} \q^{\alpha_i-1}(-1)^{\gamma_i}$. Note that sets $\{d_i\}$ satisfy the condition $r = d_1 + d_2 + \ldots + d_n$. \textcol{Further, these quiver matrix elements can be negative integers and be made non-negative by shifting all elements of the matrix $C$ by $f$ \cite{Kucharski:2017ogk}. That is,}

\begin{equation}
   \textcol{ C\mapsto C+f\left(\begin{array}{cccc}
    1&1&\cdots&1\\
    1&1&\cdots&1\\
    \vdots&\vdots&\ddots&\vdots\\
    1&1&\cdots&1
\end{array}\right),}\label{shifting-of-quiver-mat}
\end{equation}

\textcol{which changes the quadratic term in quiver generating function (\ref{KQ1}) by $(-\q)^{f\sum_{i,j}d_id_j}.$ This operation can be incorporated in knot polynomials by using the fact that the operation of framing $f\in\mathbb{Z}$ changes the colored HOMFLY-PT polynomial by a factor, which for the symmetric representation $S^r$ takes the form $ \textcol{a^{2fr}\q^{fr(r-1)}.}$}

Moreover, it is worth mentioning here that this correspondence was motivated entirely by empirical evidence: knot data and quiver data are computed separately and shown to coincide \cite{Kucharski:2017ogk}. Motivated by this correspondence, Ekholm and others \cite{Ekholm2020PhysicsAG} provided the geometric and physical meaning to this conjecture. It was shown that the quiver encodes a 3d $\mathcal{N}=2$ theory $T[Q_K]$ whose low energy dyanamics arises on the worldvolume of an M5 brane wrapping the knot conormal. Further, they showed that the spectrum of generalized \textcol{holomorphic} curves on knot conormal $L_K$ is generated by a finite set of basic disks. These disks correspond to the nodes of the quiver $Q_K$ and the linking of their boundaries to the quiver arrows.

Last but certainly not the least, this correspondence also provides a novel categorification to colored HOMFLY-PT polynomials (\ref{laurent}) in the following way. The motivic DT-invariants $\Omega_{\mathbf{d},j}$ admits geometric interpretations as either the intersection Betti numbers of the moduli space of all semi-simple representations of $Q$ with dimension vector $\mathbf{d}$, or as the Chow-Betti numbers of the moduli space of all simple representations of $Q$ with dimension vector $\mathbf{d}$ \cite{meinhardt2019donaldson,franzen2018semistable}. Hence, quiver moduli spaces themselves can be regarded as new higher categorical invariants of knots.

This fascinating relation between knot theory and quiver representation theory was explored for the simplest $(2,2p+1)$ torus knots, twist knots, and knots up to six crossings \cite{Kucharski:2017ogk}. Further, a SageMath program to generate quivers corresponding to rational tangles up to twelve crossings is available \cite{Stosic2018NO2}. 

Our aim was to find an explicit quiver data structure for a family of knots. We succeeded in finding the quiver data for a family of double-twist knots $K(p,-m)$ where $p,m\in\mathbb{Z}$ denote full-twists. In particular, we used the fact that the Alexander polynomial for this family is similar to that of the twist knots. Furthermore, there is a systematic procedure called reverse engineering of Melvin-Morton-Rozansky expansion to derive the motivic series (\ref{DTinv}) form and obtain the corresponding quiver data \cite{Banerjee_2020}.  We use this formalism in chapter \ref{chap:5} to extract the quiver data, $C^{K(p,-m)}$, for a class of double-twist knots.

Till now, we focussed on developments centred around knot polynomials.  We will now move on to discuss 
three-manifolds obtained from Dehn-surgery\cite{Dehnsurgery} of framed knots in $S^3$. Particularly, we will review the three-manifold invariants known as Witten-Reshetikhin-Turaev (WRT) invariant and their categorification in the following section.

\section{Three-Manifold}

A three-manifold is a mathematical space that locally resembles three-dimensional Euclidean space \(\mathbb{R}^3\). The study of three-manifolds is a central topic in low-dimensional topology, which concerns the properties of space preserved under continuous deformations. Some examples of 3-manifolds are \(S^3\), \(T^3 \simeq S^1 \times S^1 \times S^1\), lens space \(L(p,1) \simeq S^3/\mathbb{Z}_p\), and Poincare homology sphere $S_P^3$ \cite{Poincar1904CinquimeC,KIRBY1979113}. 

Closed 3-manifolds can be described using framed links in \( S^3 \). Specifically, there exists a procedure to construct a 3-manifold \( M \) from a framed link \( [\mathcal L,\mathbf f ]\) inside \( S^3 \). This procedure, known as surgery, involves removing a tubular neighbourhood of a knot and then gluing it back after performing a twist. The following theorem ensures that a broad class of 3-manifolds can be obtained through surgery.
\begin{theorem}[Lickorish-Wallace \cite{10.2307/1970373,wallace_1960}]
	Any closed, connected, orientable 3-manifold can be obtained from \( S^3 \) by surgery along some framed link.
\end{theorem}
Witten's approach also gives three-manifold invariant $Z_k^{SU(N)} [M;\q]$ (\ref{chern-simons-partition-function-sun}), called Chern-Simons partition function for manifold $M$, obtained from surgery of framed links on $S^3$ \cite{Witten:1988hf}. Witten-Reshitikhin-Turaev (WRT) invariants $\tau_k^{SU(N)}[M;\q]$ known in the  mathematics literature are proportional to the Chern-Simons partition function:
\begin{equation}
	Z_k^{SU(N)}[M;\q] = {\tau_k^{SU(N)}[M;\q] \over \tau_k^{SU(N)}[S^2 \times S^1;\q]}~.\label{wrtdef}
\end{equation}
We will now concentrate on WRT invariants for the $SU(2)$ gauge group for clarity.
These $SU(2)$  WRT invariants can be written in terms of colored Jones invariants of framed links\cite{Witten:1988hf,Reshetikhin:1991tc,Ramadevi:1999nd,Kaul:2000xe}:
\begin{equation}
\tau_k^{SU(2)}[M; [\mathcal L,\mathbf f];\q] \propto \sum_{r_1,r_2,\ldots =0}^k J_{r_1,r_2,\ldots} ([\mathcal L,\mathbf f];\q) ~\overline J_r[0_1;\q]~.
\label{wrtpropjones}
\end{equation}
{\bf Kirby theorem }\cite{kirby1978calculus} states that surgery of two links $[\mathcal L,\mathbf f]$, $[\mathcal L',\mathbf f']$ related by Kirby moves I \& II will give the same three-manifold $M$. Hence the above algebraic expression must remain unchanged under Kirby moves I \& II to qualify to be a three-manifold invariant.

Similar to the knot polynomials having integer coefficients(\ref{laurent}), one would expect that the three-manifold invariant $\tau_k^{SU(2)}[M;\q]$ should also be a $\q$-series with integer coefficients. Unfortunately, the three-manifold invariant (\ref{wrtpropjones}) is not a $\q$ series for a general three-manifold $M$. However, Lawrence and Zagier \cite{lawrence1999modular} found the $q$-series from WRT invariant of Poincare homology sphere. 

\begin{theorem}[Lawrence-Zagier, 99]
	Let $S^3_P$ be the Poincare homology sphere and $\q=e^{\frac{2\pi i}{k}}$. Then, there is an invariant $\widehat{Z}_k^{SU(2)}[S_P^3;q]$ in the analytic continued $q$ plane such that 
	\begin{equation}
		\tau^{SU(2)}_k[S_P^3;\q]=\lim_{q\rightarrow \q}\left(\frac{\widehat{Z}_k^{SU(2)}[S_P^3;q]}{2(q^{1/2}-q^{-1/2})}\right),
	\end{equation}
	where	
\begin{align}
		\widehat{Z}_k^{SU(2)}[S_P^3;q]=&q^{-3/2}\sum_{n\geq 0}\frac{(-1)^nq^{\frac{n(3n-1)}{2}}}{\prod_{1\leq j\leq n}(1-q^{n+j})}\\
		=&q^{-3/2}\left(1-q-q^3-q^7+q^8+q^{14}+\ldots\right).\nonumber
	\end{align}
\end{theorem}
 This result was subsequently generalized to Seifert homology spheres \cite{hikami2005quantum,hikami2005quantum1,hikami2006quantum2} and the Seifert manifolds associated with the ADE singularities \cite{hikami2006quantum}. 
Even though the WRT (\ref{wrtpropjones}) invariant for a general three-manifold $M$ can be written as a summation over all colors, taking 
$\q \rightarrow q$, which is not a root of unity, will imply $k\notin \mathbb Z$. This will make the upper bound in the summation (\ref{wrtpropjones}) as $ln~q$ hindering the computation of WRT invariant as a function of $q$. Hence, we have no clue how to obtain $q$-series with integer coefficients from $ \tau_k^{SU(2)}[M;\q]$. 

Taking insights from 
3d-3d correspondence\cite{Dimofte:2011py,Dimofte:2011ju,Dimofte:2010tz} and WRT invariant for Lens space $L(p,1)$ \cite{Gukov:2016gkn},
Gukov-Pei-Putrov-Vafa (GPPV) \cite{Gukov:2017kmk} conjectured a new three-manifold invariant $\widehat{Z}$ as $q$-series.

In this paper \cite{Gukov:2017kmk}, the exploration of $\widehat{Z}$  was performed through the analytic continuation of the WRT invariant defined for plumbed three-manifold $M(\Gamma)$ (Table \ref{Table:plumbed-3-mfd}). This investigation led to the definition of $\widehat{Z}$ for negative semidefinite plumbed 3-manifolds. Subsequent research efforts, as outlined in works\cite{2020higher,Ferrari:2020avq,chae2021towards} extended the study of $\widehat{Z}$ by examining the GPPV conjecture in the contexts of $SU(N)$, $SU(2|1)$, and $OSp(2|2)$ gauge groups.

\begin{table}[ht]
    \centering
    \begin{tabular}{|c|c|c|c|}
        \hline 
        Link, $(\mathcal{L})$ & Plumbing graph, $(\Gamma)$ & Linking matrix, $B$ & Plumbed three- \\
        & & & manifold, $M(\Gamma)$\\
        \hline
         \includegraphics[scale=0.1]{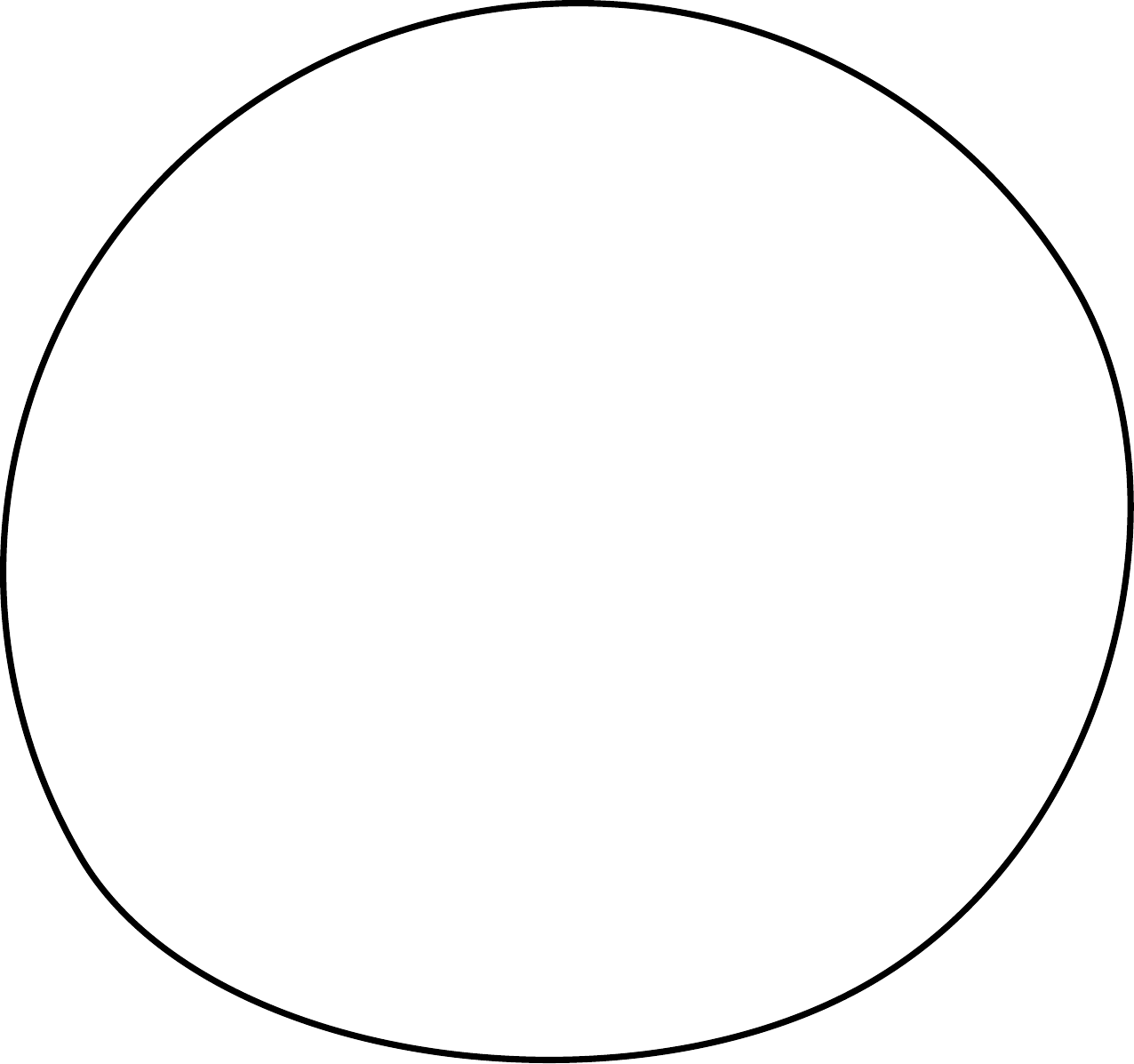}& \includegraphics[scale=0.4]{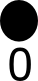}& \raisebox{0.3cm}{$\left(0\right)$} & \raisebox{0.5cm}{$S^2\times S^1$}\\ 
         \hline
         \includegraphics[scale=0.2]{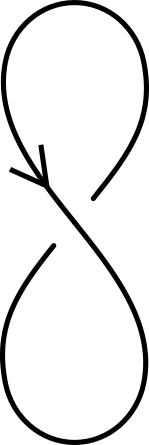}& \includegraphics[scale=0.4]{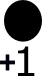}& \raisebox{0.3cm}{$\left(1\right)$} & \raisebox{0.5cm}{$S^3$}\\
         \hline
         \includegraphics[scale=0.4]{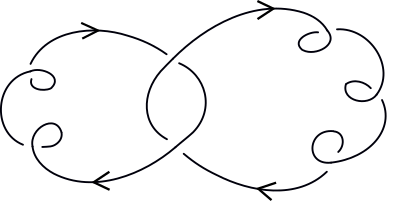}& \includegraphics[scale=0.3]{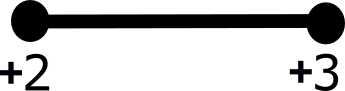}& \raisebox{0.6cm}{$\left(\begin{array}{cc}
             2 & 1  \\
             1 & 3
         \end{array}\right)$} & \raisebox{0.5cm}{$L(5,1)$}\\
         \hline
    \end{tabular}
    \caption{Some simple examples of plumbed three-manifolds. We use the standard convention in determining the framing numbers \ie, ~\raisebox{-12pt}[0.8\baselineskip]{\includegraphics[width=0.05\textwidth]{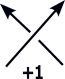}}~ and~ \raisebox{-12pt}[0.8\baselineskip]{\includegraphics[width=0.05\textwidth]{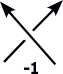}}.}
    \label{Table:plumbed-3-mfd}
\end{table}

Furthermore, it is essential to recognize that the WRT invariant for gauge group $G$, $\tau^{G}_{k'}[M(\Gamma);\q]$, is defined at a root of unity denoted as $\q\left(=\exp\left(\frac{2\pi i}{k'}\right)\right)$, where $k'$ signifies the renormalized Chern-Simons level. In contrast, the variable $q$ within $\widehat{Z}^{G}_b[M(\Gamma);q]$ can be any complex number. Interestingly, as $q$ approaches $\q$, the relationship between these two invariants expressed through a $\mathcal S$-transformation matrix\cite{Gukov:2016gkn,Gukov:2017kmk}:
\begin{equation}
	\tau^{G}_{k'}[M(\Gamma);\q]\cong \sum_{\scriptsize \begin{array}{c}a\in (P^\vee)^L/B(P^\vee)^L,\\b\in (Q^L+\delta)/BQ^L\end{array}}\mathcal S_{ab}\widehat{Z}^G_b[M(\Gamma);q]\Big\vert_{q\rightarrow\q},
\end{equation}
where $P^\vee$ is co-weight lattice associated to group $G$, $Q$ is root lattice, $L$ is the number of vertices in a plumbing graph $\Gamma$, $\delta=(\delta_1,\ldots,\delta_L)$, $\delta_v=(2-\text{deg} v)\rho$, $\rho$ is Weyl vector, and $B$ is adjacency/linking matrix associated to $\Gamma$. Further, $\widehat{Z}_b^G[M(\Gamma);q]$ admits the following physical categorification:
\begin{equation}
	\widehat{Z}_b^G[M(\Gamma);q]=\sum_{\substack{i\in \mathbb{Z}+\Delta_b\\j\in \mathbb{Z}}}q^i(-1)^j\text{dim}\;\mathcal{H}_{b,G}^{i,j}.
\end{equation}
In this equation, $\mathcal{H}_{b,G}^{i,j}$ corresponds to the BPS sector of the Hilbert space of $T[M(\Gamma);G]$ and $\Delta_b$ is a rational number specific to a 3-manifold $M(\Gamma)$. These insights provide a promising avenue for addressing the long-standing categorification problem associated with the WRT invariant.

The colored framed link invariants for groups $SO(3)$ and $OSp(1|2)$ \cite{Ennes:1997kx} can be obtained from colored Jones invariant by a change of variable. In the light of GPPV conjecture, we wanted to investigate
$\widehat{Z}$ for both $SO(3)$ and $OSp(1|2)$. This was the theme of our published paper \cite{chauhan2022hat} which is discussed in chapter \ref{chap:3}. We observed an interesting feature that $\widehat Z$ is same for both $SU(2)$ and $SO(3)$ even though the corresponding WRT invariants are different. Moreover, it is known that the Lie algebras of \(SU(2)\) and \(SO(3)\) are isomorphic, and these groups are also Langlands dual to each other. Therefore, we aimed to investigate the dependence of the \(\widehat{Z}\)-invariant on the Lie algebra or on the Langlands duality of groups. In our subsequent paper \cite{sachinramagppv}, we explored the \(\widehat{Z}\)-invariant for factor groups formed by quotienting \(SU(N)\) by a discrete group \(\mathbb{Z}_m\), where \(m\) is a divisor of \(N\). This investigation led us to conclude that \(\widehat{Z}\) depends only on the Lie algebra and not on the Lie group. We will present this work in chapter \ref{chap:4}.

In the following section, we will briefly present the plan of the thesis giving the salient features discussed in each chapter.
\section{Brief Highlights of the Thesis}
We have focused on two broad themes in this thesis:\\
$\bullet$ Gukov-Pei-Putrov-Vafa conjecture for $SO(3)$, $OSp(1|2)$, and $SU(N)/\mathbb{Z}_m$ groups.\\
$\bullet$ Knots-quivers correspondence for double twist knots.\\
The  chapter-wise highlights of the thesis are as follows:
\begin{itemize}[topsep=4pt, partopsep=4pt, itemsep=4pt, parsep=4pt]
    \item In Chapter $\ref{chap:3}$, we first give a brief introduction to knot, link, and three-manifold invariants. This includes Chern-Simons theory and colored link invariants with explicit results for $SU(2)$ gauge group. We also indicate how colored \(SO(3)\) and \(OSp(1|2)\) link invariants can be obtained from the colored \(SU(2)\) polynomials. This step would be crucial in setting up the WRT invariant for \(SO(3)\) and \(OSp(1|2)\) groups. Then, we summarise the developments of the homological invariants. 

    Following this, we briefly review the $\widehat{Z}$-invariant for $SU(2)$ group for the negative definite plumbed three-manifolds. Finally, we evaluate the \(\widehat{Z}\)-invariant for the \(SO(3)\) and \(OSp(1|2)\) groups by following the similar steps as were done for $\widehat{Z}^{SU(2)}$.

    As a result of the work done in this chapter, we discovered a surprising relationship between \(\widehat{Z}^{SU(2)}\) and \(\widehat{Z}^{OSp(1|2)}\). Additionally, we found that \(\widehat{Z}^{SU(2)} = \widehat{Z}^{SO(3)}\), which prompted us to investigate the \(\widehat{Z}\)-invariant for the \(SU(N)/\mathbb{Z}_m\) group. Consequently, we explore this topic in the next chapter. 
	\item In Chapter \ref{chap:4}, we begin by giving a quick recapitulation of $\widehat{Z}$-invariant for $SU(N)$ group. Then we establish the WRT invariant for the \( SU(N)/\mathbb{Z}_m \) group. This process involves determining the Chern-Simons level for the \( SU(N)/\mathbb{Z}_m \) gauge group. Additionally, we must consider the representations of the \( SU(N)/\mathbb{Z}_m \) group.

    Subsequently, we conduct the GPPV analysis on the \( SU(N)/\mathbb{Z}_m \) WRT invariant. Our findings indicate that the \(\widehat{Z}\)-invariant is independent of \( m \). The dependence on \( m \) manifests as an overall factor through the sublattice \( P' \) and the Chern-Simons level \( k' \). We provide explicit examples for sublattice $P'$ and Chern-Simons level $k'$ for some non-simply connected groups $SU(N)/\mathbb{Z}_m$ in appendix \ref{appendB}.
	\item Chapter \ref{chap:5} addresses the second major topic of this thesis: the knots-quivers correspondence. This chapter begins with a concise introduction to the knots-quivers correspondence. Following this, we provide a succinct review of the reverse engineering of the Melvin-Morton-Rozansky (MMR) expansion.

    Subsequently, we offer a brief introduction to double-twist knots and their colored HOMFLY-PT polynomials. Finally, we present our findings on the knots-quivers correspondence for double-twist knots, derived through the reverse engineering of the MMR expansion. We also validate these results with several examples.
    \item Finally, in Chapter \ref{chap:6}, we conclude by summarizing the results of this thesis. We also outline some open problems and suggest directions for future research.
 \end{itemize}
\section{\underline{Notation Guide}}

\renewcommand{\arraystretch}{0.8} 
\begin{tabular}{r l}
	$M$:& 3-manifold\\
	$G$:& Gauge group\\
	$G_{\mathbb{C}}$:& Complex gauge group\\
	$\mathfrak{g}$:& Lie algebra\\
	$P$:& Weight lattice\\
    $P^\vee$:& Co-weight lattice\\
	$P_+$:& Cone of dominant integer weights\\
	$\Lambda_{i}$:& Fundamental weight vector where $i=1,2,\ldots,r$\\
	$Q$:& Root lattice\\
	$P'$:& Intermediate lattice between root and weight lattice\\
	$(P')^\bullet$:& Dual of lattice $P'$\\
	$(\lambda,\mu)$:& Denotes the inner product between any two weight vectors, $\lambda$ and $\mu$\\
	$\Gamma$:& Plumbing graph of tree type\\
    $L$:& Number of vertices in a plumbing graph $\Gamma$\\
	$B$:& Linking matrix associated to plumbing graph $\Gamma$\\ 
	$b_{\pm}$:& Number of positive and negative eigenvalues of $B$\\
    $\sigma$:& Signature of linking matrix $B$ \ie~ $\sigma=b_+-b_-$\\
    $W$:& Weyl group\\
    $|W|$:& Order of the Weyl group\\
    $\omega_i$:& $i^{\text{th}}$ element of the Weyl group\\
	$\ell(\omega)$:& Length of Weyl group element $\omega$\\
	$\rho$:& Weyl vector\\
	$M(\Gamma)$:& Plumbed 3-manifold\\
	$k'$:& Renormalized Chern-Simons level\\
	$k$:& Bare Chern-Simons level\\
	$\q$:& Root of unity, $\exp(\frac{2\pi i}{k'})$\\
	$q$:& An arbitrary complex number inside the unit circle\\
	$\text{deg }v$:& Denotes the degree of vertex $v$ in a plumbing graph $\Gamma$\\ 
	$\tau^{G}_{k'}[M(\Gamma);\q]$:& WRT invariant for gauge group $G$\\ 
	$Z^G_{k'}[M;\q]$:& Chern-Simons partition function which is related to\\
	 &WRT invariant as $Z_{k'}^G[M;\q] = {\tau_{k'}^G[M;\q] \over \tau_{k'}^G[S^2 \times S^1;\q]}$\\
	$\hat{Z}^{G}_b[M(\Gamma);q]$:& $\hat{Z}$-invariant labelled by index $b$ for gauge group $G$\\
	$P_r(K;a,q)$:& Reduced $r$-colored HOMFLY-PT polynomial\\
	$\overline{P}_r(K;a,q)$:& Unreduced $r$-colored HOMFLY-PT polynomial
  \end{tabular}
  
   \begin{tabular}{r l}
    $\tilde{P}_r(K;a,q)$:& Intermediate normalised $r$-colored HOMFLY-PT polynomial, $\frac{P_r(K;a,q)}{(q^2;q^2)_r}$\\
	$K$:& Knot\\
	$C^{(K)}$:& Quiver matrix for knot $K$\\
	$\Omega_{\mathbf{d},j}$:& Motivic Donaldson-Thomas invariants where $\mathbf{d}=(d_1,d_2,\ldots,d_n)$\\
	$J_r(K;q)$:& Reduced $r$-colored Jones polynomial of knot $K$\\
    $\overline{J}_r(K;q)$:& Unreduced $r$-colored Jones polynomial of knot $K$\\
	$K(p,m)$:& Double twist knot with full twists $p$ and $m$\\
    $\mathcal{L}_m$:& $m$-component link\\
    $S,T$:& Modular matrices\\
    $\mathcal{S}$:& S-transformation relating WRT and $\widehat{Z}$ invariants in GPPV conjecture
\end{tabular}
\chapter{$\widehat{Z}$-Invariant for $SO(3)$ and $OSp(1|2)$ Groups}
\label{chap:3}

In this chapter, we will present the motivations and necessary review to discuss the WRT and $\widehat{Z}$ invariants for $SO(3)$ and $OSp(1|2)$ groups.
Further, as discussed in the previous introductory chapter, we will now recapitulate the essential points for clarity and completeness. This will provide continuity for understanding the rest of the chapter.

\textbf{\underline{Recapitulation}}

Knot theory has attracted attention from both mathematicians and physicists during the last 40 years. The seminal work of Witten\cite{Witten:1988hf} giving a three-dimensional definition for Jones polynomials of knots and links, using $SU(2)$ Chern-Simons theory on $S^3$,  triggered  a tower of new colored link invariants. Such new invariants are given by expectation value of  Wilson loops  carrying higher dimensional representation $R \in G$  in Chern-Simons theory where $G$ denotes gauge group. These link invariants  are in variable $\q$ which depends on the rank of the gauge group $ G$ and the Chern-Simons coupling constant $k \in \mathbb Z$ \Big(For eg: when $ G=SU(N)$ then $\q=\text{exp}\left(\frac{2\pi i}{k+N}\right)$\Big). Witten's approach also gives three-manifold invariant $Z_k^{ G} [M;\q]$ (\ref{chern-simons-partition-function-sun}), called Chern-Simons partition function for manifold $M$, obtained from surgery of framed links on $S^3$(Lickorish-Wallace theorem\cite{10.2307/1970373,wallace_1960}). Witten-Reshitikhin-Turaev (WRT)  invariants $\tau_k^{ G}[M;\q]$ known in the  mathematics literature are proportional to the Chern-Simons partition function (\ref{wrtdef}). These WRT invariants can be written in terms of the colored invariants of framed links\cite{10.2307/1970373,wallace_1960,Kaul:2000xe,Ramadevi:1999nd}.

It was puzzling observation that the reduced colored knot polynomials appear as  Laurent series with integer coefficients (\ref{laurent}). There must be an underlying topological interpretation of such integer coefficients. This question was answered both from mathematics and physics perspective.  Initial work of  Khovanov\cite{khovanov2000categorification} titled `categorification' followed by other papers on bi-graded homology theory including Khovanov-Rozansky homology led to new homological invariants (\ref{sl2homo},\ref{sl2colorhomo}). Thus the integer coefficients of the colored knot polynomials are interpreted as the dimensions of vector space of homological theory. From topological strings and intersecting branes\cite{Ooguri:1999bv,Gopakumar:1998ii,Gopakumar:1998jq}, the integers of HOMFLY-PT polynomials are interpretable as counting of BPS states (\ref{ovconj},\ref{ovconjfundamental}).  Further the  connections to knot homologies within topological string context was initiated in \cite{Gukov:2004hz} resulting in concrete predictions of homological invariants (\ref{gukovkhovrojtopstring}) for some knots (see review \cite{Nawata:2015wya} and references therein). Such a  physics approach involving brane set up in $M$-theory\cite{Gukov:2017kmk,Gukov:2016gkn,Mikhaylov:2014aoa,Ferrari:2020avq} suggests the plausibility of categorification of WRT invariants $\tau_k^{ G}[M;\q]$ for three-manifolds.  However,  the WRT invariants for simple three-manifolds are not a Laurent series with integer coefficients.   

\textbf{\underline{Categorification of three-manifold invariants}}

The detailed discussion on $U(N)$ Chern-Simons partition function on Lens space $L(p,1)\equiv S^3/\mathbb Z_p$ (see section 6 of \cite{Gukov:2016gkn}) shows a  basis transformation $Z_k^{ G}[M;\q] \underrightarrow{~~\mathcal S~~} {\widehat Z}^{ G}[M;q]$ so that  $\widehat Z$  are $q$-series(where variable $q$ is an arbitrary complex number inside a unit disk) with integer coefficients (GPPV conjecture\cite{Gukov:2017kmk}). These $\widehat Z$  are called the homological blocks of WRT invariants of three-manifolds $M$. Physically, the new three-manifold invariants  $\widehat Z^{ G}[M;q]$ is the partition function $Z_{T^{ G}[M]}[D^2 \times S^1]$ for simple Lie groups. Here $T^{ G}[M]$ denote the effective 3d $\mathcal N=2$ theory on $D^2 \times S^1$ obtained by reducing 6d $(2,0)$ theory (describing dynamics of coincident $M5$ branes) on $M$. 

For a class of negative definite plumbed three-manifolds as well as link complements \cite{Gukov:2017kmk,Gukov:2019mnk,park2020higher,Chung:2018rea}, $\widehat Z^{SU(N)}$ has been calculated.  Further, $\widehat Z$ invariants for super unitary group $SU(n\vert m)$ supergroup with explicit $q$-series for  $SU(2\vert 1)$ is presented in \cite{Ferrari:2020avq}. Generalisation to  orthosymplectic supergroup $OSp(2|2n)$ with explicit $q$-series for $Osp(2|2)$\cite{chae2021towards} motivates us to look at $\widehat Z$ for other gauge groups.

Our goal in this chapter is to extract $\widehat Z$ for the simplest orthogonal group $SO(3)$ and the simplest odd orthosymplectic supergroup $OSp(1|2)$.  We take the route of relating $SU(2)$ colored link invariants to the link invariants for these  two groups to obtain $\widehat Z$ invariants. 

The chapter is organised as follows. In section 2, we will review the developments on the invariants of knots, links and three-manifolds. We will first briefly present Chern-Simons theory and colored link invariants with explicit results for $SU(2)$ gauge group and indicate how colored $SO(3)$ and $OSp(1|2)$ link invariants can be obtained from the colored $SU(2)$ polynomials. Then, we will summarise the developments of the homological invariants. In section \ref{reviewsu2}, we  briefly  review 
$\widehat Z$-series invariant for $SU(2)$ group for the negative definite plumbed three-manifolds. This will serve as a warmup to extend  to $SO(3)$ and $OSp(1|2)$ group which we will present in section \ref{zhatnew}. We summarize  the results in the concluding section \ref{conclusionchap2}.

\section{Knots, Links and Three-manifold Invariants}
In this section, we will briefly summarise new invariants in knot theory from the physics approach as well as from the mathematics approach.
\subsection {Chern-Simons Field Theory Invariants}
Chern-Simons theory based on gauge group $ G$ is a Schwarz type topological field theory which provides a natural framework for study of knots, links and three-manifolds $M$. Chern-Simons action $S_{CS}^{{G}}(A)$ is explicitly metric independent:
\begin{equation}
	S_{CS}^{{G}}(A)=\frac{k}{4\pi}\int_M Tr\left(A\wedge dA+\frac{2}{3}A\wedge A\wedge A\right)~.
\end{equation}
Here, $A$ is the matrix valued gauge connection based on gauge group $ G$ and $k\in \mathbb{Z}$ is the coupling constant. In chapter \ref{chap:1}, we focused on knot and link invariants for $SU(N)$ group (\ref{expectval},\ref{linkinvariant}). For any gauge group $G$, the invariant for a link $\mathcal{L}_m$ is
\begin{multline} 
	V_{R_1,R_2, \ldots R_m}^{ G} [\mathcal L_m;\q]= \langle W_{R_1,R_2,\ldots R_m}[\mathcal L_m]\rangle= 
	{\int {\mathcal D}A \exp(iS_{CS})  \overbrace{P\left(\prod_i  \Tr_{R_i} exp \oint_{K_i} A\right)}^{W_{R_1,R_2,\ldots R_m}[\mathcal L_m]} \over {\underbrace{\int {\mathcal D} A \exp(iS_{CS})}_{Z^{ G}_k[M;\q]}}}~,
\end{multline}
where $K_i$'s denote the component knots of link $\mathcal L_m$ carrying representations $R_i$'s of gauge group $ G$ and $Z^{ G}_k[M;\q]$ defines the Chern-Simons partition function encoding the topology of the three-manifold $M$.

Exploiting the connection between Chern-Simons theory, based on  group $ G$, and the corresponding Wess-Zumino-Witten (WZW) conformal field theory with the affine Lie algebra symmetry $\mathfrak g_k$,  the  invariants of these links embedded in a three-sphere $M=S^3$ can be explicitly written in variable $\q$ :
\begin{equation} 
	\q= \exp\left({2\pi i \over k+C_v}\right)~,
\end{equation}
which depends on the coupling constant $k$ and the dual Coxeter number $C_v$ of the group $ G$. These link invariants  include the well-known polynomials in the knot theory literature. 
\renewcommand{\arraystretch}{1.2}
\begin{table}[htp]
	\begin{center}
		\begin{tabular}{|c|c|c|c|} \hline
		$ G$&$R$& Link invariant\\
		\hline
		$SU(2)$ &$\yng(1)$ & Jones, $\overline{J}(\q)$ \\
        \hline
		$SU(N)$ &$\yng(1)$ & HOMFLY-PT, $\overline{P}(a,\q)$ \\
        \hline
		$SO(N)$ & defining & Kauffman \\
		\hline
	\end{tabular}
    \end{center}\caption{Knot invariants corresponding to different gauge groups in Chern-Simons theory.}
\end{table}
\vspace*{-\baselineskip}
\subsubsection{Link Invariants}
\label{sec2.1.1}
\noindent
As indicated in the above table, unreduced Jones polynomial corresponds to the fundamental representation $R=\tiny {\yng(1)}\equiv 1 \in SU(2)$ placed on all the component knots:
\begin{equation}
	V_{1,1,1,\ldots 1}^{SU(2)} [\mathcal L_m;\q] \equiv \overline{J}\left[\mathcal L_m; \q=\exp\left({2\pi i \over k+2}\right)\right]~.
\end{equation}
Higher dimensional representations placed on the component knots $R_i =\underbrace{\tiny {\yng(4)}}_{n_i}\equiv n_i\in SU(2)$  are the colored Jones invariants:
\begin{equation}
	V_{n_1,n_2,n_3,\ldots n_m}^{SU(2)} [\mathcal L_m;\q] \equiv \overline{J}_{n_1,n_2,\ldots n_m}\left[\mathcal L_m; \q=\exp\left({2\pi i \over k+2}\right)\right]~,
\end{equation} 
and the invariants with these representations belonging to $SU(N)$ ($(SO(N)$)  are known as colored HOMFLY-PT (colored Kauffman)  invariants. For clarity, we will restrict to $SU(2)$ group to write the invariants explicitly in terms of $\q$ variable. 

We work with the following unknot ($\textcol{0\bullet}$) normalisation:
\begin{equation} 
	\overline{J}_{n} [ \textcol{0\bullet};\q] = {\rm dim}_\q \underbrace{\yng(4)}_n={ \q^{(n+1)\over 2 } - \q^{-{(n+1) \over 2}} \over \q^{1\over 2} - \q^{-{1\over 2}}}= {\sin({ \pi (n+1) \over k+2})\over \sin({ \pi  \over k+2})}=
	{S_{0 n}\over S_{00}},
\end{equation}
where ${\rm dim}_\q \underbrace{\yng(3)}_n $ denotes quantum dimension of the representation $\underbrace{\yng(3)}_n$ and $S_{n_1 n_2}$ are the modular transformation matrix elements of the $\mathfrak {su}(2)_k$ WZW conformal field theory whose action on the characters is $\chi_{n_1} (\tau)  ~~\underrightarrow{~~~S~~~}~~ \chi_{n_2} \left(-{1\over \tau}\right)~,$ where $\tau$ denotes the modular parameter. 

\begin{figure}[ht]
	\centering
	\includegraphics[scale=0.25]{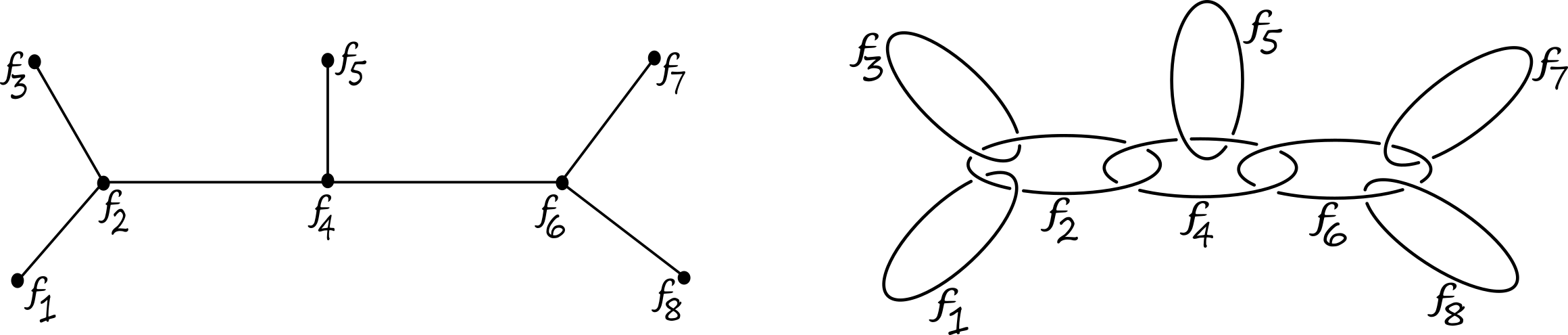}
	\caption{An example of a plumbing graph $\Gamma$ (left) and the corresponding link $\CL(\Gamma)$ of framed unknots in $S^3$ (right).}
	\label{fig:plumbing-example}
\end{figure}

For framed unknots with framing number $f$, the invariant will be 
\begin{equation} 
	\overline{J}_{n} [ \textcol{f\bullet};\q] = \q^{f \left({(n+1)^2-1 \over 4}\right)} { \q^{(n+1)\over 2 } - \q^{-{(n+1) \over 2}} \over \q^{1\over 2} - \q^{-{1\over 2}}}\propto (T_{nn})^f	{S_{0 n}\over S_{00}},
\end{equation}
where the action of the modular transformation matrix $T$ on characters is \\$ \chi_n(\tau) ~~\underrightarrow{~~~T~~~~} \chi_n(\tau+1)~.$
The colored Jones invariant for the Hopf link  can also be written in terms of $S$ matrix:
\begin{equation} 
	\overline{J}_{n_1, n_2} [H;\q] =  \left({\q^{\frac{(n_1+1)(n_2+1)}{2}} - \q^{-\frac{(n_1+1)(n_2 +1)}{2}}\over \q^{\frac{1}{2}} - \q^{-\frac{1}{2}}}\right)= 
	%{\sin({ \pi (n_1+1) (n_2+1)\over k+2})\over \sin({ \pi  \over k+2})}=
	{S_{n_1 n_2}\over S_{00}}.
\end{equation}
The invariant for a  framed Hopf link $H(f_1,f_2)$, with framing numbers $f_1$ and $f_2$ on the two component knots, in terms of $T$ and $S$ matrices is 
\begin{equation} 
	\overline{J}_{n_1, n_2} [H(f_1,f_2);\q] \propto   (T_{n_1 n_1})^{f_1} (T_{n_2 n_2})^{f_2}	{S_{n_1 n_2}\over S_{00}}~ \label {hopf}~.
\end{equation}
We will look at a class of links obtained as a connected sum of framed Hopf links. For instance,  the invariant for the connected sum of two framed Hopf links $H(f_1,f_2) \#H(f_2,f_3)$ will be
\begin{eqnarray}
	\overline{J}_{n_1,n_2,n_3} [H(f_1,f_2)\# H(f_2,f_3);\q]& \propto&
	{ \prod_{i=1}^3T_{n_i n_i}^{f_i}} {S_{n_1 n_2}\over S_{00}} {S_{n_2 n_3}\over S_{n_2 0}} \label {conn}\\
	& =& {\prod_{i=1}^2\overline{J}_{n_i, n_{i+1}} [H(f_i,f_{i+1});\q]  \over \overline{J}_{n_2} [\textcol{0\bullet};\q]}~.\nonumber~
\end{eqnarray}
Such a connected sum of two framed  Hopf links, which is a 3-component link,  can be denoted as a  linear  graph 
\begin{figure}[ht]
    \centering
    \includegraphics[scale=0.8]{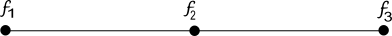}
\end{figure}
with three vertices labeled by the framing numbers and the edges connecting the adjacent vertices. These are known as `plumbing graphs'.  Another  plumbing graph $\Gamma$  with 8 vertices denoting the link $\mathcal L(\Gamma)$  (the connected sum of many framed Hopf links)  is illustrated in Figure \ref{fig:plumbing-example}. The colored  invariant for these links $\mathcal L(\Gamma)$ can be written  in terms of $S$ and $T$ matrices.
For a general  $m$ vertex plumbing graph with vertices $v_1,v_2, \ldots v_m \in V$  labelled by framing numbers $f_1,f_2, \ldots f_m$,  there can be one or more edges connecting a vertex $v$ with the other vertices. The degree of any vertex $v$ ($\text{deg}(v)$) is equal to the total number of edges intersecting $v$.  For the graph in Figure \ref{fig:plumbing-example},   $\text{deg}(2)=\text{deg}(4)=\text{deg}(6)=3$. The colored Jones' invariant for any plumbing graph $\Gamma$ is
\begin{equation}
	\overline{J}_{n_1, n_2, \ldots n_m}[\mathcal L;\q] \propto {1 \over S_{00}} \prod_{i=1}^m \{(T_{n_i n_i})^{f_i} (S_{0 n_i})^{1-\text{deg}(v_i)}\}
	\prod_{(v_1,v_2)\;\in\;\text{Edges} } (S_{n_{v_1} n_{v_2}})~. \label{conn1}
\end{equation}
Even though we have presented the colored Jones
invariants (\ref{hopf}, \ref{conn}, \ref{conn1}),  the formal expression of these link invariants in terms of $S$ and $T$ matrices are applicable for any arbitrary gauge group $G$.

$\bullet$ \underline{\bf{$SO(3)$ and $OSp(1|2)$ Link invariants}}

\noindent
Using group theory arguments, it is possible to relate colored link invariants between different groups. For instance,  the representations of the $SO(3)$  can be identified with a subset of  $SU(2)$ representations. As a consequence, the $SO(3)$ link invariants  can be related to the colored Jones invariants as follows:
\begin{equation}
	V_{n_1,n_2,n_3,\ldots n_m}^{SO(3)} \left[\mathcal L_m;Q=\exp\left({2\pi i \over K+1}\right)\right] = J_{2n_1,2n_2,\ldots 2n_m}[\mathcal L_m; \q]{\big\vert}_{\q^2=Q}~,\label{so3}
\end{equation}
where the level $K$ of the affine $\mathfrak {so}(3)_K$  Lie algebra must be an even integer ($K\in 2 \mathbb Z$). 

Similarly, the  representations of the orthosymplectic supergroup $OSp(1|2)$ can be related to the representations of the $SU(2)$ group  from the  study of $\mathfrak{osp}(1|2)_{\hat K}$ WZW conformal field theory  and the link invariants $V_{n_1,n_2,n_3,\ldots n_m}^{OSp(1|2)} \left[\mathcal L_m;\hat Q=\exp\left({2\pi i \over 2{\hat K}+3}\right)\right]$ \cite{Ennes:1997kx}.  
Particularly, there is a precise identification of the polynomial variable $\hat Q$ to $SU(2)$ variable $\q$. Further, the fusion rules of the primary fields of  $\mathfrak{osp}(1|2)_{\hat K}$ WZW conformal field theory can be compared to integer spin primary fields of the $\mathfrak{su}(2)_k$.  Particularly, the $\hat S$  and $\hat T$-matrices of $\mathfrak{osp}(1|2)_{\hat K}$ :
\begin{eqnarray}
\hat S_{n_1 n_2} &=& \sqrt{4 \over 2\hat K+3} (-1)^{n_1+n_2} \cos\left[{ (2n_1+1) (2n_2+1) \over 2(2\hat K+3)} \pi \right]~~,\\
\hat T_{n_1,n_2}&\propto&  \delta_{n_1,n_2} {\hat Q}^{[\frac{(2n_1+1)^2-1]}{4}}~~~,
\end{eqnarray}
are related  to the $S$ and $T$ matrices of  $\mathfrak{su}(2)_k$ in the following way:
\begin{equation} 
\hat S_{n_1,n_2} = S_{2n_1,2n_2}{\big\vert}_{\q = -\hat Q}~;~
\hat T_{n_1,n_1} = T_{2n_1,2n_1}{\big\vert}_{\q = -\hat Q}
\end{equation}
Using these relations, we can show that the   $OSp(1|2)$ colored invariant match the colored Jones invariant for any arbitrary link $\mathcal L_m$ in the following way:
\begin{equation}
	V_{n_1,n_2,n_3,\ldots n_m}^{OSp(1|2)} \left[\mathcal L_m;\hat Q=\exp\left({2\pi i \over 2{\hat K}+3}\right)\right] = \epsilon \overline{J}_{2n_1,2n_2,\ldots 2n_m}[\mathcal L_m; \q]{\big\vert}_{\q=-\hat Q}~,\label{osp}
\end{equation}
where $\epsilon$ could be $\pm 1$ depending on the link $\mathcal L$ and the representations $n_i$'s.  For example, the colored $OSp(1|2)$ invariant for framed Hopf link is 
\begin{eqnarray} 
	V_{n_1, n_2 } ^{OSp(1|2)} [H(f_1,f_2);\hat Q] &=& 
	{\hat Q}^{\frac{f_1((2n_1+1)^2-1)}{4}}  {\hat Q}^{\frac{f_2((2n_1+1)^2-1)}{4}}(-1)^{(n_1+n_2)}\times\nonumber\\
	~&~&\left({{\hat Q}^{\frac{(2n_1+1)(2n_2+1)}{2}} + {\hat Q}^{-\frac{(2n_1+1)(2n_2 +1)}{2}}  \over {\hat Q}^{\frac{1}{2}} +{\hat Q}^{-\frac{1}{2}}}\right)\\
	&=& \overline{J}_{2n_1, 2n_2} [H(f_1,f_2),-\hat Q]~.\nonumber
\end{eqnarray}
In fact, for any link $\mathcal L(\Gamma)$ denoted  by the graph $\Gamma$, the invariants will be 
\begin{eqnarray}
	V_{n_1, n_2,\ldots n_m } ^{OSp(1|2)} [\mathcal L(\Gamma);\hat Q] &=& {1  \over {\hat Q}^{\frac{1}{2}} +{\hat Q}^{-\frac{1}{2}}}\prod_{i=1} ^m (-1)^{n_i}
	{\hat Q}^{\frac{f_i((2n_i+1)^2-1)}{4}}\nonumber\\
	~&~&\left( {\hat Q}^{\frac{2n_i+1}{2} } + {\hat Q}^{-{\frac{2n_i+1}{2}}} \right)^{\text{deg} (v_i) -1}\nonumber\\
	~&~&\prod_{(v_1,v_2)\;\in\; {\rm Edges} } {\left({\hat Q}^{\frac{(2n_{v_1}+1)(2n_{v_2}+1)}{2}} + {\hat Q}^{-\frac{(2n_{v_1}+1)(2n_{v_2} +1)}{2}}\right)}~.
	\label{eqn2.15}
\end{eqnarray}
As three-manifolds can be constructed by a surgery procedure on any framed link, 
the Chern-Simons partition function/WRT invariant (\ref{wrtdef}) can be written in terms of link invariants\cite{10.2307/1970373,wallace_1960,Kaul:2000xe,Ramadevi:1999nd}. We will now present the salient features of such WRT invariants.
\vspace*{-\baselineskip}

\subsubsection{Three-Manifold Invariants}
\label{sec2.1.2}
\begin{figure}[ht]
	\centering
	\includegraphics[width=16cm, height=4.5cm]{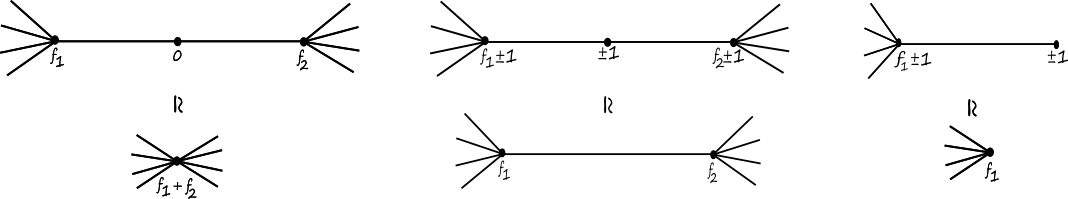}
	\caption{Kirby-Neumann moves that relate plumbing graphs which result in homeomorphic 3-manifolds.}
	\label{fig:moves}
\end{figure}

Let us confine to the   three-manifold $M[\Gamma]$ obtained from surgery of framed link associated with $L$-vertex graph (an example illustrated in Figure \ref{fig:plumbing-example}). These kind of manifolds are known in the literature as plumbed three-manifolds. The linking matrix $B$ is defined as 
\begin{equation}
	B_{v_1,v_2}=
	\begin{cases}
		1,& v_1,v_2\text{ connected}, \\
		f_v, & v_1=v_2=v, \\
		0, & \text{otherwise}.
	\end{cases}
	\qquad v_i \in \text{Vertices of }\Gamma \;\cong\;\{1,\ldots,L\}.
\end{equation}
The algebraic expression for the WRT invariant $\tau_k^{G}[M(\Gamma);\q]$ is
\begin{equation}
	\tau_k^{G}[M(\Gamma);\q]=\frac{F^{G}[\CL(\Gamma);\q]}{F^{G} [\CL(+1\bullet);\q]^{b_+}F^{G} [\CL(-1\bullet);\q]^{b_-}}
	\label{WRT}
\end{equation} 
where $b_{\pm}$ are the number of positive and negative eigenvalues of a linking matrix $B$, respectively. \textcol{Also, note that $(\pm f\bullet)$ represents the unknot with $(\pm f)$ framing}, and $F^{G}[\mathcal{L}(\Gamma);\q]$ is defined as
\begin{equation}
	F^{G}[\CL(\Gamma);\q] = \sum_{R_1,R_2,\ldots R_L}\left( \prod_{i=1}^L V_{R_i}^{G} [\textcol{0\bullet};\q]\right)  V_{R_1,R_2\ldots R_L}^{G} [\mathcal L(\Gamma);\q]~,
\end{equation}
where the summation \textcol{$\{R_i\}$} indicates all the allowed integrable representations of affine $\mathfrak{g}_k$ Lie algebra. By construction, any two homeomorphic manifolds must share the same three-manifold invariant. There is a prescribed set of moves called Kirby moves on links which gives the same three-manifold. For framed links depicted as plumbing graphs, these moves are known as Kirby-Neumann moves as shown in Figure~\ref{fig:moves}.
Hence,  the three-manifold invariant must obey
\begin{equation}
	\tau_k^{G}[M(\Gamma);\q]=
	\tau_k^{G}[M(\Gamma');\q]~,
\end{equation}
where the plumbing graph \textcol{$\Gamma$ can be transformed to $\Gamma'$ using the following set of Kirby-Neumann moves.}
\vspace*{-\baselineskip}

The impact of knot homology (section \ref{subsectionknothomologies}) on the categorification of the WRT invariants has been studied in the last eight years. We now present a concise summary of the recent developments in this direction.
\vspace*{-\baselineskip}

\subsubsection{Three-Manifold Homology}
\label{sec2.2.3}
As WRT invariants (\ref{WRT}) of three-manifolds  involves  invariants of framed links, logically we would expect the homology of three-manifold  $\mathcal H^{G;M}$  such that
\begin{equation}
	\tau_k^{G}[M;\q] 
	\stackrel{\text{?}}{=}
	\sum_{i,j} (-1)^j \q^i {\rm dim} \mathcal H^{G; M}_{i,j}~.\label {qhatz}
\end{equation}
However, the WRT invariants known for many three-manifolds are not seen as $\q$-series (\ref{qhatz}). 
We will now review the necessary steps \cite{Gukov:2017kmk} of obtaining a new three-manifold invariant $\widehat Z$, as $\q$-series, from  $U(N)$ Chern-Simons partition function for Lens space $M=L(p,1)\equiv S^3 / \mathbb Z_p$. The space of flat connections  $\{a\}$ denoted by $\pi_1[{S^3 \over \mathbb Z_p}]\equiv \mathbb Z_p$. Hence $Z_k^{U(N)}[L[p,1];\q]$  can be decomposed as  a sum of perturbative Chern-Simons $Z_a^{U(N)}[L[p,1];\q]$  around these abelian flat connections $a$ \cite{Gukov:2017kmk}:
\begin{equation}
	Z_k^{U(N)}[L[p,1];\q]= \sum_a \exp [i S_{CS}^{(a)} ]Z_a^{U(N)} [L[p,1];\q]~,
	\label{eqn2.26}
\end{equation}
where $S_{CS}^{(a)}$ is the corresponding classical Chern-Simons action. The following change of basis by $\mathcal S$ matrix of $\mathfrak u(1)^N_p$ affine algebra:
\begin{equation} 
	Z_a^{U(N)}[L(p,1);\q] = \sum_b \mathcal S_{ab} \widehat Z_b^{U(N)}[\mathcal L[p,1];q]\Big\vert_{q\rightarrow \q}~,
	\label{eqn2.27} 
\end{equation} 
is required  so that 
\begin{equation} 
	\widehat Z_b^{U(N)}[\mathcal L[p,1];q] \in q^{\Delta_b} \mathbb Z[[q]]~,~~ \Delta_b \in \mathbb Q~.
	\label{eqn2.28}
\end{equation}
Physically,  the  $\widehat Z_b[\mathcal L[p,1];q]$  is  also the vortex partition function $\widehat Z_{T[L[p,1]]}[D^2 \times_q S^1]$ obtained from reducing 
6d $(2,0)$ theory (describing dynamics of $N$-coincident $M5$ branes on $L[p,1] \times D^2 \times_q S^1$) on $L[p,1]$. The effective 3-d $\mathcal N=2$ theory on $D^2 \times_q S^1$ (cigar geometry) is denoted as $T^{U(N)} [L[p,1]]$.  

For other three-manifolds $M$, $\mathcal S$ matrix depends only on $H_1(M,\mathbb{Z})$. 
Further the  Hilbert space of BPS states $\mathcal H^{i,j}_{BPS}$  on the M5 brane system, in the ambient space-time $T^*M  \times TN \times S^1$,  where $i,j$ gradings will keep track of both spins  associated with the rotational symmetry $U(1)_q \times U(1)_R$ on $D^2\subset TN$. The Hilbert space of states for the theory $T^{G}[M]$ with boundary condition at $\partial D^2= S^1$  labeled by $a \in ({\rm Tor} H_1(M, \mathbb Z))^N/S_N$ leads to  bi-graded homological invariants of $M$:
\begin{equation}
	{\mathcal H}_a^{U(N)} [M]= {\mathcal H}_{T_{L[p,1]}^{U(N)}} [D^2; a]
	= \bigoplus_{\substack{i \in \mathbb Z+\Delta_a,\\j \in \mathbb Z}}  {\mathcal H}_a^{i,j}~.
\end{equation}
Note that the grading $i$ counts the charge under $U(1)_q$ rotation of $D^2$ and homological grading $j$  is the R-charge of the $U(1)_R$ R-symmetry. In the following section, we will review the necessary steps of obtaining $\widehat Z$ invariants for $SU(2)$ group. This will provide clarity of notations to investigate $\widehat Z$ for  $SO(3)$ and $OSp(1|2)$ group.

\section{Review of $SU(2)$ $\widehat{Z}$ invariant}
\label{reviewsu2}
As discussed in subsection \ref{sec2.2.3}\cite{Gukov:2016gkn},  the expression for Lens space partition function using eqns.(\ref{eqn2.26}-\ref{eqn2.28})
\begin{eqnarray}
	Z_k^{U(N)}[L(p,1),\q]&=&\sum_{a,b\in \mathbb{Z}_p}\mathcal{S}_{ab} \exp[iS_{CS}^{(a)}]\widehat{Z}_b^{U(N)}[\mathcal{L}(p,1);q] \Big\vert_{q \rightarrow \q}~,
\end{eqnarray}
led to the following  conjecture \cite{Gukov:2017kmk,Gukov:2019mnk} for any closed oriented  three manifold $M$ known as GPPV conjecture:
\begin{eqnarray}
	Z^{SU(2)}_k[M;\q]&=&(i\sqrt{2(k+2}))^{b_1(M)-1}\sum_{a,b\;\in \;\atop \text{Spin}^c(M)/\Z_2}\exp[2\pi i(k+2) \lk(a,a)]
	\,\times\nonumber\\
	~&~&~~~~~~~|\mathcal{W}_b|^{-1}\mathcal{S}_{ab}
	\widehat{Z}^{SU(2)}_b[M;q]|_{q\rightarrow \q=\exp({\frac{2\pi i }{k+2}})}
	\label{WRT-decomposition3.2}
\end{eqnarray}
where
\begin{equation}
	\widehat{Z}^{SU(2)}_b[M;q] \in \, 2^{-c} q^{\Delta_b} \Z[[q]]\qquad \Delta_b\in \Q,\qquad c\in\Z_+
	\label{Block-qSeries}
\end{equation}
is convergent for $|q|<1$ and
\begin{equation}
	\mathcal{S}_{ab}=\frac{e^{2\pi i\lk(a,b)}+e^{-2\pi i\lk(a,b)}}{|\CW_a|\sqrt{|H_1(M,\Z)|}}.
	\label{Sab}
\end{equation}
Here $\mathcal{W}_a$ is the stabilizer subgroup defined as 
\begin{equation}
	\CW_a \; \equiv \; \text{Stab}_{\Z_2}(a) \; = \; 
	\begin{cases}
		\Z_2, & a=-a \,, \\
		1, & \text{otherwise,}
	\end{cases}
\end{equation}
and $\ell k$ denotes the  linking pairing on $H_1(M,\mathbb{Z})$:
\begin{equation}
	\begin{array}{cccc}
		\lk: & H_1(M,\mathbb{Z})\otimes H_1(M,\mathbb{Z}) & \longrightarrow & \Q/\Z \\
		& [a]\otimes [b] & \longmapsto & {\#(a\cap \hat b)}/{n} \\
	\end{array}
\end{equation}
where $\hat b$ is a two-chain complex such that $\partial \hat b = nb$ with $n\in \Z$. Such a 
$\hat b$ and $n$ exists because $[b] \in H_1(M,\mathbb{Z})$.  The number $\#(a\cap \hat b)$ counts the intersection points with signs determined by the orientation. The set of orbits is the set of $\text{Spin}^c$ structures on $M$, with the action of $\Z_2$ by conjugation.

Although the relation (\ref{WRT-decomposition3.2}) is true for any closed oriented three-manifold $M$, the explicit $q$ series expression for $\widehat{Z}$ is waiting to be discovered for a general three-manifold. 

In the following subsection, we will review the $\widehat{Z}^{SU(2)}$  for the plumbed manifolds, \textcol{mostly following section (3.4) and appendix (A) of reference \cite{Gukov:2017kmk}.} We begin with the WRT invariant for a plumbing graph, of the type shown in Figure. \ref{fig:plumbing-example}, discussed in section (\ref{sec2.1.2}). Then analytically continue $\q \rightarrow q$ to get the $\widehat{Z}^{SU(2)}$-invariant. We will see that the analytic continuation procedure is doable only for negative definite plumbed manifolds({\it i.e.,} the signature of linking matrix $B$, $\sigma= b_+ - b_-= -L$)\footnote{In principle, this procedure is also doable when $B$ is negative on a certain subspace of $\mathbb{Z}^L$.}. Moreover, as explained in\cite{Gukov:2019mnk}, the $\text{Spin}^c$-structure in case of plumbed 3-manifold with $b_1(M)=0$, is given by $H_1(M,\mathbb{Z})\cong \text{Coker} B=\mathbb{Z}^L/B\mathbb{Z}^L$.
\subsection{$\widehat{Z}_b^{SU(2)(q)}$}
The WRT invariant $\tau_k^{SU(2)}[M(\Gamma);\q]$,\footnote{normalized such that $\tau_k^{G}[S^3;\q]=1$ and $k$ is the bare level for $SU(2)$ Chern-Simons theory} for plumbed three-manifold $M(\Gamma)$(\ref{WRT}), obtained from surgery of framed link $\CL(\Gamma)$ in $S^3$, is
\begin{eqnarray}
	\tau_k^{SU(2)}[M(\Gamma);\q]&=&\frac{F^{SU(2)}[\CL(\Gamma);\q]}{F^{SU(2)} [\CL(+1\bullet);\q]^{b_+}F^{SU(2)}[\CL(-1\bullet);\q]^{b_-}}
	\nonumber\\
	{\rm where}~F^{SU(2)}[\CL(\Gamma);\q]&=& \sum_{{n}\in \{1,\ldots,k+1\}^{L}} \overline{J}_{n_1-1,\ldots,n_L-1}[\CL(\Gamma)]
	\prod_{v=1}^L \frac{\q^{n_v/2}-\q^{-n_v/2}}{\q^{1/2}-\q^{-1/2}}.
\end{eqnarray}
Note $b_{\pm}$ are the number of positive and negative eigenvalues of a linking matrix $B$ respectively  and the colored Jones polynomial of link $\mathcal{L}(\Gamma)$ (\ref{conn1}) in variable $\q = \exp({2 i \pi /(k+2)})$ is
\begin{eqnarray}
	\overline{J}_{n_1,\ldots,n_L}[\CL(\Gamma)]&=&\frac{2i}{\q^{1/2}-\q^{-1/2}}\prod_{v\;\in\; \text{Vertices}\;\cong\;\{1,\ldots,L\}}
	\q^{\frac{f_v(n_v^2-1)}{4}}
	\,
	\times \\
	~&~&\left(\frac{2i}{\q^{n_v/2}-\q^{-n_v/2}}\right)^{\text{deg}(v)-1}\prod_{(v_1,v_2)\;\in\;\text{Edges}}
	\frac{\q^{n_{v_1}n_{v_2}/2}-\q^{-n_{v_1}n_{v_2}/2}}{2i}.\nonumber
\end{eqnarray}
Using the following Gauss sum reciprocity formula 
\begin{multline}
	\sum_{n\;\in\;\Z^L/2k\Z^L}
	\exp\left(\frac{\pi i}{2k}(n,Bn)+\frac{\pi i}{k}(\ell,n)\right)
	=\\
	\frac{e^{\frac{\pi i\sigma}{4}}\,(2k)^{L/2}}{|\det B|^{1/2}}
	\sum_{a\;\in\; \Z^L/B\Z^L}
	\exp\left(-2\pi i k\left(a+\frac{\ell}{2k},B^{-1}\left(a+\frac{\ell}{2k}\right)\right)\right)
	\label{reciprocity11}
\end{multline}
where $\ell \in \Z^L$, $(\cdot,\cdot)$ is the standard pairing on $\Z^L$ and $\sigma=b_+-b_-$ is the signature of the linking matrix $B$, we can sum 
\begin{equation}
	F^{SU(2)}[\CL(\pm 1\bullet); \q]=\sum_{n=1}^{k+1}\q^{\pm\frac{n^2-1}{4}}\,
	\left(\frac{\q^{n/2}-\q^{-n/2}}{\q^{1/2}-\q^{-1/2}}\right)^2=
	\frac{[2(k+2)]^{1/2}\,e^{\pm\frac{\pi i}{4}}\,\q^{\mp\frac{3}{4}}}{\q^{1/2}-\q^{-1/2}},
\end{equation}
for the unknot with framing $\pm 1$.  Incorporating the above equation and the fact that $L-|\text{Edges}|=1$ for  the framed link $\CL(\Gamma)$, the WRT invariant simplifies to 
\begin{multline}
	\tau_k^{SU(2)}[M(\Gamma);\q]=\frac{e^{-\frac{\pi i\sigma}{4}}\,\q^{\frac{3\sigma}{4}}}{2\,(2(k+2))^{L/2}\,(\q^{1/2}-\q^{-1/2})}
	\times\\
	{\sum_{{n}\in \Z^L/2(k+2)\Z^L}}'\prod_{v\;\in\; \text{Vertices}}
	\q^{\frac{f_v(n_v^2-1)}{4}}
	\,\left(\frac{1}{\q^{n_v/2}-\q^{-n_v/2}}\right)^{\text{deg}(v)-2}
	\times \\
	\prod_{(v',v'')\;\in\;\text{Edges}}
	\frac{\q^{n_{v'}n_{v''}/2}-\q^{-n_{v'}n_{v''}/2}}{2}
	\label{WRT-Gamma-01}
\end{multline}
where we used invariance of the summand under $n_v \rightarrow -n_v$. The prime $'$ in the sum means that the singular values $n_v=0,\,k+2$ are omitted. Let us focus on the following factor for general plumbed graph:
\begin{eqnarray}
	\prod_{(v',v'')\;\in\;\text{Edges}}
	(\q^{n_{v'}n_{v''}/2}-\q^{-n_{v'}n_{v''}/2})
	&=&\sum_{{p}\in\{\pm 1\}^\text{Edges}}
	\prod_{(v',v'')\;\in\;\text{Edges}}p_{(v',v'')} \nonumber\\
	~&~&\q^{p_{(v',v'')}n_{v'}n_{v''}/2}.
\end{eqnarray}
Note that, under $n_v \rightarrow -n_v$ on any vertex  $v$ of degree ${\rm deg}(v)$,
the factor  with a given configuration of signs associated to edges ({\it i.e.,} $p\in\{\pm 1\}^\text{Edges}$) will transform into a term with a different configuration times $(-1)^{\text{deg}(v)}$.  For the class of graphs $\Gamma$ (like Figure. \ref{fig:plumbing-example}),  the  sequence of such transforms can be finally  brought to the configuration with all signs $+1$.  Hence, the WRT invariant (\ref{WRT-Gamma-01}) for these plumbed three-manifolds can be reduced to this form:
\begin{multline}
	\tau_k^{SU(2)}[M(\Gamma)]=\frac{e^{-\frac{\pi i\sigma}{4}}\,\q^{\frac{3\sigma-\sum_v f_{v}}{4}}}{2\,(2(k+2))^{L/2}\,(\q^{1/2}-\q^{-1/2})}
	\times\\
	{\sum_{{n}\in \Z^L/2(k+2)\Z^L}}'
	\;\; \q^{\frac{(n,Bn)}{4}}
	\prod_{v\;\in\; \text{Vertices}}
	\,\left(\frac{1}{\q^{n_v/2}-\q^{-n_v/2}}\right)^{\text{deg}(v)-2}.
	\label{WRT-Gamma-11}
\end{multline}

In the above expression, the points $0$ and $k+2$ are excluded in the summation but in the reciprocity formula (\ref{reciprocity11}) no point is excluded. So, to apply the reciprocity formula  we have to first regularize the sum. This is achieved by  introducing the following regularising parameters:
\begin{eqnarray}
	\Delta_v\in\Z_+:\; \Delta_v&=&\text{deg}(v) -1\mod 2,\qquad \forall v\;\in\; \text{Vertices},
	\label{Delta-def}\\
	\omega\in \C:&\;\;&0<|\omega|<1.\nonumber
\end{eqnarray}
so that the sum in eqn.(\ref{WRT-Gamma-11}) is rewritable as  $\omega \rightarrow  1$:
\begin{multline}
	{\sum_{{n}\in \Z^L/2(k+2)\Z^L}}'
	\;\; \q^{\frac{(n,Bn)}{4}}
	\prod_{v\;\in\; \text{Vertices}}
	\,\left(\frac{1}{\q^{n_v/2}-\q^{-n_v/2}}\right)^{\text{deg}(v)-2}=\\
	\lim_{\omega\rightarrow 1} \frac{1}{2^L}\sum_{{n}\in \Z^L/2(k+2)\Z^L}
	\q^{\frac{(n,Bn)}{4}}
	F_\omega(x_1,\ldots,x_L)|_{x_v=\q^{n_v/2}},
	\label{WRT-omega-limit}
\end{multline}
where
\begin{eqnarray}
	F_\omega(x_1,\ldots,x_L)&=&
	\prod_{v\;\in\; \text{Vertices}}\left({x_v-1/x_v}\right)^{\Delta_v}
	\times\,\\
	\label{F-omega}
	~&~&\left\{
	\left(\frac{1}{x_v-\omega/x_v}\right)^{\text{deg}(v)-2+\Delta_v}
	+ \left(\frac{1}{\omega x_v-1/x_v}\right)^{\text{deg}(v)-2+\Delta_v}
	\right\}\nonumber
	\end{eqnarray}
Note that, we can perform a binomial expansion taking $(\omega/x_v^2)$ small in the first term and 
$(\omega x_v^2)$ small in the second term to rewrite $F_\omega(x_1,\ldots,x_L)$ as a formal power series:
\begin{equation}
	F_\omega(x_1,\ldots,x_L)=\sum_{\ell\in 2\Z^L +\delta} F_\omega^\ell\prod_v x_v^{
		\ell_v}
	\qquad \in \Z[\omega][[x_1^{\pm1},\ldots,x_1^{\pm L}]]~,
\end{equation}
where $\delta\in \Z^L/2\Z^L,~ \delta_v \equiv \text{deg}(v)\mod 2$ and 
 \begin{equation}
	F_\omega^\ell=\sum_{m:\,\ell\in \CI_m}N_{m,\ell}\,\omega^m \;\in \Z[\omega]
\end{equation}
with $\CI_m$ being a finite set of elements from $\Z^L$.
	%\vspace{3em}
	%\stackrel{\omega\approx 0}{=} \sum_{m\ge;q 0}\sum_{\ell\in \CI_m}N_{m,\ell}\,\prod_v x_v^{
	%	\ell_v} \;\cdot\omega^m
	%\;\;\in\;\Z[x_1^{\pm1},\ldots,x_L^{\pm1}][[\omega]]	
%\end{multline}
By definition, ${\rm lim}_{\omega \rightarrow 1} F_{\omega}^\ell$ is not dependent on $\Delta\in \Z^L$  (\ref{Delta-def}). 
However this $\omega \rightarrow 1$ limit in eqn. (\ref{F-omega}) will restrict the binomial expansion range of the first term to be $x \rightarrow \infty$ and that of the second term to $x \rightarrow 0$:
\begin{eqnarray}
	F_{\omega \rightarrow 1} (x_1,\ldots,x_L)=\sum_{\ell\in 2\Z^L +\delta} F_{\omega \rightarrow 1}^\ell\prod_v x_v^{\ell_v}&=& \label{F1lsu2}\\
	~~{\rm lim}_{\omega \rightarrow 1}\prod_{v\,\in\,\text{Vertices}}\left\{
	{\scriptsize \begin{array}{c} \text{Expansion} \\ \text{as} x\rightarrow \infty \end{array} }
	\frac{1}{(x_v-\omega/x_v)^{\text{deg}\,v-2}}\right.
	&+& \left . {\scriptsize \begin{array}{c} \text{Expansion} \\ \text{as }  x\rightarrow 0 \end{array} }
	\frac{1}{(\omega x_v-1/x_v)^{\text{deg}\,v-2}}\right\}.\nonumber
\end{eqnarray}
Now let us assume that the quadratic form $B:\Z^L\times \Z^L\rightarrow \Z$ is negative definite {\it i.e., } $\sigma=-L$. Then we can define the following series in $q$ which is convergent for $|q|<1$:
\begin{equation}
	\widehat Z_b^{SU(2)}[M(\Gamma);q]\stackrel{\text{Def}}{=\joinrel=}
	2^{-L} q^{-\frac{3L+\sum_v f_{v}}{4}}
	\sum_{\ell \in 2B\Z^L+b}F^\ell_{\omega \rightarrow 1}\,q^{-\frac{(\ell,B^{-1}\ell)}{4}}
	\;\in\; 2^{-c}q^{\Delta_b}\Z[[q]]
	\label{Z-hat-def}
\end{equation}
where $c\in \Z_+, c\leq L$ and 
\begin{eqnarray}
	b&\in& (2\Z^L+\delta)/2B\Z^L\,/\Z_2 \cong (2\text{Coker}\,B+\delta)\,/\Z_2
	\stackrel{\text{Set}}{\cong} H_1(M_3,\Z)\,/\Z_2,\\
	\Delta_b&=&-\frac{3L+\sum_v f_{v}}{4}-\max_{\ell \in 2M\Z^L+b}\frac{(\ell,B^{-1}\ell)}{4}\,\in \Q
\end{eqnarray}
where $\Z_2$ action takes $b\rightarrow -b$ and is the symmetry of (\ref{Z-hat-def}).
Using relation (\ref{WRT-omega-limit}) and applying Gauss reciprocity formula (\ref{reciprocity11}) we arrive at the following expression for the WRT invariant:
\begin{multline}
	\tau_{k}^{SU(2)}[M(\Gamma);\q]=\frac{e^{-\frac{\pi iL}{4}}\,\q^{-\frac{3L+\sum_v f_{v}}{4}}}{2\,(2(k+2))^{L/2}\,(\q^{1/2}-\q^{-1/2})}\,
	\lim_{\omega\rightarrow 1}\sum_{{n}\in \Z^L/2(k+2)\Z^L}
	\q^{\frac{(n,Bn)}{4}}
	F_\omega(x_1,\ldots,x_L)|_{x_v=\q^{n_v/2}}
	\label{WRT-Gamma-2}\\\\
	~~=\frac{2^{-L} \q^{-\frac{3L+\sum_v f_{v}}{4}}}{2\,(\q^{1/2}-\q^{-1/2})\,|\det B|^{1/2}}
	\sum_{\scriptsize\begin{array}{c}a\in \mathrm{Coker}\,B \\ b \in 2\mathrm{Coker}\, B+\delta \end{array}}
	e^{-2\pi i(a,B^{-1}b)} e^{-2\pi i(k+2)(a,B^{-1}a)}\times\\
	\lim_{\omega\rightarrow 1}
	\sum_{\ell \in 2B\Z^L+b}F^\ell_\omega\,\q^{-\frac{(\ell,B^{-1}\ell)}{4}}.~~~~~~~~~
\end{multline}
Assuming that the limit $\lim_{q\rightarrow \q}\widehat{Z}_b^{SU(2)}(q)$ exists, where  $q$ approaches $(k+2)$-th primitive root of unity from inside of the unit disc $|q|<1$, we expect
\begin{equation}
	\lim_{\omega\rightarrow 1}
	\sum_{\ell \in 2B\Z^L+b}F^\ell_\omega\,\q^{-\frac{(\ell,B^{-1}\ell)}{4}}
	=\lim_{q\rightarrow \q}
	\sum_{\ell \in 2B\Z^L+b}F^\ell_{\omega \rightarrow 1}\,q^{-\frac{(\ell,B^{-1}\ell)}{4}}.
	\label{limit-exchange}
\end{equation}
Thus we obtain GPPV conjecture form:
\begin{multline}
	\tau_{k}^{SU(2)}[M(\Gamma),\q]=\frac{1}{2\,(\q^{1/2}-\q^{-1/2})\,|\det B|^{1/2}}
	\,\times\\
	\sum_{a\in \mathrm{Coker}\,B}e^{-2\pi i(k+2)(a,B^{-1}a)}
	\sum_{b \in 2\mathrm{Coker}\, B+\delta} e^{-2\pi i(a,B^{-1}b)}\lim_{q\rightarrow \q} \widehat{Z}_b^{SU(2)}[M(\Gamma);q].
	\label{WRT-prop-1}
\end{multline}
There is also an equivalent  contour integral form for  the homological blocks(\ref{Z-hat-def}):\begin{equation}
	\widehat{Z}_b^{SU(2)}[M(\Gamma);q]=q^{-\frac{3L+\sum_v f_{v}}{4}}\cdot\text{v.p.}\int\limits_{|z_v|=1}
	\prod_{v\;\in\; \text{Vertices}}
	\frac{dz_v}{2\pi iz_v}\,
	\left({z_v-1/z_v}\right)^{2-\text{deg}(v)}\cdot\Theta^{-B}_b(z),
\end{equation}
where $\Theta^{-B}_b(x)$ is the theta function of the lattice corresponding to minus the linking form $B$:
\begin{equation}
	\Theta^{-B}_b(x)=\sum_{\ell \in 2B\Z^L+b}q^{-\frac{(\ell,B^{-1}\ell)}{4}}
	\prod_{i=1}^Lx_i^{\ell_i},
\end{equation}
and ``v.p.'' refers to principle value integral (\ie~take half-sum of contours $|z_v|=1\pm \epsilon$). This prescription corresponds to the regularization by $\omega$ made in eqn.(\ref{F-omega}).

Thus we can obtain explicit  $SU(2)$ $q$-series for any negative definite plumbed three-manifolds. For completeness, we  present the $q$-series for some examples. 
%%%%%%%%%%%%%%%%%%%
\subsection{Examples}
$\bullet$ Poincare homology sphere is a well-studied three-manifold corresponding to the graph:
\begin{equation}
	\includegraphics{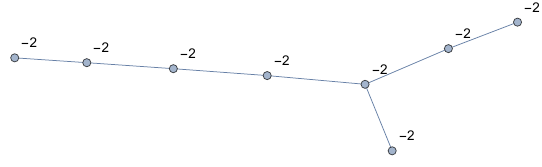}
	\label{eqn3.31new}
\end{equation}
As $H_1(M,\Z)=0$, we obtain only single homological block $\widehat Z_{b_1}$. Solving eqns.(\ref{F1lsu2},\ref{Z-hat-def}), we get
\begin{equation}
	\widehat{Z}_{b_1}^{SU(2)}=q^{-3/2}(1-q-q^3-q^7+q^8+q^{14}+q^{20}+q^{29}-q^{31}-q^{42}+\cdots).
\end{equation}
$\bullet$ The next familiar example with $H_1(M,\Z)=0$ is 
Brieskorn homology sphere. A particular example of this class is $\Sigma(2,3,7)$ with the following equivalent graphs:
\begin{equation}
	{\,\raisebox{-2.25cm}{\includegraphics[width=4.5cm]{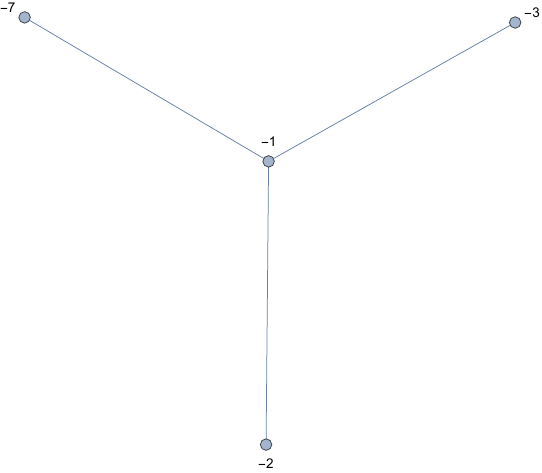}}\,}
	\stackrel{\text{Kirby}}{\sim}
	{\,\raisebox{-2.1cm}{\includegraphics[width=3.75cm]{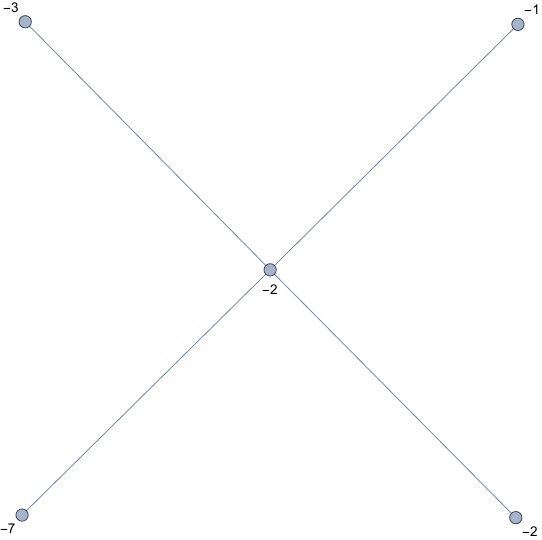}}\,}
	\label{eqn3.33new}
\end{equation}
The homological block turns out to be
\begin{equation}
	\widehat{Z}_{b_1}^{SU(2)}=q^{1/2}(1 - q - q^5 + q^{10} - q^{11} + q^{18} + q^{30} - q^{41} + q^{43} - q^{56} - 
	q^{76}\cdots).
\end{equation}
$\bullet$ For a three-manifold with non-trivial $H_1(M,\Z)=\Z_{3}$ as drawn below, 
\begin{equation}
	{\,\raisebox{-2.0cm}{\includegraphics[width=8.0cm]{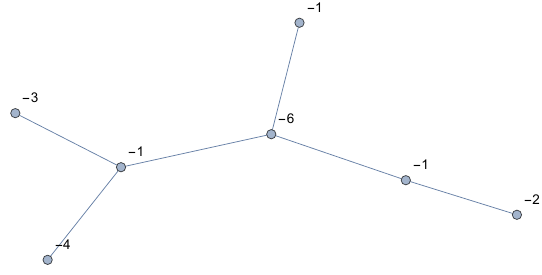}}\,}
	\stackrel{\text{Kirby}}{\sim}
	{\,\raisebox{-2.0cm}{\includegraphics[width=4.0cm]{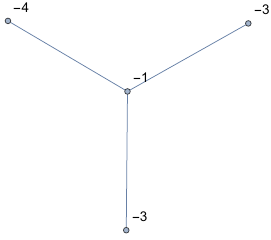}}\,}
	\label{eqn3.35new}
\end{equation}
the three homological blocks are
\begin{equation}
	\widehat{Z}^{SU(2)}=
	\left(
	\begin{array}{c}
		1-q+q^6-q^{11}+q^{13}-q^{20}+q^{35}+O\left(q^{41}\right) \\
		q^{5/3}\left(-1+q^3-q^{21}+q^{30}+O\left(q^{41}\right)\right) \\
		q^{5/3}\left(-1+q^3-q^{21}+q^{30}+O\left(q^{41}\right)\right) \\
	\end{array}
	\right),
\end{equation}
where two of them are equal.

Our focus is to obtain explicit $q$-series for $SO(3)$ and $OSp(1|2)$ groups.
Using the relation between $SU(2)$ and $SO(3)$, $SU(2)$ and $OSp(1\vert 2)$ link invariants(\ref{sec2.1.1}),  we will investigate the necessary steps starting from the WRT invariant for $SO(3)$ and $OSp(1|2)$ eventually leading to the  $\widehat Z$-invariant. This will be the theme of the following section.

\section{$\widehat{Z}$ for $SO(3)$ and $OSp(1\vert2)$}
\label{zhatnew}
\vspace*{-\baselineskip}

Our aim is to  derive the $\widehat{Z}$-invariant for $SO(3)$ and $OSp(1\vert 2)$ groups. 
We will first look at  the WRT invariants $\tau_K^{SO(3)}[M(\Gamma); Q]$ for plumbed three-manifolds written in terms of colored Jones invariants of framed links $\CL[\Gamma]$ in the following subsection and then discuss  $OSp(1|2)$ $\widehat Z$ in the subsequent section.

\subsection{$SO(3)$ WRT invariant and $\widehat{Z}^{SO(3)}$ invariant}
Recall that the framed link invariants are written in  variable $\q$ which is dependent on Chern-Simons coupling and the rank of the gauge group $G$. For $SO(3)$ Chern-Simons with coupling $K \in 2 \mathbb Z$, the variable $Q= e^{\frac{2\pi i}{K+1}}$. Hence $F^{SO(3)}[\mathcal{L}(\Gamma);Q]$ in WRT $\tau_K^{SO(3)}[M(\Gamma); Q]$ is 
\begin{multline}
	F^{SO(3)}[\mathcal{L}(\Gamma);Q]=\\ \sum_{n_1,n_2,\dots,n_L\in \{0,1,\dots,K\}}V_{n_1,n_2,\dots,n_L}^{SO(3)}(\mathcal{L}(\Gamma);Q)\prod_{v=1}^{L}V_{n_1,n_2,\dots,n_L}^{SO(3)}(\bigcirc;Q)~=\\\\
	\sum_{n_1,n_2,\ldots,n_L\in\{0,2,\dots,2K\}}\overline{J}_{n_1,n_2,\dots,n_L}^{SU(2)}\left(\mathcal{L}(\Gamma); \q=e^{\frac{2\pi i}{2K+2}}\right)\prod_{v=1}^{L}\frac{\q^{(n_v+1)/2}-\q^{-(n_v+1)/2}}{\q^{1/2}-\q^{-1/2}}\bigg\vert_{\q^2 \rightarrow Q}~,
	\label{eqn4.2}
\end{multline}
where  we have used the relation (\ref{so3}) to write $SO(3)$ link invariants in terms of the colored Jones invariants. Notice that the summation is over only even integers and hence WRT invariant for $SO(3)$ is different from the WRT for $SU(2)$ group. Further,
the highest integrable representation in the summation indicates that the Chern-Simons coupling for  $SU(2)$ group is $2K+2$. After performing the summation, we can convert 
the $\q= Q^{1/2}$(\ref{so3}) to obtain $SO(3)$ WRT invariant. We need to modify the Gauss sum reciprocity formula to incorporate the summation over odd integers in $F^{SO(3)}[\mathcal{L}(\Gamma);Q]$.

Using the following  Gauss sum reciprocity formula 
\begin{multline}
	\sum_{n\;\in\;\Z^L/k\Z^L}
	\exp\left(\frac{2\pi i}{k}(n,Bn)+\frac{2\pi i}{k}(\ell,n)\right)
	=\\
	\frac{e^{\frac{\pi i\sigma}{4}}\,(k/2)^{L/2}}{|\det B|^{1/2}}
	\sum_{a\;\in\; \Z^L/2B\Z^L}
	\exp\left(\frac{-\pi ik}{2}\left(a+\frac{\ell}{k},B^{-1}\left(a+\frac{\ell}{k}\right)\right)\right)~,
	\label{reciprocity1}
\end{multline}
for $k=2K+2$,  we can obtain the summation over even integers by replacing $n \longrightarrow \frac{n+1}{2}$ :
\begin{multline}
	\sum_{n_1,n_2,\dots,n_L\;\in\;\{1,3,\dots,4K+3\}}
	\q^{\frac{(n,Bn)}{4}+\frac{(n,d)}{2}} = \frac{e^{\frac{\pi i\sigma}{4}}\,(K+1)^{L/2}}{|\det B|^{1/2}}
	\q^{-\frac{(d,B^{-1}d)}{4}}\times\\\\ \sum_{a\;\in\; \Z^L/2B\Z^L}
	\exp\left[-\pi i(K+1)(a,B^{-1}a)\right]\exp\left[-\pi i(a,B^{-1}(d+BI))\right],
	\label{reciprocity2}
\end{multline}
where $d = \ell-B I$ with $I$ denoting $L$ component  vector with entry $1$ on all the components. That is, the transpose of the vector $I$ is
\begin{equation}
	I^T=[1,1,\ldots,1]~.\label{eqn4.4ii}
\end{equation}

For unknot with framing $\pm 1$, the $F^{SO(3)}[\CL(- 1\bullet); Q=\q^2]$ involving summation over odd integers simplifies to 
\begin{equation}
	F^{SO(3)}[\CL(\pm1\bullet); Q=\q^2]=\frac{\sqrt{K+1}\;e^{\pm\pi i/4}\;\q^{\mp 3/4}}{\q^{1/2}-\q^{-1/2}}\underbrace{(1+e^{\pi i K})}_2~,
\end{equation}
as the coupling $K \in 2 \mathbb Z$ for the $SO(3)$ Chern-Simons theory.
Hence, the WRT invariant takes the following form:
\begin{multline}
	\tau_K^{SO(3)}[M(\Gamma);Q=\q^2]=\frac{e^{-\frac{\pi i\sigma}{4}}\,\q^{\frac{3\sigma}{4}}}{2^L(K+1)^{L/2}\,(\q^{1/2}-\q^{-1/2})}
	\times\\
	{\sum_{{n}\in \{1,3,\ldots,2K+1\}^L}}\prod_{v\;\in\; \text{Vertices}}
	\q^{\frac{f_v(n_v^2-1)}{4}}
	\,\left(\frac{1}{\q^{n_v/2}-\q^{-n_v/2}}\right)^{\text{deg}(v)-2}
	\times \\
	\prod_{(v',v'')\;\in\;\text{Edges}}
	\Big(\q^{n_{v'}n_{v''}/2}-\q^{-n_{v'}n_{v''}/2}\Big).
	\label{WRT-Gamma-4.8}
\end{multline}
In above equation, the terms involving edges of the graph $\Gamma$
\begin{equation*}
	\prod_{(v',v'')\;\in\;\text{Edges}}
	\Big(\q^{n_{v'}n_{v''}/2}-\q^{-n_{v'}n_{v''}/2}\Big)=2^{L-1}		\prod_{(v',v'')\;\in\;\text{Edges}}
	\frac{\Big(\q^{n_{v'}n_{v''}/2}-\q^{-n_{v'}n_{v''}/2}\Big)}{2},
\end{equation*}
can also be rewritten as
\begin{equation*}
	\prod_{(v',v'')\in \text{Edges}} (\q^{n_{v'}n_{v''}/2}-\q^{-n_{v'}n_{v''}/2})=\sum_{p\in\{\pm 1\}^{\text{Edges}}}\prod_{(v',v'')\in \text{Edges}}p_{(v',v'')}\q^{p_{(v',v'')}n_{v'}n_{v''}/2}~.
\end{equation*}
Here again, if we make a change of variable as $n_v\longrightarrow -n_v$ at any vertex, a term in the sum with a given configuration of signs associated to edges (that is $p\in \{\pm 1\}^{\text{Edges}}$) will transform into a term with a different configuration times $(-1)^{\text{deg}(v)}$. However, for these plumbing graphs $\Gamma$, the signs of such configuration can be brought to the configuration with all signs +1. Incorporating this fact, the  WRT invariant(\ref{WRT-Gamma-4.8}) simplifies to 
\begin{multline}
	\tau_{K}^{SO(3)}[M(\Gamma);Q=\q^2]=\frac{e^{-\frac{\pi i\sigma}{4}}\,\q^{\frac{3\sigma-\sum_v f_{v}}{4}}}{2\,(K+1)^{L/2}\,(\q^{1/2}-\q^{-1/2})}
	\times\\
	{\sum_{{n}\in \{1,3,\ldots,2K+1\}^L}}
	\;\; \q^{\frac{(n,Bn)}{4}}
	\prod_{v\;\in\; \text{Vertices}}
	\,\left(\frac{1}{\q^{n_v/2}-\q^{-n_v/2}}\right)^{\text{deg}(v)-2}.
	\label{WRT-Gamma-1}
\end{multline}
\\
Further, we double the range of summation so as to use the reciprocity formula(\ref{reciprocity2})

\begin{multline}
	\tau_{K}^{SO(3)}[M(\Gamma);Q=\q^2]=\frac{e^{-\frac{\pi i\sigma}{4}}\,\q^{\frac{3\sigma-\sum_v f_{v}}{4}}}{4\,(K+1)^{L/2}\,(\q^{1/2}-\q^{-1/2})}
	\times\\
	{\sum_{{n}\in \{1,3,\ldots,4K+3\}^L}}
	\;\; \q^{\frac{(n,Bn)}{4}}
	\prod_{v\;\in\; \text{Vertices}}
	\,\left(\frac{1}{\q^{n_v/2}-\q^{-n_v/2}}\right)^{\text{deg}(v)-2}.
	\label{WRT-Gamma-1so3}
\end{multline}

The steps discussed in the $SU(2)$ context to extract $\widehat{Z}$ can be similarly followed  for $SO(3)$. This procedure leads to 
\begin{multline}
	\tau_K^{SO(3)}[M(\Gamma);Q=\q^2]=\frac{1}{2\,(\q^{1/2}-\q^{-1/2})\,|\det B|^{1/2}}
	\,
	\sum_{a\in \mathrm{Coker}\,B}e^{-\pi i(K+1)(a,B^{-1}a)}\\\\
	\sum_{b \in 2\mathrm{Coker}\, B+\delta} e^{-\pi i\big(a,B^{-1}(b{\color{blue}+BI})\big)}\lim_{q\rightarrow \q} \widehat{Z}^{SO(3)}_b[M(\Gamma);q]~.
\end{multline}
We observe that the $SO(3)$ WRT invariant is different from the $SU(2)$ invariant due to the factor highlighted in blue color in the summand whereas the $\widehat Z^{SO(3)}_b[M(\Gamma);q]$ is exactly same as the $SU(2)$ $q$-series. Even though $SO(3) \equiv SU(2)/\mathbb Z_2$,  it is surprising to see that the factor group $SO(3)$ shares the same $\widehat Z$ as that of the parent group $SU(2)$. The case of $\widehat{Z}^{SO(3)}$ was also considered in \cite{Costantino:2021yfd} but they took a different route by considering the refined WRT invariant which is consistent with our result.

In the following subsection, we will extract $\widehat Z$ from the WRT invariant $\tau_{\hat K}^{OSp(1|2)}[M(\Gamma); \hat Q]$  for $OSp(1|2)$ supergroup. We will see that the $OSp(1|2)$ $q$-series are related to $\widehat Z^{SU(2)}[M(\Gamma);q]$.
\subsection{$OSp(1|2)$ WRT and $\widehat{Z}^{OSp(1|2)}$ invariant}
Using the relation between $OSp(1|2)$ and $SU(2)$ link invariants (\ref{osp}), the WRT invariant can be written for plumbed manifolds $M(\Gamma)$ as
\begin{multline}
	\tau_{\hat{K}}^{OSp(1|2)}[M(\Gamma);\hat Q = \q]=\frac{e^{-\frac{\pi i\sigma}{4}}\,\q^{\frac{3\sigma}{4}}}{(2\hat{K}+3)^{L/2}\,(\q^{1/2}+\q^{-1/2})}
	\times\\
	{\sum_{{n_1,n_2,\ldots,n_L}\in \{1,3,\ldots,2\hat{K}+1\}}}\;\;\prod_{v\;\in\; \text{Vertices}}
	\q^{\frac{f_v(n_v^2-1)}{4}}
	\,\left(\frac{1}{\q^{n_v/2}+\q^{-n_v/2}}\right)^{\text{deg}(v)-2}
	\times \\
	\prod_{(v',v'')\;\in\;\text{Edges}}
	\Big(\q^{n_{v'}n_{v''}/2}+\q^{-n_{v'}n_{v''}/2}\Big) \Big \vert_{\q = \hat Q}~.
	\label{WRT-Gamma-0}
\end{multline}
Here again we use the Gauss reciprocity(\ref{reciprocity2}) as the summation is over odd integers to work out the steps leading to $\widehat{Z}^{OSp(1|2)}[M(\Gamma);q]$. Note that, the highest integrable representation $2\hat K +1$ which fixes the $\q$ as $(2\hat K+2)$-th root of unity. However to compare the result with $OSp(1|2)$ WRT, we have to replace $\hat K +1 \rightarrow 2\hat K+3$ which is equivalent to  $\q = \hat Q$. 

Following similar steps performed for $SU(2)$, we find the following expression for $OSp(1|2)$ WRT invariant:
\begin{multline}
	\frac{1}{2\,(\q^{1/2}+\q^{-1/2})\,|\det B|^{1/2}}
	\,
	\sum_{a\in \mathrm{Coker}\,B}e^{-\pi i(2\hat{K}+3)(a,B^{-1}a)}\times\\
	\sum_{b \in 2\mathrm{Coker}\, B+\delta} e^{-\pi i\big(a,B^{-1}(b{\color{red}+BI})\big)}\lim_{q\rightarrow \hat Q} \widehat{Z}_b^{OSp(1|2)}[M(\Gamma);q],
\end{multline}
where $I$ is again the column vector (\ref{eqn4.4ii}) and $\widehat{Z}_b^{OSp(1|2)}[M(\Gamma);q]$ is given by the following algebraic expression:
\begin{equation}
	\widehat Z_b^{OSp(1|2)}[M(\Gamma);q]\;\;=\;\;
	2^{-L} q^{-\frac{3L+\sum_v f_{v}}{4}}
	\sum_{d \;\in\; 2B\Z^L+b}F^d_1\,q^{-\frac{(d,B^{-1}d)}{4}},
	\label{Z-hat-defosp}
\end{equation}
with coefficient $F_1^d$ is obtained by following relation
\begin{multline}
	\sum_{d\;\in\;2\Z^L +\delta} F_1^d\prod_v x_v^{d_v}=\\
	\prod_{v\,\in\,\text{Vertices}}\left\{
	{\scriptsize \begin{array}{c} \text{Expansion} \\ \text{at } x\rightarrow 0 \end{array} }
	\frac{1}{(x_v+1/x_v)^{\text{deg}\,v-2}}
	+
	{\scriptsize \begin{array}{c} \text{Expansion} \\ \text{at } x\rightarrow \infty \end{array} }
	\frac{1}{(x_v+1/x_v)^{\text{deg}\,v-2}}\right\}.
	\label{eqn4.13new}
\end{multline}
Equivalently, $\widehat{Z}^{OSp(1|2)}[M(\Gamma);q]$(\ref{Z-hat-defosp}) can also represented as the following contour integral:
\begin{equation}
	\widehat{Z}_b^{OSp(1|2)}[M(\Gamma);q]=q^{-\frac{3L+\sum_v f_{v}}{4}}\cdot\text{v.p.}\int\limits_{|z_v|=1}
	\prod_{v\;\in\; \text{Vertices}}
	\frac{dz_v}{2\pi iz_v}\,
	\left({z_v+1/z_v}\right)^{2-\text{deg}(v)}\cdot\Theta^{-B}_b(z)~.
\end{equation}
Here $\Theta^{-B}_b(x)$ is the theta function of the lattice corresponding to minus the linking form $B$:
\begin{equation}
	\Theta^{-B}_b(x)=\sum_{d\;\in\;2B\Z^L+b}q^{-\frac{(d,B^{-1}d)}{4}}
	\prod_{i=1}^Lx_i^{d_i}
\end{equation}
and ``v.p.'' again means that we take principle value integral (\ie~take half-sum of contours $|z_v|=1\pm \epsilon$). Comparing eqns.(\ref{Z-hat-defosp},\ref{eqn4.13new}) with the $SU(2)$ expressions(\ref{F1lsu2},\ref{Z-hat-def}), we can see that the $\widehat Z$ for $OSp(1|2)$ are different from $SU(2)$ q-series. We will now present explicit $q$-series for some examples.

\subsection{Examples}
$\bullet$ For the Poincare homology sphere(\ref{eqn3.31new}), we find the following $OSp(1|2)$ $q$-series
\begin{equation}
	\widehat{Z}_{b_1}^{OSp(1|2)}=q^{-3/2}(1 + q + q^3 + q^7 + q^8 + q^{14} + q^{20} - q^{29} + q^{31} - q^{42} - 
	q^{52} +\cdots).
\end{equation}
$\bullet $ In the case of Brieskorn homology sphere(\ref{eqn3.33new}), the $OSp(1|2)$ $q$-series is  
\begin{equation}
	\widehat{Z}_{b_1}^{OSp(1|2)}=q^{1/2}(1 + q + q^5 + q^{10} + q^{11} + q^{18} + q^{30} + q^{41} - q^{43} - q^{56} - 
	q^{76}\cdots). 
\end{equation}
$\bullet$ For the case of plumbing graph(\ref{eqn3.35new}), the three homological blocks are
\begin{equation}
	\widehat{Z}^{OSp(1|2)}=
	\left(
	\begin{array}{c}
		1+q+q^6+q^{11}-q^{13}-q^{20}-q^{35}+O\left(q^{41}\right) \\
		q^{5/3}\left(1+q^3-q^{21}-q^{30}+O\left(q^{41}\right)\right) \\
		q^{5/3}\left(1+q^3-q^{21}-q^{30}+O\left(q^{41}\right)\right)
	\end{array}
	\right).
\end{equation}
After comparing the $q$-series for $SU(2)$ and $OSp(1|2)$, we noticed that these two $q$-series are related by a simple change of variable which is $q\longrightarrow -q$. This change of variable applies only to the series not to the overall coefficient outside the series.\\ 
$\bullet$ Lens space $L(p,q)$ is a well studied three-manifold. For $L(-5,11)\sim L(-13,29)$ whose plumbing graph is shown below, we obtain the five homological blocks \\
\begin{equation*}
	{\,\raisebox{0.0cm}{\includegraphics[width=5.0cm]{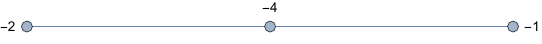}}\,}\;\;
	\stackrel{\text{Kirby}}{\sim}\;\;
	{\,\raisebox{0.0cm}{\includegraphics[width=5.0cm]{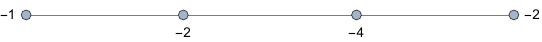}}\,}
\end{equation*}

\begin{equation}
	\widehat{Z}^{OSp(1|2)}=
	\left(
	\begin{array}{c}
		q^{1/10}\\
		q^{-1/10}\\
		0\\
		q^{-1/10}\\
		q^{1/10}
	\end{array}
	\right)~~~~\text{as}~~H_1(M,\mathbb{Z})=\Z_5.
\end{equation}
$\bullet$ For the following plumbing graph, $H_1(M,\Z)=\mathbb{Z}_{13}$~,
\begin{equation*}
	\includegraphics{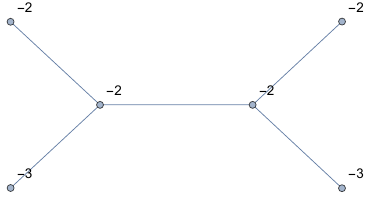}
\end{equation*}
\begin{equation}
	\widehat{Z}^{OSp(1|2)}=
	\frac{1}{4}\left(\tiny
	\begin{array}{c}
		q^{-1/2}(2+2q+2q^2-2q^4-4q^5+6q^{10}-8q^{11}+4q^{13}+2q^{14}-4q^{15}+O\left(q^{18}\right)) \\
		q^{5/26}(3+2q-2q^2-4q^3-2q^7+q^8+2q^9+q^{10}-2q^{12}-4q^{13}-2q^{16}+O\left(q^{18}\right)) \\
		q^{5/26}(3+2q-2q^2-4q^3-2q^7+q^8+2q^9+q^{10}-2q^{12}-4q^{13}-2q^{16}+O\left(q^{18}\right)) \\
		q^{7/26}(4-q-2q^3-2q^4-2q^6+3q^7-2q^8-2q^{10}+q^{11}+2q^{13}-4q^{14}+2q^{15}-4q^{16}+O\left(q^{18}\right)) \\
		q^{7/26}(4-q-2q^3-2q^4-2q^6+3q^7-2q^8-2q^{10}+q^{11}+2q^{13}-4q^{14}+2q^{15}-4q^{16}+O\left(q^{18}\right)) \\
		q^{-7/26}(3+3q^2-2q^4-2q^5+4q^7-2q^8+2q^9-2q^{10}-4q^{12}+4q^{13}-4q^{14}+2q^{15}+O\left(q^{18}\right)) \\
		q^{-7/26}(3+3q^2-2q^4-2q^5+4q^7-2q^8+2q^9-2q^{10}-4q^{12}+4q^{13}-4q^{14}+2q^{15}+O\left(q^{18}\right)) \\
		q^{-11/26}(1+2q+2q^2+4q^3+3q^6-2q^7-4q^8-2q^9+2q^{11}+2q^{13}-q^{14}+2q^{16}-2q^{17}+O\left(q^{18}\right)) \\
		q^{-11/26}(1+2q+2q^2+4q^3+3q^6-2q^7-4q^8-2q^9+2q^{11}+2q^{13}-q^{14}+2q^{16}-2q^{17}+O\left(q^{18}\right)) \\
		q^{-5/26}(2+2q^2+q^3+3q^5-2q^6-2q^7-4q^8-2q^{10}+2q^{11}-2q^{12}+2q^{13}+5q^{15}-2q^{16}+2q^{17}+O\left(q^{18}\right))\\
		q^{-5/26}(2+2q^2+q^3+3q^5-2q^6-2q^7-4q^8-2q^{10}+2q^{11}-2q^{12}+2q^{13}+5q^{15}-2q^{16}+2q^{17}+O\left(q^{18}\right))\\
		q^{-15/26}(1-2q-2q^2+q^4-2q^6-2q^7-2q^8-4q^{10}-2q^{12}+2q^{13}+2q^{15}+4q^{17}+O\left(q^{18}\right)) \\
		q^{-15/26}(1-2q-2q^2+q^4-2q^6-2q^7-2q^8-4q^{10}-2q^{12}+2q^{13}+2q^{15}+4q^{17}+O\left(q^{18}\right))
	\end{array}
	\right)
\end{equation}

\vspace{0.5cm}

We have checked for many examples that under $q\rightarrow -q$ in the $OSp(1|2)$ $q$-series (not affecting the overall coefficient), we obtain the $SU(2)$ $q$-series.

\section{Conclusions}
\label{conclusionchap2}

Our goal was to investigate $\widehat Z$ for $SO(3)$ and $OSp(1|2)$ groups for negative definite plumbed three-manifolds. The change of  variable and color indeed relates invariants of framed links $\CL[\Gamma] $(\ref{so3},\ref{osp}) of $SO(3)$ and $OSp(1|2)$ to colored Jones. Such a relation allowed us to  go through  the steps of GPPV conjecture  to extract $\widehat Z$ from  WRT invariants. 

Interestingly, we observe that the $\widehat Z^{SO(3)}$ is same as $\widehat Z^{SU(2)}$ even though the WRT invariants are different.  We know that $SU(2) /\mathbb Z_2 \equiv SO(3)$ and it is not at all obvious that the homological blocks are same for both the groups. It is important to explore other  factor groups and the corresponding $\widehat  Z$ invariants, which we do in the next chapter \ref{chap:4}. 

For the odd orthosympletic supergroup $OSp(1|2)$, we observe from our computations for many negative definite plumbing graph $\Gamma$:
\begin{equation}
\widehat Z^{OSp(1|2)}_b(\Gamma; q) = 2^{-c} q^{\Delta_b} \left (\sum_n a_n q^n \right),\label{chap2zhatosp12}
\end{equation}
whereas  their $SU(2)$  q-series is
\begin{equation}
\widehat Z^{SU(2)}_b(\Gamma;q) = 2^{-c} q^{\Delta_b} \left (\sum_n a_n (-q)^n \right),\label{chap2zhatsu2}
\end{equation}

where $c\in \mathbb{Z}_+$, $\Delta_b\in \mathbb{Q}$. We do not have a proper understanding of this relation between $\widehat{Z}$ for $SU(2)$ and $OSp(1|2)$ groups. \textcol{Interestingly, this relation has recently been proved by Costantino et al. in ref.\cite{costantino2024nonsemisimpletopologicalfieldtheory}.}

\chapter{Gukov-Pei-Putrov-Vafa Conjecture for $SU(N)/\mathbb{Z}_m$}
\label{chap:4}

In chapter \ref{chap:3}, the $q$-series valued invariant called $\widehat{Z}$ for $SO(3)$ group was investigated by performing an analytic continuation of WRT invariant $\tau_{k}^{SO(3)}[M(\Gamma);\q]$ within a unit circle. Remarkably, it was discovered that $\widehat{Z}^{SO(3)}_b[M(\Gamma);q]$ is equivalent to $\widehat{Z}^{SU(2)}_b[M(\Gamma);q]$. This finding implies that the $\widehat{Z}$-invariant depends on the Lie algebra rather than the Lie group, as $SU(2)$ and $SO(3)$ share the same Lie algebra. Furthermore, it is worth noting that $SO(3)\cong SU(2)/\mathbb{Z}_2$ is the Langlands dual group of $SU(2).$ Therefore, it raises the question of whether the equality $\widehat{Z}^{SU(2)}_b[M(\Gamma);q] =\widehat{Z}^{SO(3)}_b[M(\Gamma);q]$ can be attributed to this Langlands dual group correspondence. Additionally, for higher rank gauge group $SU(N)$ with $N>2$ and $N$ not being a prime number ($N\notin \mathbb{P}$), there exists gauge groups between $SU(N)$ and $SU(N)/\mathbb{Z}_N$. These groups are formed by taking a quotient of $SU(N)$ with a subgroup $\mathbb{Z}_m$ of $\mathbb{Z}_N$. All these quotient groups share the same $\mathfrak{su}(N)$ Lie algebra. Exploring the $\widehat{Z}$-invariant for $SU(N)/\mathbb{Z}_m$ quotient groups will eventually answer whether $\widehat{Z}$-invariant depends on $m$.

While the physics perspective suggests that $\widehat{Z}$-invariant should be Lie algebra dependent only as 3d $\mathcal{N}=2$ $T[M;G]$ obtained by compactifying 6d $\mathcal{N}=(2,0)$ SCFT of type ADE Lie algebra on 3-manifold $M$, 
\begin{equation}
	\text{6d}\; (2,0) \;\text{SCFT}\;\text{of type ADE Lie algebra}\;\stackrel{\text{compactification on M}}{\xrightarrow{\hspace{3.0cm}}} \;\text{3d} \;\mathcal{N}=2\;\; T[M;G].
\end{equation} 
But in certain cases, compactified theories do have Lie group dependence instead of Lie algebra\cite{freed2014relative}. The data of 3d $\mathcal{N}=2$ $T[M;G]$ theory is given by manifold $M$. In other words, for every 3-manifold $M$ there would be a corresponding 3d $\mathcal{N}=2$ $T[M;G]$ theory encoding the geometry and topology of $M$. Many numerical and homological invariants of $M$ have been predicted by studying $T[M;G]$ on various backgrounds\cite{Gukov:2017kmk,Gukov:2016gkn}. Further, the topology and geometry of 3-manifolds is fairly well understood now\footnote{complete topological classification of 3-manifolds is still an open problem}, but still there is no known way to explicitly identify 3d $\mathcal{N}=2$ $T[M;G]$ for a general $M$.

In this chapter, our primary objective is to address the question of whether the $\widehat{Z}$-invariant exhibits dependence on Lie group or Lie algebra. To achieve this, we explicitly study the Gukov-Pei-Putrov-Vafa conjecture for gauge groups of the form $SU(N)/\mathbb{Z}_m$. For this, we must first define the WRT invariant for $SU(N)/\mathbb{Z}_m$ gauge group and then proceed with an analysis similar to that conducted in Ref\cite{2020higher}.

The organization of this chapter is as follows: In section (\ref{review_3d3d}), we present the formula of $\widehat{Z}$-invariant for $SU(N)$ group for comparison. Section (\ref{wrtandlevel}) introduces the appropriate formula for the WRT invariant for the quotient group $SU(N)/\mathbb{Z}_m$. In section (\ref{refinement}), we demonstrate how to decompose the WRT invariant into $\widehat Z$-invariant. Finally, we conclude in section (\ref{conclusion}).

\section{$\widehat{Z}$-invariant for $SU(N)$ group}
\label{review_3d3d}
In chapter \ref{chap:3}, we reviewed the $\widehat{Z}$-invariant for negative-definite plumbed three-manifolds associated to $SU(2)$ group (\ref{reviewsu2}). For higher rank $SU(N)$ group, the $\widehat{Z}$-invariant was studied in \cite{2020higher,Cheng:2018vpl} by decomposing the WRT invariant, $\tau_k^{SU(N)}[M(\Gamma);\q]$. In fact, $\widehat{Z}$-invariant for $SU(N)$ group admits the following integral form:
\begin{gather}%\label{Zhatb}
	\widehat{Z}_b^{{\rm SU}(N)}[M(\Gamma);q] = (-1)^{\frac{N(N-1)}{2}b_+}q^{\frac{3\sigma -\operatorname{Tr}{\rm B}}{2}\frac{N^3-N}{12}}\nonumber\label{sunzhat}\\
	\hphantom{\widehat{Z}_b^{{\rm SU}(N)}(Y;q) =}{} \times {\rm v.p.}\oint_{|z_{vj}|=1}\prod_{v\in V}\prod_{1\leq j\leq N-1}\frac{{\rm d}z_{vj}}{2\pi{\rm i} z_{vj}} F_{3d}(z)\Theta_{2d}^{b}(z,q)\\
	\text{with,}~~~~F_{3d}(z) := \prod_{v\in V}\left(\sum_{\omega\in W}(-1)^{\ell(\omega)} \prod_{1\leq j\leq N-1}z_{vj}^{(\Lambda_{j},\omega(\rho))}\right)^{2-\deg v}\nonumber\\
	\hphantom{F_{3d}(z)}{} = \prod_{v\in V}\left( \prod_{1\leq j < k\leq N}\big(y_{vj}^{1/2}y_{vk}^{-1/2}-y_{vj}^{-1/2}y_{vk}^{1/2}\big) \right)^{2-\deg v},\nonumber\\
	\Theta_{2d}^{b}(z,q) :=\sum_{s\in {\rm B} Q^{L}+b}q^{-\frac{1}{2}(s,{\rm B}^{-1}s)}\prod_{v\in V}\prod_{1\leq j\leq N-1}z_{vj}^{-(\Lambda_j,s_{v})},\nonumber\label{sunzhat}
\end{gather}
where $z_{j} = \frac{y_{j}}{y_{j+1}}$, $B$ is the linking matrix associated to a plumbing graph (\ref{fig:plumbing-example}), $\sigma$ denotes the signature of linking matrix, $W$ represent the Weyl group, $\Lambda$ is the fundamental weight vector, $Q$ is the root lattice, and $\ell(\omega)$ represents the length of the Weyl group element $\omega$. Further, the principal value integral ``v.p.'' implies the average over $|W|$ number of deformed contours, each associated with a Weyl chamber.

In the following section, we will focus on the WRT invariant for quotient group $SU(N)/\mathbb{Z}_m$ which is necessary to study the corresponding $\widehat{Z}$-invariant.

\section{WRT invariant for $SU(N)/\mathbb{Z}_m$}
\label{wrtandlevel}

For a 3-manifold $M(\Gamma)$, we define the WRT invariant\footnote{we use the normalization $\tau^G_{k'}[S^3,M]=1$} for quotient group $SU(N)/\mathbb{Z}_m$ as follows\cite{Dedushenko:2018bpp,Jeffrey:1992tk,Kaul:2000xe}:
\begin{equation}
	\tau^{SU(N)/\mathbb{Z}_m}_{k'}[M(\Gamma);\q]=\widetilde{S}_{\rho\rho}^{L-1}\frac{\sum_{C^L}\prod_{v\in V}\mathcal V_v\prod_{e\in E}\mathcal E_e}{\left(\sum_{C}\mathcal V(+1\bullet)\right)^{b_+}\left(\sum_{C}\mathcal V(-1\bullet)\right)^{b_-}},
	\label{wrtA1}
\end{equation}
where $\mathcal{V}$, $\mathcal{E}$ denotes the vertex and edge factors of $L$-component plumbing graph $\Gamma$, $b_{\pm}$ represents the number of positive and negative eigenvalues of the linking matrix $B$ and $\pm1 \bullet$ denotes the single vertex with $\pm 1$ framing. The summation is performed over the set of allowed representations of the $SU(N)/\mathbb{Z}_m$ group, which are:
\begin{equation}
	C=\{\lambda\in (P_+\cap P')+\rho\;\vert\;(\lambda,\theta^\vee)<k'\}.
\end{equation}
Here, $P_+$ represents the set of dominant weights, $\theta^\vee$ refers to the maximal root, $\rho$ denotes the Weyl vector and $P'$ is a sublattice of $P$ such that there is an isomorphism between abelian group $P/P'$ and cyclic group $\mathbb{Z}_m(Q\subseteq P'\subseteq P)$. Furthermore, when $m=N$, then $P'$ is simply the root lattice $Q$, and when $m=1$, then $P'=P$. The vertex and edge factor can be expressed in terms of $\widetilde{S}$ and $\widetilde{T}$ matrices:
\begin{equation}
	\mathcal V_v=\widetilde{T}_{\lambda\lambda}^{f_v}\widetilde{S}_{\rho\lambda}^{2-\text{deg} v}\;,\;\;\mathcal E=\widetilde{S}_{\lambda\mu}.
\end{equation}
The $\widetilde{S}$ and $\widetilde{T}$ matrices are:\footnote{these matrices are exactly the usual modular transformation matrices when $m=1$}
\begin{equation}
\widetilde{S}_{\lambda\mu}=\frac{i^{|\Delta_+|}}{|P'/k'Q|^{1/2}}\sum_{\omega\in W}(-1)^{\ell(\omega)}\q^{(\omega(\lambda),\mu)},\;\;\;\;\;\widetilde{T}_{\lambda\mu}=\delta_{\lambda\mu}\q^{\frac{1}{2}(\lambda,\lambda)}\q^{-\frac{1}{2}(\rho,\rho)}
	\label{STmat}
\end{equation}
with
\begin{equation}
	\q=\exp\left(\frac{2\pi i}{k'}\right)~~~\text{and}~~~k'\in\mathbb{Z}^+,
\end{equation}
 $W$, $Q$, and $k'$ denote the Weyl group, root lattice and renormalized Chern-Simons level respectively. 
  \vspace*{-1cm}

\paragraph{\underline{Chern-Simons level for $SU(N)/\mathbb{Z}_m$ WRT invariant}}
In Ref\cite{Dijkgraaf:1989pz}, it was shown that the three-dimensional Chern-Simons gauge theories with compact gauge group $G$ are classified by fourth cohomology group of the classifying space of the gauge group: $H^4(BG;\mathbb{Z})$. The classification parameter is the Chern-Simons level of the theory which is most commonly denoted by $k$. The level $k'$ is the renormalised Chern-Simons level which is related to the bare Chern-Simons level $k$ for $SU(N)$ gauge group as follows:
\begin{equation}
	k'=k+N.
\end{equation}
However for $SU(N)/\mathbb{Z}_m$ group which is non-simply connected, the relation between $k$ and $k'$ is as follows:
\begin{equation}
	k'=\gamma k+N,
\end{equation}
where $\gamma$ is some integer which can be calculated by considering the following short exact sequence:
\begin{equation}
	1\longrightarrow \mathbb{Z}_m\longrightarrow SU(N)\stackrel{\pi}{\longrightarrow} SU(N)/\mathbb{Z}_m\longrightarrow 1,
	\label{shortexseq}
\end{equation}
where $\mathbb{Z}_m$ is the subgroup of $\mathbb{Z}_N$. Let $\alpha$ and $\tilde{\alpha}$ be the generators of $H^4(B(SU(N));\mathbb{Z})$ and $H^4(B(SU(N)/\mathbb{Z}_m);\mathbb{Z})$ respectively. Then we have the following relation:
\begin{equation}
	B\pi^*(\tilde{\alpha})=\gamma\alpha,
\end{equation}
where $\pi^*$ is the pullback map of $\pi$ in equation (\ref{shortexseq}). The factor $\gamma$ is simply determined by comparing the images of $\tilde{\alpha}$ and $\alpha$ in the cohomology group $H^*(BT;\mathbb{Z})$ where $T$ is the maximal torus of rank $N-1$. So, the factor $\gamma$ is found to be the smallest integer for which the following equation is satisfied\cite{Dijkgraaf:1989pz}:
\begin{equation}
	\frac{\gamma}{2}(\Lambda_a,\Lambda_a)\in\mathbb{Z}, \;\;\forall \;a,
\end{equation}
where $\Lambda_a$'s are the fundamental weight vectors corresponding to the subgroup $\mathbb{Z}_m$ of $\mathbb{Z}_N$. For $SU(N)/\mathbb{Z}_N$ group, $\gamma$ is determined to be $2N$ when $N$ is even and $N$ when $N$ is odd:
\begin{equation}
	k' = \begin{cases}
		Nk+N,\;\;\;\;\;\;\text{when $N$ is odd}
		\\
		2Nk+N,\;\;\;\;\text{when $N$ is even}
	\end{cases}.
\end{equation}
For clarity, we have included the computation of sublattice $P'$ and Chern-Simons level $k'$ for certain non-simply connected groups in Appendix (\ref{appendB}). With this prescription of WRT for $SU(N)/\mathbb{Z}_m$, we will now focus on the corresponding $\widehat{Z}$ by studying the GPPV conjecture.

\section{GPPV conjecture for $SU(N)/\mathbb{Z}_m$}
\label{refinement}

As discussed in the previous section, the WRT invariant $\tau_{k'}^{SU(N)/\mathbb{Z}_m}[M(\Gamma);\q]$ associated with $M(\Gamma)$ is given by
\begin{equation}
	\tau^{SU(N)/\mathbb{Z}_m}_{k'}[M(\Gamma);\q]=\widetilde{S}_{\rho\rho}^{L-1}\frac{\sum_{C^L}\prod_{v\in V}\mathcal V_v\prod_{e\in E}\mathcal E_e}{\left(\sum_{C}\mathcal V(+1\bullet)\right)^{b_+}\left(\sum_{C}\mathcal V(-1\bullet)\right)^{b_-}}.
	\label{wrtA}
\end{equation}
For the sake of convenience, let's express the above equation in the following manner:
\begin{equation}
	\tau^{SU(N)/\mathbb{Z}_m}_{k'}[M(\Gamma);\q]=\frac{ F[M(\Gamma);\q]}{\left( F[M(+1\bullet);\q]\right)^{b_+}\left(F[M(-1\bullet);\q]\right)^{b_+}},
\end{equation}
where
\begin{equation}
	 F[M(\Gamma);\q]=\frac{1}{(\widetilde{S}_{\rho\rho})^{L+1}}\sum_{\substack{C^L}}\prod_{v\in V}\mathcal V_v\prod_{e\in E}\mathcal E_e.
	\label{F(Mgamma)}
\end{equation}
Similar to the $SU(2)$ group, we will have to perform Gauss decomposition of eqn.(\ref{wrtA}) to extract the homological blocks from it. Hence we rewrite the above equation (\ref{F(Mgamma)}) in a form so that we can use Gauss sum reciprocity formula\cite{DELOUP2007153}. We achieve this by extending the summation range $C^L$ over all Weyl chambers $W(C)^L$. Note that the matrices are invariant under the action of Weyl group elements.\footnote{upto a sign but that will not affect our final answer for $\tau^{SU(N)/\mathbb{Z}_m}_{k'}[M(\Gamma);\q]$} Hence we can sum over all the Weyl chambers and divide by the number of Weyl chambers to rewrite (\ref{F(Mgamma)}) as
\begin{equation}
	 F[M(\Gamma);\q]=\frac{1}{(\widetilde{S}_{\rho\rho})^{L+1}|W|^L}\sum_{\substack{W(C)^L}}\prod_{v\in V}\mathcal V_v\prod_{e\in E}\mathcal E_e,~~~~~~~~~~~~~~~~~~~~~~~~~~~~~~~~~~~~~~~~~~~~~~~~~~~~~~~~~~~~~~~~~~
\end{equation}
\begin{multline}
	~~~~~~~~~~~~~~~~~~~~=\frac{1}{(\widetilde{S}_{\rho\rho})^{L+1}|W|^L}\q^{-\frac{\sum_{j=1}^{L}f_j}{2}(\rho,\rho)}\left(\frac{i^{|\Delta_+|}}{|P'/k'Q|^{\frac{1}{2}}}\right)^{L+1}\sum_{\lambda\in W(C)^L}\underbrace{\prod_{v\in V}\left(\sum_{\omega\in W}(-1)^{\ell(\omega)}\q^{(\lambda_v,\omega(\rho))}\right)^{2-\text{deg}\;v}}_{\text{linear in }\lambda_v}\\\\
	\times\underbrace{\q^{\frac{f_v}{2}(\lambda_v,\lambda_v)}\prod_{(e_1,e_2)\in E}\sum_{\omega\in W}(-1)^{\ell(\omega)}\q^{(\omega(\lambda_{e_1}),\lambda_{e_2})}}_{\q^{\frac{1}{2}(\lambda,B\lambda)}}\;\;\;\;\;\;\;\;\;\;\;\;\;\;\;\;\;\;\;\;\;\;\;\;\;\;\;\;\;\;\;\;\;\;\;\;\;\;\;\;\;\;\;\;\;\;\;\;\;\;\;\;
	\label{wrtdec1}
\end{multline}
where we have used the fact that sum and product can be interchanged. The set over which the summation is being performed in equation (\ref{wrtdec1}) has now become $W(C)$. We further extend it to the whole lattice $((P\cap P')+\rho)/k'Q$ which is just $(P'+\rho)/k'Q$. In doing so, we observe for some representations $\lambda$, the term linear in $\lambda_v$ (\ref{wrtdec1}) will be zero:
\begin{equation}
	\prod_{v\in V}\left(\sum_{\omega\in W}(-1)^{\ell(\omega)}\q^{(\lambda_v,\omega(\rho))}\right)^{2-\text{deg}\;v}=0.~~~~~~~~~~~~~~~~~~~~~~~~~~~~~~~~~~
	\label{linearterm}
\end{equation}
Using Weyl denominator formula, the expression can be rewritten as
\begin{equation}
	\prod_{v\in V}\left(\prod_{\alpha\in \Delta_+}\left(\q^{\frac{(\lambda_v,\alpha)}{2}}-\q^{-\frac{(\lambda_v,\alpha)}{2}}\right)\right)^{2-\text{deg}\;v}=0.~~~~~~~~~~~~~~~~~~~~~~~~~~~~\nonumber 
\end{equation}		
Further, expressing $\lambda_v$ in terms of fundamental weight vectors $\Lambda_{i}$ \ie $\lambda_v=\sum_{j=1}^rn_{v_j}\Lambda_j$, the above equation becomes
\begin{equation}
~~~~~~~~\prod_{v\in V}\left(\prod_{\alpha\in\Delta_+}\left(\prod_{1\leq j\leq N-1}x_{v_j}^{\frac{(\Lambda_j,\alpha)}{2}}-\prod_{1\leq j\leq N-1}x_{v_j}^{-\frac{(\Lambda_j,\alpha)}{2}}\right)\right)^{2-\text{deg}\;v}\Bigg\vert_{x_{v_j}=\q^{n_{v_j}}}=0.
\end{equation}
 Hence, the points for which linear term in $\lambda_v$ becomes zero satisfy the following equation
 \begin{equation}
 	\sum_{1\leq j\leq N-1}n_{v_j}(\Lambda_j,\alpha)=0.
 \end{equation}
These points causes the singularity when $\text{deg }v>2$. Hence we first need to regularise the sum over these points. We introduce a parameter $\beta$ such that
\begin{equation}
	\beta\in\mathbb{C}~~~\text{and}~~~0< |\beta|< 1\nonumber.
\end{equation}
Using this parameter, we can rewrite the linear term in $\lambda_v$ as:
\begin{multline}
	~~~~~~~\prod_{v\in V}\left(\sum_{\omega\in W}(-1)^{\ell(\omega)}\q^{(\lambda_v,\omega(\rho))}\right)^{2-\text{deg}\;v}=
	\frac{1}{|W|^L}\prod_{v\in V}\Bigg[\left(\sum_{\omega\in W}(-1)^{\ell(\omega)}\q^{(\lambda_v,\omega(\rho))}\beta^{f(\omega,\omega_1)}\right)^{2-\text{deg}\;v}+\\\\ 
	\left(\sum_{\omega\in W}(-1)^{\ell(\omega)}\q^{(\lambda_v,\omega(\rho))}\beta^{f(\omega,\omega_2)}\right)^{2-\text{deg}\;v}+ \ldots+\left(\sum_{\omega\in W}(-1)^{\ell(\omega)}\q^{(\lambda_v,\omega(\rho))}\beta^{f(\omega,\omega_{|W|})}\right)^{2-\text{deg}\;v}\Bigg]\Bigg\vert_{\beta\rightarrow 1},
	\label{lineartermexpn}
\end{multline}
in which the function $f(\omega,\omega_i)$ is defined as follows:
\begin{equation}
	f(\omega,\omega_i):=\begin{cases}
		1, ~~~~~\text{when } \omega=\omega_i\\
		0, ~~~~~\text{otherwise}.
	\end{cases}
\end{equation}
The RHS of equation (\ref{lineartermexpn}) can be expanded as $|\beta|<1$:
\begin{equation}
	\frac{1}{|W|^L}\sum_{m\ge 0}\left(\sum_{s}\chi_s^m\q^{(\lambda,s)}\right)\beta^m,	
	\label{reg1}
\end{equation}
where $s=\{s_1,s_2,\ldots,s_L\}$ is some subset of $P^L$. Further, we interchange the summation to rewrite equation (\ref{reg1}) as
\begin{equation}
	\frac{1}{|W|^L}\sum_{s\in Q^L+\delta}\underbrace{\left(\sum_{m\ge 0}\chi_s^m\beta^m\right)}_{\xi^\beta_s}\q^{(\lambda,s)}=\frac{1}{|W|^L}\sum_{s\in Q^L+\delta}\xi_s^\beta \q^{(\lambda,s)},
\end{equation}
where $\chi_s^m\in \mathbb{Z}$. Hence, we can rewrite the linear term as series in $\q$, and  its coefficients $\xi_s^1$ can be determined by the following equation:
\begin{equation}
	\prod_{v\in V}\left(\sum_{\omega\in W}(-1)^{\ell(\omega)}\q^{(\lambda_v,\omega(\rho))}\right)^{2-\text{deg}\;v}=\frac{1}{|W|^L}\sum_{s\in Q^L+\delta}\xi_{s}^\beta\q^{(\lambda,s)}\Big\vert_{\beta\longrightarrow 1},
	\label{regularisation}
\end{equation}
where $\delta_v= (2-\text{deg}~v)\rho~ \text{mod}~Q$ and $\xi_s^1\in\mathbb{Z}$. In summary, we have rewritten the linear term as some series in $\q$. This series is obtained by taking an average of individual geometric series in $\q$, each determined by a specific selection of Weyl chamber. This completes our regularization of linear term. For clarity we have provided a detailed example in the appendix (\ref{appenda}). This led us to the following equation (\ref{beforereciprocity1}):
\begin{multline}
	~~~~~~~~~F[M(\Gamma);\q]=\frac{1}{(\widetilde{S}_{\rho\rho})^{L+1}|W|}q^{-\frac{\sum f_j}{2}(\rho,\rho)}\left(\frac{i^{|\Delta_+|}}{|P'/k'Q|^{\frac{1}{2}}}\right)^{L+1}\times\\\\
	\sum_{\lambda\in (P'+\rho)^{L}/k'Q^L}\left(\frac{1}{|W|^L}\sum_{s\in Q^L+\delta}\xi_{s}^\beta\q^{(\lambda,s)}\right)\q^{\frac{1}{2}(\lambda,B\lambda)}\Big\vert_{\beta\longrightarrow 1}.~~~~~~~~
	\label{beforereciprocity1}
\end{multline}
Now, in order to use the reciprocity formula we replace $Q$ with $\eta P'$ for some positive integer $\eta$ as $\eta P'\subseteq Q\subseteq P'\subseteq P$ and subsequently multiply it by the suitable factor given by the order of quotient of these two lattices. Hence, the above equation becomes:
\begin{multline}
	~~~~~~~~~F[M(\Gamma);\q]=\frac{1}{(\widetilde{S}_{\rho\rho})^{L+1}|W|}q^{-\frac{\sum f_j}{2}(\rho,\rho)}\left(\frac{i^{|\Delta_+|}}{|P'/k'Q|^{\frac{1}{2}}}\right)^{L+1}\underbrace{\left(\frac{1}{|Q/\eta P'|^{L}}\right)}_{\text{factor which compensates for replacing}~Q~\text{with}~\eta P'}\times\\\\
	\sum_{\lambda\in (P'+\rho)^{L}/\eta k'(P')^L}\left(\frac{1}{|W|^L}\sum_{s\in Q^L+\delta}\xi_{s}^\beta\q^{(\lambda,s)}\right)\q^{\frac{1}{2}(\lambda,B\lambda)}\Big\vert_{\beta\longrightarrow 1}.~~~~~~~~
	\label{beforereciprocity2}
\end{multline}
Since we are interested in non-simply connected group $SU(N)/\mathbb{Z}_m$ for which $\rho\notin P'$, we have to do the following shift in $\lambda$: $\lambda\longrightarrow \lambda+\boldsymbol{\rho}$, where $\boldsymbol{\rho}=(\underbrace{\rho,\rho,\ldots,\rho}_{L\text{-times}})$. Subsequently, we get the following:
\begin{multline}
	~~~~~~~~F[M(\Gamma);\q]=\frac{1}{(\widetilde{S}_{\rho\rho})^{L+1}|W|^{L+1}}\q^{-\frac{\sum f_j}{2}(\rho,\rho)}\left(\frac{i^{|\Delta_+|}}{|P'/k'Q|^{\frac{1}{2}}}\right)^{L+1}\left(\frac{1}{|Q/\eta P'|^{L}}\right)\q^{\frac{1}{2}(\boldsymbol\rho,B\boldsymbol\rho)}\times\\\\
	\sum_{s\in Q^L+\delta}\xi_s^\beta \q^{(\boldsymbol\rho,s)}\sum_{\lambda\in (P')^L/\eta k'(P')^L}\q^{\frac{1}{2}(\lambda,B\lambda)}\q^{(\lambda,s+B\boldsymbol\rho)}\Big\vert_{\beta\longrightarrow 1}.~~~~~~~~~~~
\end{multline}
\\
Now using Gauss sum reciprocity formula\cite{DELOUP2007153} and with the assumption that the quadratic form, $B:\mathbb{Z}^L\times \mathbb{Z}^L\longrightarrow \mathbb{Z}$ is negative definite\footnote{that is $\sigma=-L$},\footnote{in following equation $l$ denotes the rank of the lattice $(P')^L$ and $\ell$ represents the length of the Weyl group element}, $F[M(\Gamma);\q]$ equals:
\begin{gather}
	=\left(\frac{1}{|W|\sum_{\omega\in W}(-1)^{\ell(\omega)}\q^{(\rho,\omega(\rho))}}\right)^{L+1}\q^{-\frac{\sum f_j}{2}(\rho,\rho)}\left(\frac{1}{|Q/\eta P'|^L}\right)\left(\frac{\exp(\frac{\pi i\sigma}{4})\;(\eta k')^{l/2}}{|\text{det}(\eta B)|^{\frac{N-1}{2}}\;\;\text{Vol}[((P')^\bullet)^L]}\right)\times\nonumber\\ \nonumber\\
	\sum_{a\in ((P')^\bullet)^L/\eta B((P')^\bullet)^L}\exp[-\pi ik'(a,B^{-1}a)]\sum_{b\in (Q^L+\delta)/BQ^L}\exp[-2\pi i(a,B^{-1}(b+B\boldsymbol\rho))]\sum_{s\in BQ^L+b}\xi_s^\beta \q^{-\frac{(s,B^{-1}s)}{2}}\Big\vert_{\beta\longrightarrow 1},
\end{gather}
where $(P')^\bullet$ denotes the dual lattice of $P'$. The WRT invariant $\tau^{SU(N)/\mathbb{Z}_m}_{k'}[M(\Gamma);\q]$ including the framing factor reduces to
\begin{multline}
	=\frac{1}{|W|^{L+1}}\q^{-\frac{(\rho,\rho)}{2}(3L+\text{Tr} B)}\underbrace{\left(\sum_{a\in (P')^\bullet/\eta(P')^\bullet}\exp(\pi ik'(a,a)-2\pi i(a,\rho))\right)^{-L}}_{|(P')^\bullet/\eta(P')^\bullet|^{-L}}
	\frac{1}{|\text{det} B|^{\frac{N-1}{2}}\sum_{\omega\in W}(-1)^{\ell(\omega)}\q^{(\rho,\omega(\rho))}}\times\\\\
	\underbrace{\sum_{a\in ((P')^\bullet)^L/\eta B((P')^\bullet)^L}}_{|(P')^\bullet/\eta(P')^\bullet|^L\sum_{a\in ((P')^\bullet)^L/B((P')^\bullet)^L}}\exp(-\pi ik'(a,B^{-1}a))\sum_{b\in (Q^L+\delta)/BQ^L}\exp(-2\pi i(a,B^{-1}(b+B\rho)))\times\\\\
	\sum_{s \in BQ^L+b}\xi_s^\beta\q^{-\frac{(s,B^{-1}s)}{2}}\Big\vert_{\beta\longrightarrow 1},~~~~~~~~~~~~~~~~~~~~~~~~~~~~~~~~~~~~~~~~~~~~~~~~~~~~~~~~~~~~~~~~~~~~~~~~~~~~
\end{multline}
which simplifies to the following:
\begin{multline}
		\tau^{SU(N)/\mathbb{Z}_m}_{k'}[M(\Gamma);\q]=\frac{1}{|W|^{L+1}}\;
		\frac{q^{-\frac{(\rho,\rho)}{2}(3L+\text{Tr} B)}}{|\text{det} B|^{\frac{N-1}{2}}\sum_{\omega\in W}(-1)^{\ell(\omega)}q^{(\rho,\omega(\rho))}}\sum_{a\in ((P')^\bullet)^L/B((P')^\bullet)^L}\exp(-\pi ik'(a,B^{-1}a))\\\\
		\times\sum_{b\in (Q^L+\delta)/BQ^L}\exp(-2\pi i(a,B^{-1}(b+B\boldsymbol\rho)))\sum_{s \in BQ^L+b}\xi_s^\beta \q^{-\frac{(s,B^{-1}s)}{2}}\Big\vert_{\beta\longrightarrow 1}.~~~~~~~~~~~~~
\end{multline}
Now, assuming that the following holds:
\begin{equation}
	\lim_{\beta\longrightarrow 1}\sum_{s \in BQ^L+b}\xi_s^\beta \q^{-\frac{(s,B^{-1}s)}{2}}=\lim_{q\rightarrow \q}\sum_{s \in BQ^L+b}\xi_s^1 q^{-\frac{(s,B^{-1}s)}{2}},
\end{equation}
we finally obtain,
\begin{multline}
	\tau^{SU(N)/\mathbb{Z}_m}_{k'}[M(\Gamma);\q]=\frac{1}{|W|^{L+1}}\;
	\frac{q^{-\frac{(\rho,\rho)}{2}(3L+\text{Tr} B)}}{|\text{det} B|^{\frac{N-1}{2}}\sum_{\omega\in W}(-1)^{\ell(\omega)}q^{(\rho,\omega(\rho))}}\sum_{a\in ((P')^\bullet)^L/B((P')^\bullet)^L}\exp(-\pi ik'(a,B^{-1}a))\\\\
	\times\sum_{b\in (Q^L+\delta)/BQ^L}\exp(-2\pi i(a,B^{-1}(b+B\boldsymbol\rho)))\sum_{s \in BQ^L+b}\xi_s^1 q^{-\frac{(s,B^{-1}s)}{2}}\Big\vert_{q\longrightarrow \q},~~~~~~~~~~~~~
	\label{gaussdecomposed}
\end{multline}

\begin{multline}
	~~~~~~~~~~~~~~~~~~~~~~~~~=\frac{1}{|W||\text{det} B|^{\frac{N-1}{2}}\sum_{\omega\in W}(-1)^{\ell(\omega)}\q^{(\rho,\omega(\rho))}}\sum_{a\in ((P')^\bullet)^L/B((P')^\bullet)^L}\exp(-\pi
	ik'(a,B^{-1}a))\times\\\\
	\sum_{b\in (Q^L+\delta)/BQ^L}\exp(-2\pi i(a,B^{-1}(b+B\boldsymbol\rho)))\lim_{q\rightarrow \q}	\underbrace{\widehat{Z}^{\mathfrak{su}(N)}_b[M(\Gamma);q]}_{\text{independent of }m}.~~~~~~~~~~~~~~~~~~~~~
\end{multline}
From equation (\ref{gaussdecomposed}) explicit expression of $\widehat{Z}$-invariant can be read off as:
\begin{equation}
	\widehat{Z}^{\mathfrak{su}(N)}_b[M(\Gamma);q]=|W|^{-L}q^{-\frac{(3L+\text{Tr}\;B)}{2}(\rho,\rho)}\sum_{s\in BQ^L+b}\xi_s^1 q^{-\frac{(s,B^{-1}s)}{2}}\;\;\in\;|W|^{-L}q^{\Delta_b}\mathbb Z[[q]],
\end{equation}
Thus we have shown that $\widehat{Z}$-invariant does not depend on $m$. The overall factor which relates the $\widehat Z$ with $\tau^{SU(N)/\mathbb{Z}_m}_{k'}[M(\Gamma);\q]$ has the $m$ dependence. This led us to the following proposition:
 \vspace*{-\baselineskip}

\begin{proposition}
	Let $M(\Gamma)$ be a negative definite plumbed 3-manifold. For non-simply connected group $SU(N)/\mathbb{Z}_m$, WRT invariant can be decomposed in the following form:
	\begin{multline}
	\tau^{SU(N)/\mathbb{Z}_m}_{k'}[M(\Gamma);\q]=\frac{1}{|W||\text{det} B|^{\frac{N-1}{2}}\sum_{\omega\in W}(-1)^{\ell(\omega)}\q^{(\rho,\omega(\rho))}}\sum_{a\in ((P')^\bullet)^L/B((P')^\bullet)^L}\exp(-\pi
	ik'(a,B^{-1}a))\times~~~~\\\\
\sum_{b\in (Q^L+\delta)/BQ^L}\exp(-2\pi i(a,B^{-1}(b\textcolor{blue}{+B\boldsymbol\rho})))\lim_{q\rightarrow \q}	\widehat{Z}^{\mathfrak{su}(N)}_b[M(\Gamma);q],~~~~~~~~~~~
\label{proposition}
	\end{multline}
where $\bullet$ denotes the dual operation on the lattice $P'$, $k'=\gamma k+N$, and $\q=\exp\left(\frac{2\pi i}{k'}\right)$.\\
\end{proposition}
Moreover, we can express the terms appearing as coefficients to $\widehat{Z}$-invariant as linking pairing and homology group. The linking pairing is defined as follows:\\
\begin{definition}[Linking pairing]
For a closed and connected 3-manifold $M$, with $\partial M=\emptyset$, we have the linking pairing($\ell k$) on the torsion part of $H_1(M;\mathbb{Z})$,
\begin{equation}
	\ell k: \text{Tor}~H_1(M;\mathbb{Z})\otimes \text{Tor}~H_1(M;\mathbb{Z})\longrightarrow \mathbb{Q}/\mathbb{Z}.
\end{equation}
For $a,b\in \text{Tor}~H_1(M;\mathbb{Z})$, $\ell k$ is given as:
\begin{equation}
\ell k(a,b)=\frac{\#(a'\cdot b)}{n}~\text{mod}~\mathbb{Z},
\end{equation}
where $n\in \mathbb{Z}_{\neq 0}$ such that $n~a=0\in H_1(M;\mathbb{Z})$ and $a'$ is a 2-chain which is bounded as $\partial a'=na$. For plumbed 3-manifold $M(\Gamma)$, $\ell k$ is simply,
\begin{equation}
	\ell k(a,b)=(a,B^{-1}b)~\text{mod}~\mathbb{Z},~~~~a,b\in \mathbb{Z}^L/B\mathbb{Z}^L.
\end{equation}
\end{definition}
\vspace{0.5cm}
Using this we write the Gukov-Pei-Putrov-Vafa conjecture for $SU(N)/\mathbb{Z}_m$ as follows:\\
\begin{conjecture}
Let $M$ be a closed 3-manifold with $b_1(M)=0$ and $\text{Spin}^c(M)$ be the set of $\text{Spin}^c$ structures on $M$. Then WRT invariant $\tau^{SU(N)/\mathbb{Z}_m}_{k'}[M;\q]$ can be decomposed as follows:
\begin{align}
\tau^{SU(N)/\mathbb{Z}_m}_{k'}[M;\q]=\frac{1}{|H_1(M;\mathbb{Z})|^{\frac{N-1}{2}}\sum_{w\in W}(-1)^{\ell(w)}\q^{(\rho,\omega(\rho))}}\sum_{a,b\in (\text{Spin}^c(M))^{(N-1)}/S_N}\exp(-2\pi i k'\sum_{i=1}^{N-1}\ell k(a_i,a_i))\times\nonumber\\\nonumber\\
\textcolor{blue}{\exp(-2\pi i\sum_{i=1}^{N-1}a_i)}\exp(-4\pi i\sum_{i=1}^{N-1}\ell k(a_i,b_i))\lim_{q\rightarrow\q}\widehat{Z}_b^{\mathfrak{su}(N)}[M;q],
\end{align}
where $\widehat{Z}_b^{\mathfrak{su}(N)}[M;q]\in |W|^{-c}q^{\Delta_b}\mathbb{Z}[[q]]$ and $S_N$ is the symmetric group of degree $N$.\\
\end{conjecture}
Note that for simply connected $SU(N)$ group, appendix(\ref{appenc}), there is no shift in $b$. The shift in eqn.(\ref{proposition}), $b\longrightarrow b+B\boldsymbol\rho$, is attributed to the non-simply connected nature of $SU(N)/\mathbb{Z}_m$ group. This introduces the term $\exp(-2\pi i\sum_{i=1}^{N-1}a_i)$ in the above conjecture.

%Further, we observe that the term $$ is arising due to the non-simply connected nature of the $SU(N)/\mathbb{Z}_m$ group.
\section{Conclusion and Discussion}
\label{conclusion}
In this chapter, we have worked out the explict form of GPPV conjecture for the case of $SU(N)/\mathbb{Z}_m$ gauge group. We have found that the $\widehat{Z}$-invariant is independent of $\mathbb{Z}_m$ factor. In fact, it turns out that the dependence of $\mathbb{Z}_m$ arises as an overall factor to WRT invariant:

\begin{equation}\boxed{
	\tau^{SU(N)/\mathbb{Z}_m}_{k'}[M;\q]=\sum_{b}c_b(\mathbb{Z}_m)\widehat{Z}_b^{\mathfrak{su}(N)}[M;q]\Big|_{q\rightarrow e^{\frac{2\pi i}{k'}}}.}
\end{equation}

Moreover, in the process of Gauss decomposition of WRT invariant, there exists singularities correspondingly to the walls of the Weyl group. For certain cases of quotient groups $SU(N)/\mathbb{Z}_m$, these singularities do not arise by definition(For eg. $SU(2)/\mathbb{Z}_2$). We are interested in observing the progression of the proof for the GPPV conjecture in these particular instances. Although, recently a proof of this conjecture appeared for simply laced Lie alegbras\cite{murakami2023homological} but the proof is not available for non-simply connected groups or quotient groups.
\chapter{Knots-Quivers Correspondence for\\ Double-Twist Knots}
\label{chap:5}

Recall that in chapter \ref{chap:1}, we introduced the knots-quivers correspondence (KQC). This correspondence allows us to express the generating function of the symmetric $r$-colored HOMFLY-PT polynomial, $P_r(K;a,\q)$, in terms of the motivic generating series associated with the symmetric quiver (\ref{DTinv}). Further, the quivers associated with knot $K$ are symmetric quivers.

Moreover, KQC has been studied for a class of torus knots $(2,2p+1)$, twist knots $K_p(p\in\mathbb{Z})$ and other knots up to seven crossings \cite{Kucharski:2017ogk}. In fact, this correspondence has been proven for arborescent knots in \cite{Stosic2018NO2}. Additionally, KQC is not a bijection \ie, except for unknot and trefoil, knots have more than one quiver presentation with the same number of nodes. In Ref.\cite{PhysRevD.104.086017}, equivalent quivers with the same number of nodes were shown as vertices on a permutohedra graph, giving a systematic enumeration of such equivalent quivers. 

There is a family of knots called arborescent knots whose $r$-colored HOMFLY-PT polynomials can be explicitly computed \cite{Mironov:2015aia,Mironov:2016deg}. The work on the  existence  of quivers for all rational knots, tangles and arborescent knots \cite{Stosic:2020xwn,Stosic2018NO2}  motivated us to deduce quivers for our arborescent knot family. Even though the problem is concrete, finding explicit quivers for this universal arborescent family appears to be a hard problem. 

As a first step, we wanted to investigate some arborescent knots whose Alexander polynomial has a structure similar to that of the twist knots $K_p$. That is., $\Delta(K_p;X) =1-pX$. In fact, there is a systematic reverse engineering approach of the Melvin-Morton-Rozansky (MMR) formalism to obtain the quiver representation for such twist knots\cite{Banerjee_2020}. We observed $\Delta(8_3;X)= 1-4X$ for knot $8_3$ is part of the family of double twist knots characterized by two variables, denoted as $K(p,-m)$, illustrated in Figure \ref{DT}. Note $p,m \in \mathbb Z_+$ denote the number of full-twists and the Alexander polynomial is $\Delta\left(K(p,-m);X\right)= 1- pm X$.  Such a form motivated us to attempt quiver representation for double twist knots.

Even though the $r$-colored HOMFLY-PT\footnote{The colored HOMFLY-PT in this chapter matches with that in chapter \ref{chap:1}, table \ref{table1} under variable changes $a\rightarrow a^{\frac{1}{2}}$, $\q\rightarrow \q^{\frac{1}{2}}$. For colored Jones polynomial, change of variable is $\q\rightarrow \q^{\frac{1}{2}}$.} for any double twist knot in the cyclotomic form are known\cite{chen2021cyclotomic,Wang:2020uz}, rewriting them in the form of motivic series is still a challenging problem. We tried to determine  the quiver representation of $P_r(K(p,-m);a=\q^N,\q)$  following the methodology in Ref.\cite{Banerjee_2020}. However, we faced computational difficulty in deducing the $a$ dependence in the quiver representation. Also, we know that the quiver matrix $C^{(K)}$  do not depend on the variable $a=\q^N$. As our aim is to conjecture the quiver matrix form for the double twist knots $K(p,-m)$,  we focus on rewriting $r$-colored Jones polynomial ($a=\q^{N=2}$) :
 $$J_r(K(p,-m);\q) \equiv P_r\left(K(p,-m);a=\q^2,\q\right)~,$$
 as a motivic series. Particularly, we  obtain  quiver  matrix $C_{i,j}^{K(p,-m)}$ associated with the $K(p,-m)$ for $m\leq 3$. We conjecture that the quiver matrix $C_{i,j}^{K(m,-m)}$  is sufficient to recursively generate the quiver matrix for all the double twist knots $K(p\neq m,-m)$.

 We follow the route of reverse engineering of MMR expansion\cite{Banerjee_2020}  to derive  the motivic series form for $P_r\left(K(p,-m);a=\q^N,\q\right)\Big |_{N=2}$.
 We will now briefly review the reverse engineering formalism, which will set the notation and procedure we follow for $K(p,-m)$ in the next section.
 
  \textbf{\underline{Reverse Engineering of Melvin-Morton-Rozansky (MMR) expansion}}\\  
  Melvin-Morton-Rozansky(MMR) expansion states that  the symmetric $r$-colored  HOMFLY-PT for knot $K$ has the following  semiclassical expansion:
\begin{eqnarray} 
\lim_{\hslash\to 0, r\to\infty}  P_r(K;a,\q=e^{\hslash})&\simeq & \frac{1}{\Delta(K;x)^{N-1}} + \nonumber\\ &&\sum_{k=1}^\infty \left(\frac{R_k(K;x,N)}{\Delta(K;x)}\right)^{N+2k-1} \hslash^k,\nonumber\\
\label{mmr-intro}
\end{eqnarray}
with the leading term being the Alexander polynomial $\Delta(K;x)$ and the variable  $x$ in terms of color $r$ is $x=\q^r={\rm ~const}$. The symbol $R_k(K;x, N)$ represent polynomials in the variable $x$. The reverse approach is to obtain 
$P_r(K;a,\q)$ using the Alexander polynomial $\Delta(K;x)$\cite{Banerjee_2020}. This approach also has obstacles to lift the $\hslash \rightarrow 0$ expansion to $\q$-dependent $P_r(K;a,\q)$  but can be fixed for some situations by comparing with the data of symmetric $r$-colored HOMFLY-PT polynomials known for $r=1,2,3$.

We will briefly highlight the steps involved in the reverse engineering formalism of MMR expansion\cite{Banerjee_2020}: 
\begin{enumerate}
    \item[i.] We rewrite the Alexander polynomial in new variable $X=\frac{(1-x)^2}{x}$. Thus, the Alexander polynomial takes the following form:
    \begin{equation*}
        \Delta(x)=1-\sum_{i=1}^{s}a_i\frac{(1-x)^{2i}}{x^i}\equiv 1-g(X)
    \end{equation*}
where the coefficients $a_i$ are integers, and $s$ is a positive integer.
    \item[ii.] Now, we use the following inverse binomial theorem \begin{equation*}
        \frac{1}{(1-u)^n}=\sum_{m=0}^{\infty}\binom{n+m-1}{m}u^m
    \end{equation*}
to write the first term of MMR expansion (\ref{mmr-intro}) as follows:
\begin{eqnarray}{\label{MMRE}}
    \frac{1}{\Delta(x)^{N-1}}&=&\sum_{m=0}^{\infty}\binom{N+m-2}{m}g(X)^m=\sum_{k=0}^{\infty}c_k X^k\nonumber\\
   &=& \sum_{k=0}^{\infty}(k! c_k)\frac{X^k}{k!}
\end{eqnarray}
\item[iii.] We make the following quantum deformation to get the quantum-deformed polynomial:\\
    \begin{equation*}
        \frac{X^k}{k!}=\frac{(1-x)^{2k}}{k!x^k}\rightsquigarrow \qbinom{r}{k} \q^{-rk}(-a \q^rt^3;\q)_k.
    \end{equation*}
Here, the variable $t$ is known as the refined parameter. In this article, we will take $t = -1$ to obtain unrefined polynomial invariants for double twist knots.  The term within parentheses represents the $\q$-Pochhammer, while square brackets correspond to the $\q$-binomials, which are defined as: 
    $$\qbinom{r}{k}=\frac{(\q^2;\q^2)_r}{(\q^2;\q^2)_k(\q^2;\q^2)_{r-k}}~~~;~~~ (x;\q)_k=\prod_{i=0}^{k-1}(1-x \q^{i}).$$
\item[iv.] Further, the coefficient $k!c_k\xrightarrow{\rm \q-deformation} \Tilde{c}^{K}_k$ depends on the knot $K$ and must be written in terms of   $\q$-Pochhammers, $\q$-binomials, and $(\q,a)$- dependent powers so that 
% Thus, we get the quantum-deformed polynomial which has been obtained from Alexander polynomial:
\begin{equation*}
    P_r(K;a,\q)=\sum_{k=0}^{r}{r \brack k}_{\q}\q^{-rk}(a \q^r;\q)_k\Tilde{c}^{K}_k
\end{equation*} 
can be transformed into the following form to deduce  the corresponding quiver $Q_{K}$:
\begin{eqnarray}
P_r(K;a,\q)&=&  
\sum_{\mathbf d}(-1)^{\sum_i\gamma_i d_i}\frac{q^{\sum_{i,j} C^{K}_{i,j} d_i d_j}(\q^2;\q^2)_r}{\prod_{i=1}^m (\q^2;\q^2)_{d_i}}  \nonumber\\
&&\q^{\sum
\alpha_i d_i} a^{\sum \beta_i d_i}.   \label{Pr-quiver1}
\end{eqnarray}
Here $C^{K}_{i,j}$ is the quiver matrix and the variables $\alpha_i,~\beta_i$ and $\gamma_i$ are integer parameters. The set $\{d_i\}\equiv \mathbf d$ must obey $r = d_1 +d_2 \ldots + d_n$ with $d_i\geq 0$.
Even though such a transformation is motivated by comparing Ooguri-Vafa partition function\cite{Ooguri:2000tr} with the motivic generating series\cite{MR2889742,MR2851153,MR2956038}, it is still a hard problem to obtain $\Tilde{c}^{K}_k$ for any knot.
\end{enumerate}
\vspace*{-\baselineskip}
Note that the quadratic power of $\q$ depends on $C^{K}_{i,j}$ and it is independent of  $a=\q^N$.
Hence, we will work with the colored Jones polynomials $J_r(K;\q)$ of a knot $K$ to extract its quiver matrix using the reverse engineering techniques of MMR formalism replacing $a\rightarrow \q^2$ in eqn.(\ref{Pr-quiver1})\footnote{ Theorem 1.1, in  Ref.\cite{Stosic2018NO2} indicates the colored Jones polynomials of rational links also admit generating functions in quiver form.}.  

The plan of the chapter is as follows: In section \ref{sec2}, we briefly discuss the colored Jones polynomials of double twist knot $K(p,-m)$ obtained from the reverse engineering techniques of MMR expansion. In section \ref{sec3}, we conjecture $C^{K}_{i,j}$ for $K = K(p,-m)$ and validate it for some double twist knots. We conclude in section \ref{sec4} by summarising our results.

%**********************************************************
%**********************************************************

%**********************************************************
\section{ Double twist knots}    \label{sec2}
\begin{figure}[ht]
    \centering
    \begin{minipage}{0.45\textwidth}
        \centering
        \resizebox{5.3cm}{3.4cm}{
        \begin{tabular}{|c|c|}
           \hline
            $\mathbf{K(p,-m)}$& \textbf{Knots} \\
            \hline
            $K(p,-1)$ & Twist Knots \\
            \hline
            K(2,-2)& $8_3$\\
            \hline
            K(3,-2)& $10_3$ \\
            \hline
        \end{tabular}}
        \caption{Examples of  double twist knots $K(p,-m)$}
        \label{table:1}
    \end{minipage}\hfill
    \begin{minipage}{0.45\textwidth}
        \centering
        \includegraphics[width=7cm,height=7cm]{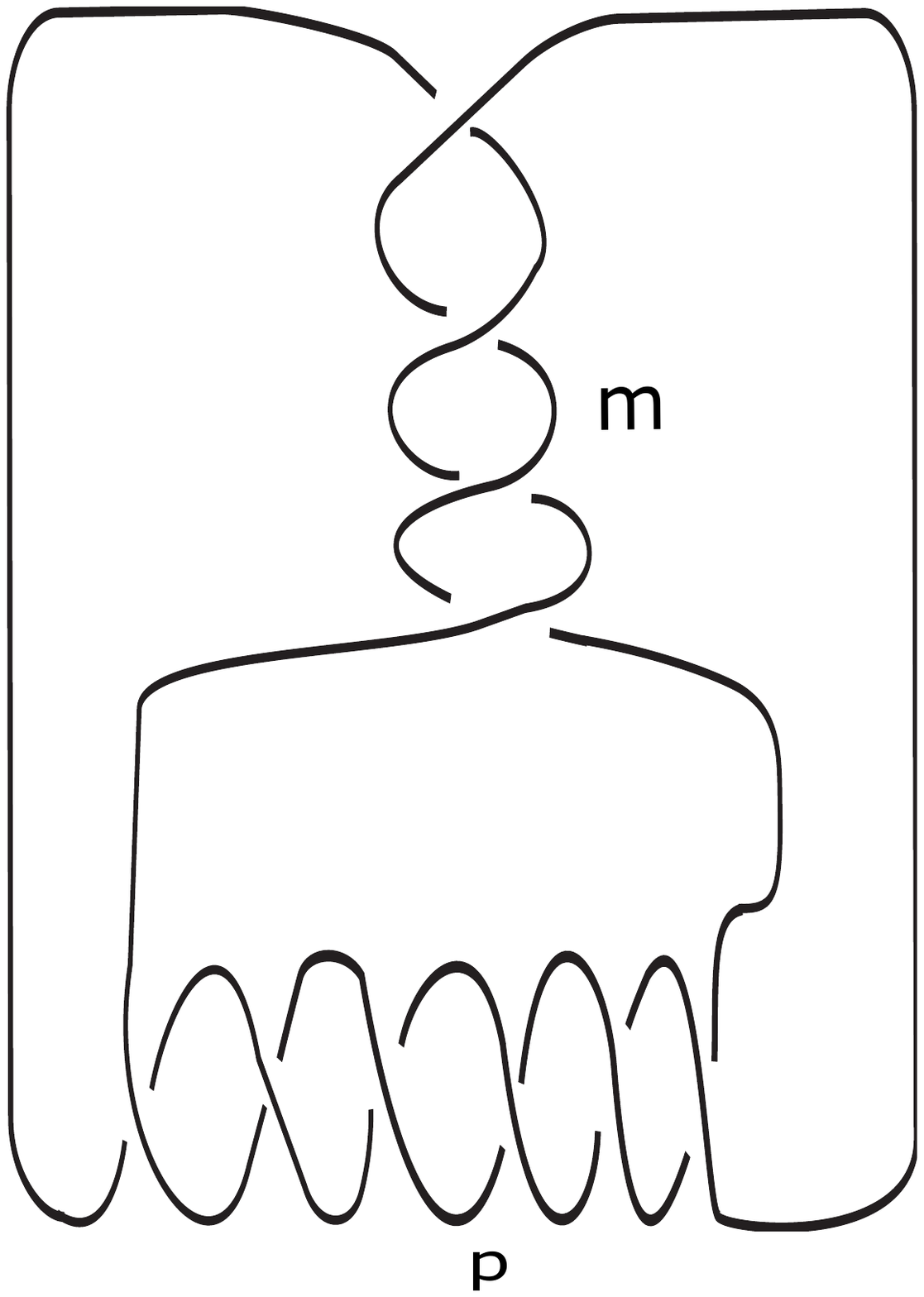}
        \caption{Double twist knots $\newline$ ($m, p$ denote number of  full-twists)}
        \label{DT}
    \end{minipage}
\end{figure}

We have listed some of the double twist knots $K(p,-m)$ in Table \ref{table:1}.
As these double twist knots belong to arborescent family, the symmetric $r$-colored HOMFLY-PT polynomials can be obtained for every $r$ from Chern-Simons theory\cite{Devi:1993ue,Nawata:2012vk}. In fact, colored HOMFLY-PT for arbitrary $r$ in closed form is given in Ref. \cite{chen2021cyclotomic}. Hence, our aim is not to reconstruct  $r$-colored HOMFLY-PT for double twist knots. We will now present the reverse engineering of MMR formalism (\ref{mmr-intro}) to rewrite $r$-colored Jones as a motivic series to extract the matrix of the quiver $Q_{K(p,-m)}$. 
\subsection{Colored HOMFLY-PT polynomials for a class of Double twist knots $K(p,-m)$ }\label{sec2}
For given positive integers $p$ and $m$, the Alexander polynomial of a double twist knot of type $K(p,-m)$ takes the form
\begin{equation}
    \Delta\left(K(p,-m);x\right) = 1 - (p ~m) X .    \label{Alexander-twist-I}
\end{equation}
Here $X=\frac{(1-x)^2}{x}$. Such a linear expression  appeared in many knots \cite{Banerjee_2020}, suggesting the
inverse binomial expansion to take the following form:
\begin{equation}
    \frac{1}{\Delta\left(K(p,-m);x\right)^{N-1}}=\sum_{0 \le k_1 \le \ldots \le k_{2 m p} }^\infty  {N+k_{2 m p}-2 \choose k_{2 m p}}
\prod_{i=1}^{2mp-1}{k_{i+1} \choose k_i}  \frac{(1-x)^{2k_{2 m p}}}{x^{k_{2 m p}}}.\label{CBE}
\end{equation}
Further, using the quantum deformation procedure discussed in \cite{Banerjee_2020} and taking $a\rightarrow \q^2 (N=2)$ in eqn.(\ref{CBE}), the colored Jones polynomial  can be written as 
\begin{equation}
J_r(K(p,-m);\q)=  
\sum_{d_1+d_2 +\ldots +d_{4mp+1}=r}(-1)^{\sum_i \gamma_i d_i} \q^{\sum
\xi_i d_i}\frac{\q^{\sum_{i,j} C^{K(p,-m)}_{i,j} d_i d_j} (\q^2;\q^2)_r}{\prod_{i=1}^m (\q^2;\q^2)_{d_i}},  \label{Pr-quiver}
\end{equation}
where $C^{K(p,-m)}_{i,j}$ is $(4 pm+1) \times (4 pm+1)$  matrix for quiver $Q_{K(p,-m)}$. It is worth noting that $\xi_i = \alpha_i + 2 \beta_i$ and $\gamma_i$ are integer parameters that can be determined by comparing them with  $r=1,2,3$ \cite{chen2021cyclotomic,lovejoy2017colored}.  By this approach, we explicitly  determined $\{\xi_i\},$
$\{\gamma_i\}$ parameters(\ref{Pr-quiver}) for $K_{(2,-2)}={\bf{8_3}}$ knot:
\begin{equation*} \label{83qui}
J_r(8_3;\q)= \sum_{d_1+d_2 +\ldots +d_{17}=r}(-1)^{\sum_i \gamma_i d_i}
(\q^2;\q^2)_r \frac{q^{\sum_{i,j} C_{i,j} d_i d_j }}{\prod_{i=1}^{17} (\q^2;\q^2)_{d_i}} \q^{\sum
\xi_i d_i},
\end{equation*}
where, 
\begin{align*}
\sum_i \gamma_i d_i & = d_{11} + d_{13} + d_{14} + d_{16} + d_{3} + d_{5} + d_{6} + d_{8},\\  
\sum_i\xi_i d_i & = (d_{11} - 2 d_{12} - d_{13} + d_{14} + 2 d_{15} + 3 d_{16} + 4 d_{17} - 2 d_2 - d_3 - 4 d_4 - 3 d_5 - d_6 + d_8 + 2d_9).
\end{align*}
The quiver  matrix $C^{K(2,-2)}=C^{8_3}$ is as follows:
\vskip.1cm
\noindent
\resizebox{\linewidth}{6cm}{$\left(
\begin{array}{c|cccc|cccc|cccc|cccc}
 0 & -1 & -1 & -1 & -1 & 0 & 0 & 0 & 0 & -1 & -1 & -1 & -1 & 0 & 0 & 0 & 0 \\
 \hline
 -1 & -2 & -2 & -3 & -3 & -2 & -2 & -1 & -1 & -2 & -2 & -3 & -3 & -2 & -2 & -1 & -1 \\
 -1 & -2 & -1 & -2 & -2 & -1 & -1 & 0 & 0 & -1 & -1 & -2 & -2 & -1 & -1 & 0 & 0 \\
 -1 & -3 & -2 & -4 & -4 & -3 & -3 & -1 & -1 & -2 & -2 & -4 & -4 & -3 & -3 & -1 & -1 \\
 -1 & -3 & -2 & -4 & -3 & -2 & -2 & 0 & 0 & -1 & -1 & -3 & -3 & -2 & -2 & 0 & 0 \\
 \hline
 0 & -2 & -1 & -3 & -2 & -1 & -1 & 0 & 0 & -1 & -1 & -2 & -2 & -1 & -1 & 0 & 0 \\
 0 & -2 & -1 & -3 & -2 & -1 & 0 & 1 & 1 & 0 & 0 & -1 & -1 & 0 & 0 & 1 & 1 \\
 0 & -1 & 0 & -1 & 0 & 0 & 1 & 1 & 1 & 0 & 0 & 0 & 0 & 1 & 1 & 1 & 1 \\
 0 & -1 & 0 & -1 & 0 & 0 & 1 & 1 & 2 & 1 & 1 & 1 & 1 & 2 & 2 & 2 & 2 \\
 \hline
 -1 & -2 & -1 & -2 & -1 & -1 & 0 & 0 & 1 & 0 & 0 & -1 & -1 & 0 & 0 & 1 & 1 \\
 -1 & -2 & -1 & -2 & -1 & -1 & 0 & 0 & 1 & 0 & 1 & 0 & 0 & 1 & 1 & 2 & 2 \\
 -1 & -3 & -2 & -4 & -3 & -2 & -1 & 0 & 1 & -1 & 0 & -2 & -2 & -1 & -1 & 1 & 1 \\
 -1 & -3 & -2 & -4 & -3 & -2 & -1 & 0 & 1 & -1 & 0 & -2 & -1 & 0 & 0 & 2 & 2 \\
 \hline
 0 & -2 & -1 & -3 & -2 & -1 & 0 & 1 & 2 & 0 & 1 & -1 & 0 & 1 & 1 & 2 & 2 \\
 0 & -2 & -1 & -3 & -2 & -1 & 0 & 1 & 2 & 0 & 1 & -1 & 0 & 1 & 2 & 3 & 3 \\
 0 & -1 & 0 & -1 & 0 & 0 & 1 & 1 & 2 & 1 & 2 & 1 & 2 & 2 & 3 & 3 & 3 \\
 0 & -1 & 0 & -1 & 0 & 0 & 1 & 1 & 2 & 1 & 2 & 1 & 2 & 2 & 3 & 3 & 4 \\
  \end{array}
\right)$}
\vskip.1cm

\noindent
\textcol{One can draw the quiver graph for this matrix by appropriately shifting by a constant matrix as shown in equation (\ref{shifting-of-quiver-mat}). However, drawing such quiver graphs is not particularly insightful.} To give clarity to the readers, we present a step-by-step procedure for determining the quiver matrix for 
the knot $8_3$ in the Appendix \ref{appendixd}. 
The polynomial invariants matches with the  closed form \cite{chen2021cyclotomic} for large value of $r$ as well confirming that the above $8_3$ quiver data is indeed correct. Such an  exercise for $K(2,-2)$  suggested that we could propose and conjecture $C^{K(p,-m)}$ for the double twist knot family. We discuss them in the following section.

%**********************************************************
\section{Knot-Quiver Correspondence of double twist knots $K(p,-m)$}   \label{sec3}
 We observe that the quiver matrix has a block structure by performing a similar analysis of the previous section for other examples of the double twist knots  $K(p, -m)$.  Our explicit computation suggests the following proposition.\\
\textbf{Proposition: }\\ \emph{The $r$-colored Jones polynomial for double twist knots $K(p,-m)$, with $p\geq m$, can be expressed in the quiver representation:
\begin{eqnarray}{\label{KQ}}
 J_{r}(K(p,-m);\q)&=&\sum_{d_{1}+d_{2}+\ldots +d_{4pm+1}=r}(-1)^{\Lambda_{(p,-m)}}\q^{\Xi_{(p,-m)}}\nonumber\\&&\frac{(\q^2;\q^2)_{r}}{\prod_{i=1}^{4pm+1}(\q^2;\q^2)_{d_i}}  \q^{\sum_{i,j} C^{K(p,-m)}d_{i}d_j}\nonumber\\ \label{proposition-knot-quiver}
\end{eqnarray}
where the linear term $\Xi_{(p,-m)}\equiv \sum_i \xi_i d_i$, phase factor $\Lambda_{(p,-m)}\equiv \sum_i \gamma_i d_i$.}\\
The block structure of the matrix $C^{K(p,-m)}$ for some examples lead to the following conjecture:\\
{\bf Conjecture:} \emph{ The generic structure of  the quiver  matrix will take the form}
\begin{equation}
\label{QUIVERDT}
C^{K(p,-m)}=\left(\arraycolsep=1.8pt\def\arraystretch{0.9}\begin{array}{c|c|c|c|c|c|c|c|c|c}
F_{0} & F_{k} & \Tilde{F}_{k} &\cdots & \cdots &F_{k} &\Tilde{F}_{k} &\cdots &F_{k} & \Tilde{F}_{k}\\
\hline
F_{k}^{\top} & U_{1} & R_1 &\Tilde{R}_1  & \cdots &\Tilde{R}_1&R_1  &\cdots&\Tilde{R}_1 & {R}_1  \\
\hline
\Tilde{F}_{k}^{\top} & R_{1}^{\top} & \Tilde{U}_{1} &T_1&\Tilde{T}_1  &\vdots &\vdots & \cdots &T_1 &\Tilde{T}_1 \\
\hline
{F}_{k}^{\top} & \Tilde{R}_{1}^{\top} & T_{1}^{\top} &U_2&\Tilde{R}_2  & \vdots&\vdots & \cdots &\Tilde{R}_2 &{R}_2 \\
\hline
\vdots & \vdots & \vdots & \vdots & \ddots &\vdots &\vdots &\cdots & \vdots & \vdots\\
\hline
F_{k}^{\top} & R_{1}^{\top} & \cdots & \cdots & \vdots &U_{i} &\cdots &\vdots & \Tilde{R}_{i} & {R}_{i}\\
\hline
\Tilde{F}_{k}^{\top} & \Tilde{R}_{1}^{\top} &  \cdots &R_{i}^{\top}  &  \cdots&&\Tilde{U}_{i} & \cdots& T_{i} & \tilde{T}_{i}\\
\hline
\vdots & \vdots & \vdots & \vdots & \ddots &\vdots &\vdots &\ddots & \vdots & \vdots\\
\hline
F_{k}^{\top} &\Tilde{R}_{1}^{\top} & T_{1}^{\top} & \Tilde{R}_{2}^{\top}&\cdots&\vdots & \vdots& \cdots & U_{p} & R_{p}\\
\hline
\vspace{.2cm}
\Tilde{F}_{k}^{\top} & {R}_{1}^{\top} & \Tilde{T}_{1}^{\top} & {R}_{2}^{\top} & \cdots& \cdots&\vdots & \cdots & R_{p}^{\top} & \Tilde{U}_{p}
\end{array}\right),  
\end{equation}
where $X^{\top}$ stands for transposition of matrix $X$, the row matrices $$F_{k}=\left(-1,-1,\ldots -1,-1\right), \Tilde{F}_k=\left(0,0,\ldots 0,0\right),$$ of size $1\times 2m$ and $F_0=0$. 
%Surprisingly, 
Let $X_{k}$  denote the following set of  $2m \times 2m$ matrices :
\begin{equation}{\label{G}}
X_{k}=\{U_k,\Tilde{U}_{k},R_k,\Tilde{R}_{k},T_{k},\Tilde{T}_{k}\},
\end{equation}
where $k=1,2\ldots p$. All these matrices can be recursively obtained using
$$X_{k}=X_{k-1}+2 (k-1) J,~~\forall k\geq 2,$$ where $J$ is a matrix of size $2m \times 2m$ where all the elements  are one. So, knowing the set $X_1$ is sufficient to determine the full quiver matrix $C^{K(p,-m)}$. 

It appears that the set $X_1$ for the simplest twist knot $K(p=1,-m=-1)\equiv 4_1$ will suffice to obtain $X_1$ for double twist knots $K(p,-m)$ as $K(p,-m) = K^*(m,-p)$ ($K^*$ denotes mirror image of the knot $K$).  However, our matrix conjecture 
assumes $p \geq m$.  Hence, our explicit computations of set $X_1$ for $m=2,3$ is not derivable  from the $C^{K(p,-1)}$. 

In the following subsections, we will give some examples to validate our proposition and conjecture. 
Specifically, we work out the $X_1$ set matrices for double twist knots $K(m,-m)$ for $m =1,2,3$. This set is sufficient to 
obtain the explicit quiver presentations for all the double twist knots $K(p, -m)$ where $m =1,2,3$. 
\subsection{Knot-Quiver correspondence for  $K(p,-1)$}
$K(p,-1)$ are known in the literature as `twist knots' which is the simplest class of double twist knots.
In this case, we fix the parameter $m=1$ and vary the other parameter $p$. 
% Note that theorem 1.1, in \cite{Stosic:2017wno},  implies that the colored Jones polynomials of rational links also admit generating functions in quiver form}. Thus, to avoid computational complexity, we consider colored jones polynomials i.e $A\rightarrow q^2$ limit of Eqn.(\ref{KQ}).  
The simplest example, we consider $p=1$, i.e. $K(1,-1)={\bf 4_1}$ knot. Using eqn.(\ref{KQ}), we obtained the quiver form of ${\bf 4_1}$ as
\begin{align*}
J_{r}(4_1;\q) &= \sum_{d_1+\ldots+d_5=r}\left((-1)^{d_3+d_4} \q^{-2 d_2 - 2 d_1 d_2 - 2 d_2^2 - d_3 + 2 d_5^2}\right)\times\\
&~~~~~~~~~~~~~~~~~~~\left(\frac{(\q^2;\q^2)_{r}}{(\q^2;\q^2)_{d_1}(\q^2;\q^2)_{d_2}(\q^2;\q^2)_{d_3}(\q^2;\q^2)_{d_4}(\q^2;\q^2)_{d_5}}\right)\times\\\\
&~~~~~~~~~~~~~~~~~~~\left(\q^{- 2 d_1 d_3 - 4 d_2 d_3 - d_3^2 + d_4 - 
 2 d_2 d_4 + d_4^2 + 2 d_5 - 2 d_2 d_5 + 2 d_4 d_5}\right). 
\end{align*}
Thus, the quiver matrix

\begin{eqnarray*} C^{4_1}=
\resizebox{5.5cm}{2.75cm}{$\left(
\begin{array}{c|cc|cc}
 0 & -1 & -1 & 0 & 0  \\
 \hline
 -1 & -2 & -2 & -1 & -1 \\
 -1& -2 & -1 & 0& 0  \\
 \hline
 0 & -1 & 0 & 1 & 1  \\
 0 & -1 & 0 & 1 & 2  \\
 \end{array}
\right)$}.
\end{eqnarray*}

Similarly, we obtained other matrices for $p=2, 3$ i.e

\begin{eqnarray*}\centering C^{6_1}=
\resizebox{11.0cm}{5.5cm}{$\left(
\begin{array}{c|cc|cc|cc|cc}
 0 & -1 & -1 & 0 & 0 & -1 & -1 & 0 & 0 \\
 \hline
 -1 & -2 & -2 & -1 & -1 & -2 & -2 & -1 & -1 \\
 -1 & -2 & -1 & 0 & 0 & -1 & -1 & 0 & 0 \\
 \hline
 0 & -1 & 0 & 1 & 1 & 0 & 0 & 1 & 1 \\
 0 & -1 & 0 & 1 & 2 & 1 & 1 & 2 & 2 \\
 \hline
 -1 & -2 & -1 & 0 & 1 & 0 & 0 & 1 & 1 \\
 -1 & -2 & -1 & 0 & 1 & 0 & 1 & 2 & 2 \\
 \hline
 0 & -1 & 0 & 1 & 2 & 1 & 2 & 3 & 3 \\
 0 & -1 & 0 & 1 & 2 & 1 & 2 & 3 & 4 \\
\end{array}
\right)$},
\end{eqnarray*}

\begin{center}
\resizebox{\linewidth}{8cm}{$C^{8_1}=\left(
\begin{array}{c|cc|cc|cc|cc|cc|cc}
 0 & -1 & -1 & 0 & 0 & -1 & -1 & 0 & 0 & -1 & -1 & 0 & 0 \\
 \hline
 -1 & -2 & -2 & -1 & -1 & -2 & -2 & -1 & -1 & -2 & -2 & -1 & -1 \\
 -1 & -2 & -1 & 0 & 0 & -1 & -1 & 0 & 0 & -1 & -1 & 0 & 0 \\
 \hline
 0 & -1 & 0 & 1 & 1 & 0 & 0 & 1 & 1 & 0 & 0 & 1 & 1 \\
 0 & -1 & 0 & 1 & 2 & 1 & 1 & 2 & 2 & 1 & 1 & 2 & 2 \\
 \hline
 -1 & -2 & -1 & 0 & 1 & 0 & 0 & 1 & 1 & 0 & 0 & 1 & 1 \\
 -1 & -2 & -1 & 0 & 1 & 0 & 1 & 2 & 2 & 1 & 1 & 2 & 2 \\
 \hline
 0 & -1 & 0 & 1 & 2 & 1 & 2 & 3 & 3 & 2 & 2 & 3 & 3 \\
 0 & -1 & 0 & 1 & 2 & 1 & 2 & 3 & 4 & 3 & 3 & 4 & 4 \\
 \hline
 -1 & -2 & -1 & 0 & 1 & 0 & 1 & 2 & 3 & 2 & 2 & 3 & 3 \\
 -1 & -2 & -1 & 0 & 1 & 0 & 1 & 2 & 3 & 2 & 3 & 4 & 4 \\
 \hline
 0 & -1 & 0 & 1 & 2 & 1 & 2 & 3 & 4 & 3 & 4 & 5 & 5 \\
 0 & -1 & 0 & 1 & 2 & 1 & 2 & 3 & 4 & 3 & 4 & 5 & 6 \\
\end{array}
\right).$}
\end{center}

\newpage
These three examples confirm our conjecture for $m=1$.
 For clarity, the explicit quiver matrix for any  twist knot $K(p,-1)$ is
\begin{equation}
C^{K(p,-1)}=\left(\arraycolsep=8.4pt\def\arraystretch{1.5}\begin{array}{c|c|c|c|c|c|c|c|c|c}
F_{0} & F_{k} & \Tilde{F}_{k} &\cdots & \cdots &F_{k} &\Tilde{F}_{k} &\cdots &F_{k} & \Tilde{F}_{k}\\
\hline
F_{k}^{\top} & U_{1} & R_1 &\Tilde{R}_1  & \cdots &\Tilde{R}_1&R_1  &\cdots&\Tilde{R}_1 & {R}_1  \\
\hline
\Tilde{F}_{k}^{\top} & R_{1}^{\top} & \Tilde{U}_{1} &T_1&\Tilde{T}_1  &\vdots &\vdots & \cdots &T_1 &\Tilde{T}_1 \\
\hline
{F}_{k}^{\top} & \Tilde{R}_{1}^{\top} & T_{1}^{\top} &U_2&\Tilde{R}_2 & \vdots&\vdots & \cdots &\Tilde{R}_2 &{R}_2 \\
\hline
\vdots & \vdots & \vdots & \vdots & \ddots &\vdots &\vdots &\cdots & \vdots & \vdots\\
\hline
F_{k}^{\top} & R_{1}^{\top} & \cdots & \cdots & \vdots &U_{i} &\cdots &\vdots & \Tilde{R}_{i} & {R}_{i}\\
\hline
\Tilde{F}_{k}^{\top} & \Tilde{R}_{1}^{\top} &  \cdots &R_{i}^{\top}  &  \cdots&\vdots&\Tilde{U}_{i} &\cdots & T_{i} & \tilde{T}_{i}\\
\hline
\vdots & \vdots & \vdots & \vdots & \ddots &\vdots &\vdots &\ddots & \vdots & \vdots\\
\hline
F_{k}^{\top} &\Tilde{R}_{1}^{\top} & T_{1}^{\top} & \Tilde{R}_{2}^{\top}&\cdots&\vdots & \vdots& \cdots & U_{p} & R_{p}\\
\hline
\Tilde{F}_{k}^{\top} & {R}_{1}^{\top} & \Tilde{T}_{1}^{\top} & {R}_{2}^{\top} & \cdots& \cdots&\vdots & \cdots & R_{p}^{\top} & \Tilde{U}_{p}
\end{array}\right),  
    \label{twist}
\end{equation}

where the generators $X_1\equiv (U_1, \tilde U_1,R_1, \tilde R_1, T_1, \tilde T_1)$ are as follows:

\begin{align*}
    U_1&=\left(\arraycolsep=8.4pt\def\arraystretch{1.5}
\begin{array}{cc}
 -2 & -2 \\
 -2 & -1 
\end{array}
\right),~\Tilde{U}_1=\left(\arraycolsep=8.4pt\def\arraystretch{1.5}
\begin{array}{cc}
 1&1  \\
 1& 2 \\
 \end{array}
\right),~R_1=\left(\arraycolsep=8.4pt\def\arraystretch{1.5}
\begin{array}{cc}
 -1 & -1 \\
 0& 0 \\
 \end{array}
\right),\\
\Tilde{R}_1&=\left(\arraycolsep=8.4pt\def\arraystretch{1.5}
\begin{array}{cc}
 -2 & -2   \\
 -1  &  -1\\
 \end{array}
\right),~T_1=\left(\arraycolsep=8.4pt\def\arraystretch{1.5}
\begin{array}{cc}
 0 & 0   \\
 1 & 1 \\
 \end{array}
\right),~\Tilde{T}_1=\left(\arraycolsep=8.4pt\def\arraystretch{1.5}
\begin{array}{cc}
 1 &1   \\
 2 & 2\\
 \end{array}
\right),
\end{align*}

and 
$F_{k}=\left(-1,~~~-1\right),~ \Tilde{F}_{k}=\left(0,~~~0\right),~\text{and} ~F_0=\left(0\right).$ 
Based on these calculations, we can infer the general expressions for the linear term $\Xi(p,-1)$ and the phase factor $\Lambda (p,-1)$ in proposition $(\ref{twist})$ for any given value of $p$:
\begin{equation}{\label{LT1}}
\Xi_{(p,-1)}=2p d_{4p+1}+\sum_{i=1}^{p}\left(\rho_{1}(i)d_{4 i-3}+\rho_{2}(i)d_{4 i-2}+\rho_{3}(i)d_{4 i-1}+\rho_{4}(i)d_{4 i}\right),
\end{equation} 
where~ $\rho_{1}(i)=2i-2,~ \rho_{2}(i)=2i-4,~ \rho_{3}(i)=-3+2i,~ \rho_{4}(i)=-1+2i$, and the phase factor is 
\begin{eqnarray}{\label{PH1}}
\Lambda_{(p,-1)}=\sum_{i=1}^{2p}d_{\frac{1}{2}((-1)^{i+1}+4i+1)}.
\end{eqnarray} 
These results agree with the quiver matrix of twist knots obtained in Ref.\cite{Kucharski:2017ogk}. 

\subsection{ Knot-Quiver correspondence for $K(p,-2)$}\label{knot-quiver-K(p,-2)}
We have already worked out $K(2,-2)\equiv 8_3$ knot in section \ref{sec2}. Further, we explicitly worked out colored 
Jones for $K(p=3,-2) \equiv 10_3$ and the quiver matrix elements  $C^{K(p=3,-2)}$ are presented in the Appendix \ref{appendixd}.

Our matrix form for $p=3$ is consistent with our conjecture (\ref{QUIVERDT}), and the basic set of matrices  $X_1$ are:
\begin{eqnarray}{\label{GM2}} U_{1}&=&\left(\arraycolsep=8.4pt\def\arraystretch{1.5}
\begin{array}{cccc}
 -2 & -2 &-3 &-3    \\
 -2&-1 &-2&-2 \\
 -3 &-2  & -4 &-4\\
 -3 &-2 &-4&-3\\
 \end{array}
\right),~\Tilde{U}_{1}=\left(\arraycolsep=8.4pt\def\arraystretch{1.5}
\begin{array}{cccc}
 -1 & -1 & 0&0 \\
 -1 & 0 & 1&1\\
 0 & 1 & 1&1 \\
 0 & 1 &1 &2 \\
 \end{array}
\right),\nonumber\\ 
R_{1}&=&\left(\arraycolsep=8.4pt\def\arraystretch{1.5}
\begin{array}{cccc}
 -2 & -2 & -1 & -1   \\
 -1 & -1 & 0&0 \\
 -3 & -3 & -1 &-1\\
 -2 & -2 & 0 &0\\
 \end{array}
\right),~\Tilde{R}_{1}=\left(\arraycolsep=8.4pt\def\arraystretch{1.5}
\begin{array}{cccc}
 -2 & -2 & -3&-3 \\
 -1 & -1 & -2& -2\\
 -2 & -2 & -4&-4 \\
 -1 & -1 & -3&-3 \\
 \end{array}
\right),\nonumber\\ T_{1}&=&\left(\arraycolsep=8.4pt\def\arraystretch{1.5}
\begin{array}{cccc}
 -1 & -1 & -2 & -2   \\
 0 & 0 & -1&-1 \\
 0 & 0 & 0 &0\\
 1 &1  &1 &1\\
 \end{array}
\right),~\Tilde{T}_{1}=\left(\arraycolsep=8.4pt\def\arraystretch{1.5}
\begin{array}{cccc}
 -1 & -1 & 0&0 \\
 0 &0 & 1& 1\\
 1 & 1 & 1&1 \\
 2 & 2 & 2&2 \\
 \end{array}
\right),
\end{eqnarray}

$F_{k}=\left(
    -1, ~~ -1, ~~ -1,~~-1   
 \right)$, $\Tilde{F}_{k}=\left(0, ~~ 0, ~~ 0,~~0\right)$, ~ \text{and}~$ F_0=\left(0\right)$.
    
 We further worked out for $p=4,5$ as well and verified  our conjecture (\ref{QUIVERDT}) form obeyed. From these computations, we can deduce the general form of  the linear term $\Xi_{(p,-2)}$ and phase factor $\Lambda_{(p,-2)}$ in the proposition(\ref{KQ})  for arbitrary $p$ as:
\begin{eqnarray}{\label{LT2}}
\Xi_{(p,-2)}=\sum_{i}^{2p}\left(\tau_{1}(i)d_{4 i+1}+\tau_{2}(i)d_{4 i}+\tau_{3}(i)d_{4 i-1}\nonumber +\tau_{4}(i)d_{4 i-2}\right),
\end{eqnarray} 
where~ $\tau_{1}(i)=-2+2(-1)^i+i, ~\tau_{2}(i)=-3+2(-1)^i+i,~\tau_{3}(i)=-2+i, ~\tau_{4}(i)=-3+i$, and the phase factor is 
\be{\label{PH2}}
\Lambda_{(p,-2)}=\sum_{i=2}^{4p+1}d_{(-3 + 4 i - (-1)^{(\lfloor{\frac{i}{2}\rfloor})}\frac{1}{2})}.
\ee
Using the above data, we can write the colored Jones polynomial for any $K(p,-2)$ in quiver presentation  with the quiver matrix consistent with the conjecture (\ref{QUIVERDT}). So far, we have obtained the set of matrices $X_1$ for $m=1, 2$. With the hope of deducing the pattern for the set $X_1$ for any $m$, we will investigate double twist knots with $m=3$ in the following subsection.
%**********************************************************
\subsection{ Knot-Quiver correspondence for $K(p,-3)$}\label{knot-quiver-K(p,-3)}
Following reverse MMR, we could write the quiver presentation for $K(3,-3)$ and obtain the  quiver matrix $C^{K(3,-3)}$. The explicit matrix form is presented in the Appendix \ref{appendixd}.
The generators ($X_1$) of the quiver matrix  can be read off comparing with the conjectured form (\ref{QUIVERDT}):\\

\resizebox{\linewidth}{!}{$ U_{1}=\left(\arraycolsep=8.4pt\def\arraystretch{1.5}
\begin{array}{cccccc}
 -2 & -2 & -3 &  -3 &-3&-3  \\
 -2& -1 & -2&-2&-2&-2\\
 -3 & -2 & -4&-4&-5&-5\\
-3 & -2 & -4 &-3&-4&-4\\
 -3 &-2 & -5 &-4&-6&-6\\
-3 &  -2& -5 &-4&-6&-5\\
 \end{array}
\right),
~~\Tilde{U}_{1}=\left(\arraycolsep=8.4pt\def\arraystretch{1.5}
\begin{array}{cccccc}
 -3 & -3 & -2&-2&0&0 \\
 -3 & -2 & -1& -1&1&1\\
 -2 & -1 & -1&-1&0&0 \\
 -2 & -1 &-1&0&1&1\\
 0 & 1 & 0&1 &1&1\\
 0 & 1 & 0&1&1&2\\
 \end{array}
\right)$,}\\
\resizebox{\linewidth}{!}{$R_{1}=\left(\arraycolsep=8.4pt\def\arraystretch{1.5}
\begin{array}{cccccc}
 -2 & -2 & -2&-2&-1&-1 \\
 -1 & -1 & -1&-1&0&0\\
 -4 & -4 & -3&-3&-1& -1\\
 -3 & -3 &-2&-2 &0&0\\
 -5& -5 & -3&-3 &-1&-1\\
 -4 & -4 &-2& -2&0&0\\
 \end{array}
\right),~~~
\Tilde{R}_{1}=\left(\arraycolsep=8.4pt\def\arraystretch{1.5}
\begin{array}{cccccc}
 -2 & -2 & -3&-3&-3&-3 \\
 -1 & -1 & -2& -2&-2&-2\\
 -2 & -2 & -4&-4&-5&-5 \\
 -1 & -1 & -3&-3 &-4&-4\\
 -2 & -2 & -4&-4 &-6&-6\\
 -1 & -1 & -3&-3&-5&-5\\
 \end{array}
\right),\nonumber
$}
\resizebox{\linewidth}{!}{$ T_{1}=\left(\arraycolsep=8.4pt\def\arraystretch{1.5}
\begin{array}{cccccc}
 -1 & -1 & -3&-3&-4&-4 \\
 0 & 0 & -2&-2&-3&-3\\
 -1 & -1 & -2&-2&-2&-2 \\
 0 & 0 & -1&-1 &-1&-1\\
 0 & 0 & 0&0 &0&0\\
 1 & 1 & 1&1 &1&1\\
 \end{array}
\right),~~
\Tilde{T}_{1}=\left(\arraycolsep=8.4pt\def\arraystretch{1.5}
\begin{array}{cccccc}
 -3 & -3 & -2&-2&0&0 \\
 -2 & -2 & -1& -1&1&1\\
 -1 & -1 & -1&-1&0&0 \\
 0 & 0 & 0&0 &1&1\\
 1 & 1 & 1&1 &1&1\\
 2 & 2 & 2&2 &2&2\\
 \end{array}
\right),$}\\

$F_{k}=\left(-1,~-1,~-1,~-1,~-1,~-1\right),~\Tilde{F}_{k}=\left(0,~0,~0,~0,~0,~0\right) ,~\text{and}~ F_0=\left(0\right)$.

We have verified that our conjecture (\ref{QUIVERDT}) is true for $p=4, 5$. The linear term and phase factor in the proposition (\ref{KQ}) for $p=3,4$ are as follows:
\begin{eqnarray*}
\Xi_{(3,-3)}&=&-2d_2-d_3-4 d_4-3 d_5-6 d_6-5 d_7-3 d_8-2 d_9-d_{10}+d_{12}+2 d_{13}+d_{15}-2 d_{16}-\\&&d_{17}-4 d_{18}-
3 d_{19}-d_{20}+d_{22}+2 d_{23}+3 d_{24}+4 d_{25}+2 d_{26}+3 d_{27}+d_{29}-2 d_{30}\\&&-d_{31}+d_{32}+2 d_{33}+3 d_{34}+
4 d_{35}+5 d_{36}+6 d_{37}.\\
\Xi_{(4,-3)}&=&\Xi_{(3,-3)}+4 d_{38}+5 d_{39}+2 d_{40}+3 d_{41}+d_{43}+3 d_{44}+4 d_{45}+5 d_{46}+6 d_{47}+7 d_{48}+8 d_{49}.\\
\Lambda_{(3,-3)}&=&d_3+d_5+ d_7+d_8+d_{10}+ d_{12}+ d_{15}+d_{17}+ d_{19}+d_{20}+ d_{22}+ d_{24}+ d_{27}+ d_{29}+ d_{31}\\&& + d_{32} + d_{34}+d_{36}.\\
\Lambda_{(4,-3)}&=&\Lambda_{(3,-3)}+d_{39}+d_{41}+ d_{43}+ d_{44}+ d_{46}+d_{48}.
\end{eqnarray*}
Probably,there is a closed-form expression for $\Xi_{p,-3}$ and $\Lambda_{p,-3}$ for any $p$. We are not able
to infer the closed form from the above data. 

Ideally, it would be beneficial to find the set of matrices $X_1$ for any $m$ as well as the closed form for $\Xi_{p,-m}$ and $\Lambda_{p,-m}$. The size of the quiver matrix $(4mp+1) \times (4mp+1)$ makes the computations difficult.

%**********************************************************
%**********************************************************
%**********************************************************
%**********************************************************
%**********************************************************

\section{Conclusions}\label{sec4}
Double twist knots $K(p,-m)$ depend on two full twist parameters $p,m \in \mathbb Z_+$ belong to the arborescent  family (see Fig.\ref{DT}).  Finding a quiver with matrix (\ref{QUIVERDT}) associated to each of the double twist knots was  attempted using reverse engineering of Melvin-Morton-Rozansky expansion. We observed the Alexander polynomial form to be $\Delta(X)=1-p m X$, almost similar to twist knots $K(p,-1)$ studied in Ref.\cite{Kucharski:2017ogk}. Comparing the structure of twist knot quiver, we put forth a proposition (\ref{KQ}) for colored Jones in a quiver presentation  as well as conjectured (\ref{QUIVERDT}) the structure of the quiver  matrix $C^{K(p,-m)}$ for any double twist knot $K(p,-m)$. 
We have explicitly worked out some double twist knots  to validate our proposition and the conjecture for $m=1,2,3$. Our detailed methodology shows the complexity of the equations to deduce a concise form for $\Xi_{p,-3}$ and $\Lambda_{p,-3}$.

\chapter{Conclusions and Future Outlooks}
\label{chap:6}
\section{Conclusions}
We have focussed on WRT invariant for negative-definite plumbed 3-manifold $M(\Gamma)$ for $SO(3)$ and $OSp(1|2)$ group, involving framed link invariants. One of the reasons is that these invariants can be computed in $SU(2)$ Chern-Simons theory. Particularly, the colored link invariants for the $SO(3)$ and $OSp(1|2)$ are obtainable by a change of polynomial variable of colored Jones invariants. The $q$-series form of the $\widehat Z$ for these two groups could be systematically deduced.
We find a neat relation between $\widehat{Z}^{OSp(1|2)}_b[M(\Gamma),q]$ and $\widehat{Z}^{SU(2)}_b[M(\Gamma),q]$ (\ref{chap2zhatsu2},\ref{chap2zhatosp12})\textcol{, which was subsequently proved by Costantino et al. in ref.\cite{costantino2024nonsemisimpletopologicalfieldtheory}.} Also, we show that $\widehat{Z}^{SO(3)}_b[M(\Gamma);q]=\widehat{Z}^{SU(2)}_b[M(\Gamma);q]$. Such a result forced us to investigate the $\widehat{Z}$ for factor groups $SU(N)/\mathbb{Z}_m$ and find its relation to $\widehat{Z}^{SU(N)}$. 

Inspired by the result of our first work \cite{chauhan2022hat}, we investigated the $\widehat{Z}$-invariant for $SU(N)/\mathbb{Z}_m$ group. For this, we followed the route of GPPV conjecture, which required us to first fix the WRT invariant for $SU(N)/\mathbb{Z}_m$ group. After fixing it, we extracted the $\widehat{Z}$-invariant for $SU(N)/\mathbb{Z}_m$ group. Interestingly, we found $\widehat{Z}^{SU(N)/\mathbb{Z}_m}=\widehat{Z}^{SU(N)}$ (\ref{proposition}), showing $\widehat{Z}$ is Lie algebra dependent.

In the final piece of work, we concentrated on rewriting the colored Jones as a motivic series and extracting the quiver data for the double twist knot family, $K(p,-m)$. We found a pattern in quiver matrices for $K(p,-2)$ knots, where $p\geq 2$ (\ref{knot-quiver-K(p,-2)}). Further, we found the quiver matrix for $K(3,-3)$ (Appendix \ref{appendixd}).

\newpage

\section{Future Outlooks}
Some of the challenging problems we hope to pursue in future are:

\begin{enumerate}
\item[(a)] The brane setup in string theory for $U(N)$ gauge group gives a natural interpretation for these $q$-series as partition function of the theory 
$T[M;G]$. In principle, there should be a natural generalisation to orthogonal $SO(N)$ and symplectic group $Sp(2n)$  involving orientifolds. It will be worth investigating such a construction to obtain $\widehat Z$ for $SO(N)$ group and compare with our $SO(N=3)$ results. 
	
\item[(b)] We have conjectured the relation between WRT invariant for $SU(N)/\mathbb{Z}_m$ group and $\widehat{Z}^{\mathfrak{su}(N)}$. It would be interesting to study the $w$-refined version of $SU(N)$ WRT invariant and its relation with $\widehat{Z}^{\mathfrak{su}(N)}$ invariant. For $SU(2)$ WRT invariant, a $w$-refined WRT invariant was introduced in Ref.\cite{10.2140/agt.2015.15.1363}, and its relation with $\widehat{Z}$ was studied in Ref.\cite{Costantino:2021yfd}.

\item[(c)] Explore the explicit form of $\widehat{Z}$-invariant for other Lie groups (type $B,C,D,E$) and their Langlands dual. 
   
\item[(d)] The work on GPPV conjecture in this thesis was restricted to negative-definite plumbed 3-manifolds. It would be nice to study the GPPV conjecture and $\widehat{Z}$-invariant for more general 3-manifolds.

\item[(e)] \textcol{It would be interesting to check the superconformal index for $SU(N)/\mathbb{Z}_m$ group. However, conjecture (2.2) of GPPV \cite{Gukov:2017kmk} suggests that the superconformal index is insensitive to global aspects provided $\widehat{Z}$ is Lie algebra dependent. We would like to rigorously compute it for the $SU(N)/\mathbb{Z}_m$ group.  
}  

\item[(f)] It will be worth investigating whether our conjectured form for the block structure in $C^{K(p,-m)}$ has any connections to (i) the tangle operations \cite{Stosic2018NO2} and (ii) the holomorphic discs on the knot conormal \cite{Ekholm:2018eee}. Probably, this could be another way to tackle quiver representation form for $r$-colored HOMFLY-PT.

 There is a pretzel family of knots whose $r$-colored Jones and HOMFLY-PT are known. We are presently working on a systematic way of deducing a quiver presentation for the pretzel family. From our double twist knot results, it may be straightforward to attempt quiver presentation for knots whose   Alexander polynomial takes the form $\Delta(X)=1\pm (m_1 m_2 \ldots m_p )X~.$

\end{enumerate}

%=====================================================================
% APPENDIX
%  Appendices, if any, must precede the cited literatures.
%  Appendices shall be numbered in Roman Capitals (e.g. Appendix IV)

\begin{appendices}
    \chapter{Explicit calculation of $\hat{Z}^{\mathfrak{su}(3)}$-invariant for a particular plumbing graph}
\label{appenda} 

In this appendix, we work out explicitly the $\hat{Z}$ for $\mathfrak{su}(3)$ Lie algebra for the following plumbed 3-manifold:

\begin{figure}[h]
	\centering
	\includegraphics[scale=0.5]{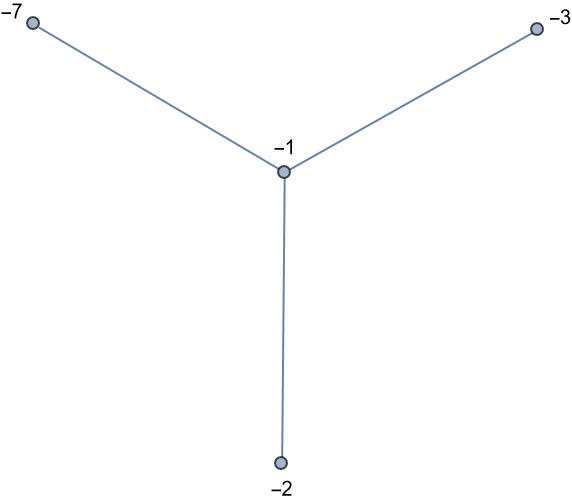}
\end{figure}

For $\mathfrak{su}(3)$ case, $W=\{1,s_1,s_2,s_1s_2,s_2s_1,s_1s_2s_1\}$, $\rho=\Lambda_1+\Lambda_2$ and $(\rho,\rho)=2$.

So, the $\hat{Z}$-invariant is:
\begin{equation}
	\hat{Z}^{\mathfrak{g}}_b[M(\Gamma),q]=|W|^{-L}q^{-\frac{(3L+\text{Tr}\;B)}{2}(\rho,\rho)}\sum_{s\in BQ^L+b}\xi_s^1 q^{-\frac{(s,B^{-1}s)}{2}}
	\label{zhatdef}
\end{equation} 

which in this case:

\begin{equation}
	\hat{Z}^{\mathfrak{su}(3)}_b[M(\Gamma),q]= \frac{q}{6^4}\sum_{s\in BQ^L+b}\xi_s^1 q^{-\frac{(s,B^{-1}s)}{2}}
\end{equation} 

where $\xi_s^1$ is determined from the following equation:

\begin{equation}
	\prod_{v\in V}\left(\sum_{\omega\in W}(-1)^{\ell(\omega)}\q^{(\lambda_v,\omega(\rho))}\right)^{2-\text{deg}\;v}=\frac{1}{6^4}\sum_{s\in Q^L+\delta}\xi_{s}^\beta\q^{(\lambda,s)}\Big\vert_{\beta\longrightarrow 1}
	\label{regularisation}
\end{equation}

We write the LHS of above equation as follows:

\begin{align}
\prod_{v\in V}\Big(\q^{(n^v_1\Lambda_1+n^v_2\Lambda_2)(\Lambda_1+\Lambda_2)}-\q^{(n^v_1\Lambda_1+n^v_2\Lambda_2)(-\Lambda_1+2\Lambda_2)}
-\q^{(n^v_1\Lambda_1+n^v_2\Lambda_2)(2\Lambda_1-\Lambda_2)}\nonumber\\+\q^{(n^v_1\Lambda_1+n^v_2\Lambda_2)(-2\Lambda_1+\Lambda_2)}
+\q^{(n^v_1\Lambda_1+n^v_2\Lambda_2)(\Lambda_1-2\Lambda_2)}-\q^{(n^v_1\Lambda_1+n^v_2\Lambda_2)(-\Lambda_1-\Lambda_2)}\Big)^{2-\text{deg }v}
\end{align}

where $\lambda_v=n^v_1\Lambda_1+n^v_2\Lambda_2$. The above equation simplies to the following using $\Lambda_1^2=\Lambda_2^2=\frac{2}{3}$ and $\Lambda_1\Lambda_2=\frac{1}{3}$:

\begin{align}
	\prod_{v\in V}\Big(\q^{(n^v_1+n^v_2)}-\q^{n^v_2}-\q^{n^v_1}+\q^{-n^v_1}+\q^{-n^v_2}-\q^{-(n^v_1+n^v_2)}\Big)^{2-\text{deg }v}
\end{align}

We write this equation using regularising parameter $\beta$ as follows,

\begin{align}
	=\lim_{\beta\rightarrow 1}\frac{1}{6^4}\prod_{v\in V}\Bigg(\Big(\beta\q^{(n^v_1+n^v_2)}-\q^{n^v_2}-\q^{n^v_1}+\q^{-n^v_1}+\q^{-n^v_2}-\q^{-(n^v_1+n^v_2)}\Big)^{2-\text{deg }v}\times\nonumber\\\Big(\q^{(n^v_1+n^v_2)}-\beta\q^{n^v_2}-\q^{n^v_1}+\q^{-n^v_1}+\q^{-n^v_2}-\q^{-(n^v_1+n^v_2)}\Big)^{2-\text{deg }v}\times\nonumber\\\Big(\q^{(n^v_1+n^v_2)}-\q^{n^v_2}-\beta\q^{n^v_1}+\q^{-n^v_1}+\q^{-n^v_2}-\q^{-(n^v_1+n^v_2)}\Big)^{2-\text{deg }v}\times\nonumber\\\Big(\q^{(n^v_1+n^v_2)}-\q^{n^v_2}-\q^{n^v_1}+\beta\q^{-n^v_1}+\q^{-n^v_2}-\q^{-(n^v_1+n^v_2)}\Big)^{2-\text{deg }v}\times\nonumber\\\Big(\q^{(n^v_1+n^v_2)}-\q^{n^v_2}-\q^{n^v_1}+\q^{-n^v_1}+\beta\q^{-n^v_2}-\q^{-(n^v_1+n^v_2)}\Big)^{2-\text{deg }v}\times\nonumber\\\Big(\q^{(n^v_1+n^v_2)}-\q^{n^v_2}-\q^{n^v_1}+\q^{-n^v_1}+\q^{-n^v_2}-\beta\q^{-(n^v_1+n^v_2)}\Big)^{2-\text{deg }v}\Bigg),
	\label{beta1}
\end{align}

now we expand the term inside the parenthesis in the above equation as $|\beta|<1$, to get the following:

\begin{equation}
	=\lim_{\beta\rightarrow 1}\left(\frac{1}{6^4}\sum_{s\in Q^4+\delta}\xi_s^\beta\q^{(\lambda,s)}\right)
\end{equation}

The above equation fixes the $\xi_s^1$. Moreover, $\delta_v=0~ \forall ~v$, for $\mathfrak{su}(3)$ Lie algebra as $\rho(=\Lambda_1+\Lambda_2=\alpha_1+\alpha_2)\in Q$. Further, for this plumbing graph, there is only one homological block, which corresponds to $b=0$. Using all this we obtain the $\hat{Z}$-invariant (\ref{zhatdef}) as follows:

\begin{align*}
	\hat{Z}_0^{\mathfrak{su}(3)}[M(\Gamma);q]=q^{3/2}\Big(&1 - 2 q + 2 q^3 + q^4 - 2 q^5 - 2 q^8 + 4 q^9 + 2 q^{10} - 4 q^{11} + 2 q^{13} - 6 q^{14} + 2 q^{15}\\
 &- 2 q^{16} + 4 q^{18} - q^{20} + 4 q^{21} - 2 q^{22} - 4 q^{23} + 2 q^{24} + 2 q^{25} - 4 q^{26} + 6 q^{30}\\
 &- 2 q^{31} + 6 q^{33} - 2 q^{34} - 2 q^{35} - 2 q^{38} + q^{40} +\mathcal{O}(q^{41})\Big).
\end{align*}

Note that the variable $q$ here is the analytically continued variable inside the unit circle.

    \chapter{Sublattice $P'$ and Chern-Simons level $k'$ for $SU(4)/\mathbb{Z}_2$, $SU(6)/\mathbb{Z}_2$ and $SU(6)/\mathbb{Z}_3$}
\label{appendB}

In this appendix, we will present the sublattice $P'$ and Chern-Simons level $k'$ for some non-simply connected groups. The $i^{\text{th}}$ fundamental weight vector for $\mathfrak{su}(N)$ Lie algebra is given by:

\begin{equation}
\Lambda_i=\frac{1}{N}(\underbrace{N-i,N-i,\ldots,N-i}_{i-\text{times}},\underbrace{-i,-i,\ldots,-i}_{(N-i)-\text{times}})~~~\text{where}~i\in\{1,2,\ldots,N-1\}.	
\end{equation}

\paragraph{\underline{$SU(4)/\mathbb{Z}_2:$}}

The center of $SU(4)$ is $\mathbb{Z}_4$ which is isomorphic to $\{e,\Lambda_1,\Lambda_2,\Lambda_3\}$. Further, the center of $SU(4)/\mathbb{Z}_2$ is $\mathbb{Z}_2$ which is isomorphic to $\{e,\Lambda_2\}$. The root lattice $Q$ of $\mathfrak{su}(4)$ Lie algebra, which also corresponds to the equivalence class for identity group element, is given by adding the following vectors:
\begin{equation}
	\{(2n_1-n_2)\Lambda_1,(-n_1+2n_2-n_3)\Lambda_2,(-n_2+2n_3)\Lambda_3|n_1,n_2,n_3\in \mathbb{Z}\},
	\label{su4z2a}
\end{equation}
where $\Lambda_1,\Lambda_2$ and $\Lambda_3$ are weight vectors. The equivalence class for the group element $\Lambda_2$ corresponds to 
\begin{equation}
	\{(2n_1-n_2)\Lambda_1,(-n_1+2n_2-n_3+1)\Lambda_2,(-n_2+2n_3)\Lambda_3|n_1,n_2,n_3\in \mathbb{Z}\}.
	\label{su4z2b}
\end{equation}

Hence the lattice $P'$ is given by taking the union of two equivalence classes (\ref{su4z2a}) and (\ref{su4z2b}).

Chern-Simons level $k'=\gamma k+4$ and factor $\gamma$ is fixed by smallest integer which satisfies the following equation:

\begin{equation}
	\frac{\gamma}{2}(\Lambda_2,\Lambda_2)\in\mathbb{Z}\implies \gamma=2,
\end{equation}

therefore, $k'=2k+4$.

\paragraph{\underline{$SU(6)/\mathbb{Z}_2:$}}

The center of $SU(6)/\mathbb{Z}_2$ is $\mathbb{Z}_3\cong\{e,\Lambda_2,\Lambda_4\}$. Therefore the sublattice $P'$ corresponding to the allowed representations of $SU(6)/\mathbb{Z}_2$ is given by:
\begin{align}
	\{(2 n_1 - n_2) \Lambda_1 , (-n_1 + 2 n_2 - n_3) \Lambda_2 , (-n_2 + 2 n_3 - n_4) \Lambda_3 , (-n_3 + 2 n_4 - n_5) \Lambda_4 , (-n_4 + 2 n_5) \Lambda_5\}\cup\nonumber\\
	\{(2 n_1 - n_2) \Lambda_1 , (-n_1 + 2 n_2 - n_3+1) \Lambda_2 , (-n_2 + 2 n_3 - n_4) \Lambda_3 , (-n_3 + 2 n_4 - n_5) \Lambda_4 , (-n_4 + 2 n_5) \Lambda_5\}\cup\nonumber\\
	\{(2 n_1 - n_2) \Lambda_1 , (-n_1 + 2 n_2 - n_3) \Lambda_2 , (-n_2 + 2 n_3 - n_4) \Lambda_3 , (-n_3 + 2 n_4 - n_5+1) \Lambda_4 , (-n_4 + 2 n_5) \Lambda_5\}.
\end{align}

Chern-Simons level $k'=\gamma k+6$ where $\gamma$ is fixed by requiring:

\begin{equation}
	\frac{\gamma}{2}(\Lambda_3,\Lambda_3)\in\mathbb{Z}\implies 4,
\end{equation}

therefore, $k'=4k+6$.

\paragraph{\underline{$SU(6)/\mathbb{Z}_3:$}}

For $\mathbb{Z}_2\cong\{e,\Lambda_3\}$, $P'$ is:
\begin{align}
	\{(2 n_1 - n_2) \Lambda_1 , (-n_1 + 2 n_2 - n_3) \Lambda_2 , (-n_2 + 2 n_3 - n_4) \Lambda_3 , (-n_3 + 2 n_4 - n_5) \Lambda_4 , (-n_4 + 2 n_5) \Lambda_5\}\cup\nonumber\\
	\{(2 n_1 - n_2) \Lambda_1 , (-n_1 + 2 n_2 - n_3) \Lambda_2 , (-n_2 + 2 n_3 - n_4+1) \Lambda_3 , (-n_3 + 
	2 n_4 - n_5) \Lambda_4 , (-n_4 + 2 n_5) \Lambda_5\}.
\end{align}

and Chern-Simons level $k'=\gamma k+6$ is fixed as follows:

\begin{equation}
	\frac{\gamma}{2}(\Lambda_2,\Lambda_2)\in \mathbb{Z}\;\; \text{and}\;\; 	\frac{\gamma}{2}(\Lambda_4,\Lambda_4)\in \mathbb{Z} \implies \gamma=3,
\end{equation}

hence, $k'=3k+6$.

    \chapter{GPPV conjecture for simply connected case: $SU(N)$}
\label{appenc}

In this appendix, we provide the GPPV conjecture for simply connected $SU(N)$ group. 
 
 For the $SU(N)$ group, with the weight lattice being $P$ and the root lattice being $Q$, the WRT invariant for plumbed 3-manifold $M(\Gamma)$ can be decomposed as follows:

\begin{align}
	\tau^{SU(N)}_{k'}[M(\Gamma);\q]=\frac{1}{|W||\text{det} B|^{1/2}\sum_{w\in W}(-1)^{\ell(w)}\q^{(\rho,w(\rho))}}\sum_{a\in (Q)^L/B(Q)^L}\exp(-\pi
	ik'(a,B^{-1}a))\nonumber\\ \nonumber\\
	\times\sum_{b\in (Q^L+\delta)/BQ^L}\exp(-2\pi i(a,B^{-1}b))\lim_{q\rightarrow \q}\hat{Z}^{\mathfrak{su}(N)}_b[M(\Gamma),q].
\end{align}

where $B$ is the linking matrix for $\Gamma$. For a general closed 3-manifold with $b_1(M)=0$, we conjecture the following:

\begin{align}
	\tau^{SU(N)}_{k'}[M;\q]=\frac{1}{|H_1(M;\mathbb{Z})|^{\frac{N-1}{2}}\sum_{w\in W}(-1)^{\ell(w)}\q^{(\rho,\omega(\rho))}}\sum_{a,b\in (\text{Spin}^c(M))^{(N-1)}/S_N}\exp(-2\pi i k'\sum_{i=1}^{N-1}\ell k(a_i,a_i))\nonumber\\\nonumber\\
	\times\exp(-4\pi i\sum_{i=1}^{N-1}\ell k(a_i,b_i))\lim_{q\rightarrow\q}\hat{Z}_b^{\mathfrak{su}(N)}[M;q].
\end{align}
    \chapter{Quiver Matrix}
\label{appendixd}

In this appendix, we provide a detailed calculation of obtaining a quiver presentation for a double-twist knot, $K(2,-2)=8_3$. Further, we also present quiver matrices for double twist knots $K(3,-2)$ and $K(3,-3)$.

The Alexander polynomial  in variable $X$ is $\Delta(8_3;X)=(1-4X)~.$
Such a form implied that we could perform the reverse  MMR method discussed\cite{Banerjee_2020} for $\Delta(K_p;X)= 1-pX$ 
with $p$ being integer. The $r$-colored Jones polynomials\footnote{Note that, the $\Tilde{J}_r(K;\q)$ is related to $J_r(K;\q)$ as follows: $\Tilde{J}_r(K;\q)=\frac{J_r(K;\q^2)}{(\q^2;\q^2)_r}$.} take the following interesting form:
\begin{align*}   
\Tilde{J}_{r}({\bf{8_3}};\q) =& \frac{1}{(\q^2;\q^2)_r}\sum_{r\geq k_8\geq \ldots \geq k_1 \geq 0} (\q^2;\q^2)_{k_8} {r \brack  k_8}_\q { k_8 \brack  k_7}_\q{ k_7 \brack  k_6 }_\q{k_6 \brack  k_5}_\q{ k_5 \brack  k_4}_\q { k_4 \brack  k_3}_\q{ k_3 \brack  k_2 }_\q{ k_2 \brack  k_1 }_\q\\
&~~~~~~~~~~~~~~~~~~\times\q^{\sum_{i,j=1}^p a_{ij} k_i k_j} 
\q^{{\sum_{i=1}^p} b_i k_i}(-1)^{{\sum_{i=1}^p} c_i k_i}\nonumber
\end{align*}

 By comparing these polynomials with known $r$-colored Jones polynomials, we have successfully determined the unknown parameters ${a_{ij}, b_i, c_i}$.
The exact expression is
\begin{eqnarray*}
\Tilde{J}_{r}({\bf{8_3}};\q) &= & \frac{1}{(\q^2;\q^2)_r}\sum_{r\geq k_8\geq\ldots \geq k_1 \geq 0} (-1)^{ k_2+k_4+ k_6}(\q^2;\q^2)_{k_8} {r \brack  k_8}_\q \nonumber\\
&& \times\prod_{i=0}^{6}{k_{8-i}\brack k_{7-i}}_\q \q^{2 k_1+3 k_2+2 r k_2-2 k_1 k_2-2 k_3 k_8+2 k_5 k_8}\nonumber\\ && \times\q^{-k_4-2 r k_4+2 k_1 k_4-2 k_3 k_4+k_4^2+2 k_5-2 k_1 k_5+2 k_3 k_5+3 k_6}\nonumber\\
&&\times\q^{+2 r k_6-2 k_1 k_6+2 k_3 k_6-2 k_5 k_6+k_6^2-2 k_7+2 k_1 k_7-2 k_3 k_7}\nonumber\\
&&\times\q^{+2 k_5 k_7-2 k_8-2 r k_8+2 k_1 k_8+k_2^2-2 k_3+2 k_1 k_3-2 k_7 k_8}.~\nonumber
\end{eqnarray*}
Using the $\q$-binomial and $\q$-Pochhammer identities discussed in Ref. \cite{Kucharski:2017ogk}, we could rewrite the $r$-colored Jones polynomial as
\begin{eqnarray}{\label{83jones1}}
\Tilde{J}_{r}({\bf{8_3}};\q)&=&\sum_{r, \mathbf k, \boldsymbol{\alpha}}\frac{(-1)^{k_2+k_4+k_6+\alpha_8}}{\prod_{i=0}^{6}(\q^2;\q^2)_{k_{8-i}-\alpha_{8-i}-k_{8-i-1}+\alpha_{8-i-1}}}\nonumber\\&&\times\frac{1}{(\q^2;\q^2)_{r-k_8}\prod_{i=0}^{6}(\q^2;\q^2)_{\alpha_{8-i}-\alpha_{8-i-1}}(\q^2;\q^2)_{k_1-\alpha_1}}\nonumber \\
&& \times\frac{1}{(\q^2;\q^2)_{\alpha_1}} \q^{ \left(3k_2+2r k_2+k_2^2-k_4-2rk_4\right)+2\left(k_1-k_3+k_5-k_7\right)}\nonumber\\
&&\times\q^{2\left(-k_1 k_2+k_1 k_3+k_1 k_4-k_3 k_4-k_1 k_5+k_3 k_5-k_1 k_6+k_3 k_6-k_5 k_6\right)}\nonumber\\
&& \times\q^{\sum_{i=0}^{6}2(\alpha_{8-i} - \alpha_{7-i}) (k_{7-i} - \alpha_{7-i})+\alpha_8(\alpha_8+1)+k_4^2+3k_6 } \nonumber\\
&& \times\q^{\left(+2r k_6+k_6^2-2 k_8-k_8^2-2 r k_8+2(+k_1 k_7-k_3 k_7) \right)}\nonumber\\
&&\times\q^{2(k_5 k_7+k_1 k_8-k_3 k_8+k_5 k_8-k_7
k_8)} ,\nonumber
\end{eqnarray}
where the summation variables in the first line must obey $r\geq k_8\geq \alpha_8 \geq k_7 \geq \alpha_7 \ldots \geq k_1\geq \alpha_1 \geq 0.$ On making the following substitutions
$$\alpha_j = \sum_{i=0}^{j-1}d_{19-2j+2i}~~~~ \text{and}~~~~ k_j = \sum_{i=0}^{2j-1}d_{18-2j+i}$$

in eqn.(\ref{83jones1}), we obtain the motivic series quiver representation form:
\begin{eqnarray}
\Tilde{J}_{r}({\bf{8_3}};\q)&=& \sum_{d_1+d_2+ \ldots+d_{17}=r}(-1)^{d_3+ d_5+ d_6+d_8+d_{11}+d_{13}+d_{14}+d_{16}}\nonumber\\
&& \times\frac{\q^{\sum C^{8_3}_{(i,j)} d_i d_j}}{\prod_{i=1}^{17}(\q^2;\q^2)_{d_{i}}}\q^{\left(-2 d_2-d_3-4 d_4-3 d_5-d_6 \right)} \nonumber\\
&& \times\q^{\left(d_8+2 d_9+d_{11}-2 d_{12}-d_{13}+d_{14}+2 d_{15} +3 d_{16}+4 d_{17} \right)}\nonumber.
\end{eqnarray}
We can read off the quiver  matrix elements $C^{8_3}$ from the above expression. See section\ref{sec2}, where we have presented the matrix.
The explicit matrix elements for $C^{K(3,-2)}$ and $C^{K(3,-3)}$ are presented below:\\

%For $K(3,-2)$,  the quiver matrix is\\
\begin{center}
\resizebox{\textwidth}{12cm}{
$C^{(3,-2)}=\left(\begin{array}{c|cccc|cccc|cccc|cccc|cccc|cccc}
 0 & -1 & -1 & -1 & -1 & 0 & 0 & 0 & 0 & -1 & -1 & -1 & -1 & 0 & 0 & 0 & 0 & -1 & -1 & -1 & -1 & 0 & 0 & 0 & 0 \\
 \hline
 -1 & -2 & -2 & -3 & -3 & -2 & -2 & -1 & -1 & -2 & -2 & -3 & -3 & -2 & -2 & -1 & -1 & -2 & -2 & -3 & -3 & -2 & -2 & -1 & -1 \\
 -1 & -2 & -1 & -2 & -2 & -1 & -1 & 0 & 0 & -1 & -1 & -2 & -2 & -1 & -1 & 0 & 0 & -1 & -1 & -2 & -2 & -1 & -1 & 0 & 0 \\
 -1 & -3 & -2 & -4 & -4 & -3 & -3 & -1 & -1 & -2 & -2 & -4 & -4 & -3 & -3 & -1 & -1 & -2 & -2 & -4 & -4 & -3 & -3 & -1 & -1 \\
 -1 & -3 & -2 & -4 & -3 & -2 & -2 & 0 & 0 & -1 & -1 & -3 & -3 & -2 & -2 & 0 & 0 & -1 & -1 & -3 & -3 & -2 & -2 & 0 & 0 \\
 \hline
 0 & -2 & -1 & -3 & -2 & -1 & -1 & 0 & 0 & -1 & -1 & -2 & -2 & -1 & -1 & 0 & 0 & -1 & -1 & -2 & -2 & -1 & -1 & 0 & 0 \\
 0 & -2 & -1 & -3 & -2 & -1 & 0 & 1 & 1 & 0 & 0 & -1 & -1 & 0 & 0 & 1 & 1 & 0 & 0 & -1 & -1 & 0 & 0 & 1 & 1 \\
 0 & -1 & 0 & -1 & 0 & 0 & 1 & 1 & 1 & 0 & 0 & 0 & 0 & 1 & 1 & 1 & 1 & 0 & 0 & 0 & 0 & 1 & 1 & 1 & 1 \\
 0 & -1 & 0 & -1 & 0 & 0 & 1 & 1 & 2 & 1 & 1 & 1 & 1 & 2 & 2 & 2 & 2 & 1 & 1 & 1 & 1 & 2 & 2 & 2 & 2 \\
 \hline
 -1 & -2 & -1 & -2 & -1 & -1 & 0 & 0 & 1 & 0 & 0 & -1 & -1 & 0 & 0 & 1 & 1 & 0 & 0 & -1 & -1 & 0 & 0 & 1 & 1 \\
 -1 & -2 & -1 & -2 & -1 & -1 & 0 & 0 & 1 & 0 & 1 & 0 & 0 & 1 & 1 & 2 & 2 & 1 & 1 & 0 & 0 & 1 & 1 & 2 & 2 \\
 -1 & -3 & -2 & -4 & -3 & -2 & -1 & 0 & 1 & -1 & 0 & -2 & -2 & -1 & -1 & 1 & 1 & 0 & 0 & -2 & -2 & -1 & -1 & 1 & 1 \\
 -1 & -3 & -2 & -4 & -3 & -2 & -1 & 0 & 1 & -1 & 0 & -2 & -1 & 0 & 0 & 2 & 2 & 1 & 1 & -1 & -1 & 0 & 0 & 2 & 2 \\
 \hline
 0 & -2 & -1 & -3 & -2 & -1 & 0 & 1 & 2 & 0 & 1 & -1 & 0 & 1 & 1 & 2 & 2 & 1 & 1 & 0 & 0 & 1 & 1 & 2 & 2 \\
 0 & -2 & -1 & -3 & -2 & -1 & 0 & 1 & 2 & 0 & 1 & -1 & 0 & 1 & 2 & 3 & 3 & 2 & 2 & 1 & 1 & 2 & 2 & 3 & 3 \\
 0 & -1 & 0 & -1 & 0 & 0 & 1 & 1 & 2 & 1 & 2 & 1 & 2 & 2 & 3 & 3 & 3 & 2 & 2 & 2 & 2 & 3 & 3 & 3 & 3 \\
 0 & -1 & 0 & -1 & 0 & 0 & 1 & 1 & 2 & 1 & 2 & 1 & 2 & 2 & 3 & 3 & 4 & 3 & 3 & 3 & 3 & 4 & 4 & 4 & 4 \\
 \hline
 -1 & -2 & -1 & -2 & -1 & -1 & 0 & 0 & 1 & 0 & 1 & 0 & 1 & 1 & 2 & 2 & 3 & 2 & 2 & 1 & 1 & 2 & 2 & 3 & 3 \\
 -1 & -2 & -1 & -2 & -1 & -1 & 0 & 0 & 1 & 0 & 1 & 0 & 1 & 1 & 2 & 2 & 3 & 2 & 3 & 2 & 2 & 3 & 3 & 4 & 4 \\
 -1 & -3 & -2 & -4 & -3 & -2 & -1 & 0 & 1 & -1 & 0 & -2 & -1 & 0 & 1 & 2 & 3 & 1 & 2 & 0 & 0 & 1 & 1 & 3 & 3 \\
 -1 & -3 & -2 & -4 & -3 & -2 & -1 & 0 & 1 & -1 & 0 & -2 & -1 & 0 & 1 & 2 & 3 & 1 & 2 & 0 & 1 & 2 & 2 & 4 & 4 \\
 \hline
 0 & -2 & -1 & -3 & -2 & -1 & 0 & 1 & 2 & 0 & 1 & -1 & 0 & 1 & 2 & 3 & 4 & 2 & 3 & 1 & 2 & 3 & 3 & 4 & 4 \\
 0 & -2 & -1 & -3 & -2 & -1 & 0 & 1 & 2 & 0 & 1 & -1 & 0 & 1 & 2 & 3 & 4 & 2 & 3 & 1 & 2 & 3 & 4 & 5 & 5 \\
 0 & -1 & 0 & -1 & 0 & 0 & 1 & 1 & 2 & 1 & 2 & 1 & 2 & 2 & 3 & 3 & 4 & 3 & 4 & 3 & 4 & 4 & 5 & 5 & 5 \\
 0 & -1 & 0 & -1 & 0 & 0 & 1 & 1 & 2 & 1 & 2 & 1 & 2 & 2 & 3 & 3 & 4 & 3 & 4 & 3 & 4 & 4 & 5 & 5 & 6 \\
\end{array}\right)$,}
\end{center}

\newpage
\begin{center}
\resizebox{\textwidth}{12cm}{$C^{(3,-3)} =\left(
\begin{array}{c|cccccc|cccccc|cccccc|cccccc|cccccc|cccccc}
 0 & -1 & -1 & -1 & -1 & -1 & -1 & 0 & 0 & 0 & 0 & 0 & 0 & -1 & -1 & -1 & -1 & -1 & -1 & 0 & 0 & 0 & 0 & 0 & 0 & -1 & -1 & -1 & -1 & -1 & -1 & 0
& 0 & 0 & 0 & 0 & 0 \\
\hline
 -1 & -2 & -2 & -3 & -3 & -3 & -3 & -2 & -2 & -2 & -2 & -1 & -1 & -2 & -2 & -3 & -3 & -3 & -3 & -2 & -2 & -2 & -2 & -1 & -1 & -2 & -2 & -3 & -3 &
-3 & -3 & -2 & -2 & -2 & -2 & -1 & -1 \\
 -1 & -2 & -1 & -2 & -2 & -2 & -2 & -1 & -1 & -1 & -1 & 0 & 0 & -1 & -1 & -2 & -2 & -2 & -2 & -1 & -1 & -1 & -1 & 0 & 0 & -1 & -1 & -2 & -2 & -2
& -2 & -1 & -1 & -1 & -1 & 0 & 0 \\
 -1 & -3 & -2 & -4 & -4 & -5 & -5 & -4 & -4 & -3 & -3 & -1 & -1 & -2 & -2 & -4 & -4 & -5 & -5 & -4 & -4 & -3 & -3 & -1 & -1 & -2 & -2 & -4 & -4 &
-5 & -5 & -4 & -4 & -3 & -3 & -1 & -1 \\
 -1 & -3 & -2 & -4 & -3 & -4 & -4 & -3 & -3 & -2 & -2 & 0 & 0 & -1 & -1 & -3 & -3 & -4 & -4 & -3 & -3 & -2 & -2 & 0 & 0 & -1 & -1 & -3 & -3 & -4
& -4 & -3 & -3 & -2 & -2 & 0 & 0 \\
 -1 & -3 & -2 & -5 & -4 & -6 & -6 & -5 & -5 & -3 & -3 & -1 & -1 & -2 & -2 & -4 & -4 & -6 & -6 & -5 & -5 & -3 & -3 & -1 & -1 & -2 & -2 & -4 & -4 &
-6 & -6 & -5 & -5 & -3 & -3 & -1 & -1 \\
 -1 & -3 & -2 & -5 & -4 & -6 & -5 & -4 & -4 & -2 & -2 & 0 & 0 & -1 & -1 & -3 & -3 & -5 & -5 & -4 & -4 & -2 & -2 & 0 & 0 & -1 & -1 & -3 & -3 & -5
& -5 & -4 & -4 & -2 & -2 & 0 & 0 \\
\hline
 0 & -2 & -1 & -4 & -3 & -5 & -4 & -3 & -3 & -2 & -2 & 0 & 0 & -1 & -1 & -3 & -3 & -4 & -4 & -3 & -3 & -2 & -2 & 0 & 0 & -1 & -1 & -3 & -3 & -4 &
-4 & -3 & -3 & -2 & -2 & 0 & 0 \\
 0 & -2 & -1 & -4 & -3 & -5 & -4 & -3 & -2 & -1 & -1 & 1 & 1 & 0 & 0 & -2 & -2 & -3 & -3 & -2 & -2 & -1 & -1 & 1 & 1 & 0 & 0 & -2 & -2 & -3 & -3
& -2 & -2 & -1 & -1 & 1 & 1 \\
 0 & -2 & -1 & -3 & -2 & -3 & -2 & -2 & -1 & -1 & -1 & 0 & 0 & -1 & -1 & -2 & -2 & -2 & -2 & -1 & -1 & -1 & -1 & 0 & 0 & -1 & -1 & -2 & -2 & -2 &
-2 & -1 & -1 & -1 & -1 & 0 & 0 \\
 0 & -2 & -1 & -3 & -2 & -3 & -2 & -2 & -1 & -1 & 0 & 1 & 1 & 0 & 0 & -1 & -1 & -1 & -1 & 0 & 0 & 0 & 0 & 1 & 1 & 0 & 0 & -1 & -1 & -1 & -1 & 0 &
0 & 0 & 0 & 1 & 1 \\
 0 & -1 & 0 & -1 & 0 & -1 & 0 & 0 & 1 & 0 & 1 & 1 & 1 & 0 & 0 & 0 & 0 & 0 & 0 & 1 & 1 & 1 & 1 & 1 & 1 & 0 & 0 & 0 & 0 & 0 & 0 & 1 & 1 & 1 & 1 & 1
& 1 \\
 0 & -1 & 0 & -1 & 0 & -1 & 0 & 0 & 1 & 0 & 1 & 1 & 2 & 1 & 1 & 1 & 1 & 1 & 1 & 2 & 2 & 2 & 2 & 2 & 2 & 1 & 1 & 1 & 1 & 1 & 1 & 2 & 2 & 2 & 2 & 2
& 2 \\
\hline
 -1 & -2 & -1 & -2 & -1 & -2 & -1 & -1 & 0 & -1 & 0 & 0 & 1 & 0 & 0 & -1 & -1 & -1 & -1 & 0 & 0 & 0 & 0 & 1 & 1 & 0 & 0 & -1 & -1 & -1 & -1 & 0 &
0 & 0 & 0 & 1 & 1 \\
 -1 & -2 & -1 & -2 & -1 & -2 & -1 & -1 & 0 & -1 & 0 & 0 & 1 & 0 & 1 & 0 & 0 & 0 & 0 & 1 & 1 & 1 & 1 & 2 & 2 & 1 & 1 & 0 & 0 & 0 & 0 & 1 & 1 & 1 &
1 & 2 & 2 \\
 -1 & -3 & -2 & -4 & -3 & -4 & -3 & -3 & -2 & -2 & -1 & 0 & 1 & -1 & 0 & -2 & -2 & -3 & -3 & -2 & -2 & -1 & -1 & 1 & 1 & 0 & 0 & -2 & -2 & -3 & -3
& -2 & -2 & -1 & -1 & 1 & 1 \\
 -1 & -3 & -2 & -4 & -3 & -4 & -3 & -3 & -2 & -2 & -1 & 0 & 1 & -1 & 0 & -2 & -1 & -2 & -2 & -1 & -1 & 0 & 0 & 2 & 2 & 1 & 1 & -1 & -1 & -2 & -2
& -1 & -1 & 0 & 0 & 2 & 2 \\
 -1 & -3 & -2 & -5 & -4 & -6 & -5 & -4 & -3 & -2 & -1 & 0 & 1 & -1 & 0 & -3 & -2 & -4 & -4 & -3 & -3 & -1 & -1 & 1 & 1 & 0 & 0 & -2 & -2 & -4 & -4
& -3 & -3 & -1 & -1 & 1 & 1 \\
 -1 & -3 & -2 & -5 & -4 & -6 & -5 & -4 & -3 & -2 & -1 & 0 & 1 & -1 & 0 & -3 & -2 & -4 & -3 & -2 & -2 & 0 & 0 & 2 & 2 & 1 & 1 & -1 & -1 & -3 & -3
& -2 & -2 & 0 & 0 & 2 & 2 \\
\hline
 0 & -2 & -1 & -4 & -3 & -5 & -4 & -3 & -2 & -1 & 0 & 1 & 2 & 0 & 1 & -2 & -1 & -3 & -2 & -1 & -1 & 0 & 0 & 2 & 2 & 1 & 1 & -1 & -1 & -2 & -2 & -1
& -1 & 0 & 0 & 2 & 2 \\
 0 & -2 & -1 & -4 & -3 & -5 & -4 & -3 & -2 & -1 & 0 & 1 & 2 & 0 & 1 & -2 & -1 & -3 & -2 & -1 & 0 & 1 & 1 & 3 & 3 & 2 & 2 & 0 & 0 & -1 & -1 & 0 &
0 & 1 & 1 & 3 & 3 \\
 0 & -2 & -1 & -3 & -2 & -3 & -2 & -2 & -1 & -1 & 0 & 1 & 2 & 0 & 1 & -1 & 0 & -1 & 0 & 0 & 1 & 1 & 1 & 2 & 2 & 1 & 1 & 0 & 0 & 0 & 0 & 1 & 1 & 1
& 1 & 2 & 2 \\
 0 & -2 & -1 & -3 & -2 & -3 & -2 & -2 & -1 & -1 & 0 & 1 & 2 & 0 & 1 & -1 & 0 & -1 & 0 & 0 & 1 & 1 & 2 & 3 & 3 & 2 & 2 & 1 & 1 & 1 & 1 & 2 & 2 & 2
& 2 & 3 & 3 \\
 0 & -1 & 0 & -1 & 0 & -1 & 0 & 0 & 1 & 0 & 1 & 1 & 2 & 1 & 2 & 1 & 2 & 1 & 2 & 2 & 3 & 2 & 3 & 3 & 3 & 2 & 2 & 2 & 2 & 2 & 2 & 3 & 3 & 3 & 3 & 3
& 3 \\
 0 & -1 & 0 & -1 & 0 & -1 & 0 & 0 & 1 & 0 & 1 & 1 & 2 & 1 & 2 & 1 & 2 & 1 & 2 & 2 & 3 & 2 & 3 & 3 & 4 & 3 & 3 & 3 & 3 & 3 & 3 & 4 & 4 & 4 & 4 & 4
& 4 \\
\hline
 -1 & -2 & -1 & -2 & -1 & -2 & -1 & -1 & 0 & -1 & 0 & 0 & 1 & 0 & 1 & 0 & 1 & 0 & 1 & 1 & 2 & 1 & 2 & 2 & 3 & 2 & 2 & 1 & 1 & 1 & 1 & 2 & 2 & 2 &
2 & 3 & 3 \\
 -1 & -2 & -1 & -2 & -1 & -2 & -1 & -1 & 0 & -1 & 0 & 0 & 1 & 0 & 1 & 0 & 1 & 0 & 1 & 1 & 2 & 1 & 2 & 2 & 3 & 2 & 3 & 2 & 2 & 2 & 2 & 3 & 3 & 3 &
3 & 4 & 4 \\
 -1 & -3 & -2 & -4 & -3 & -4 & -3 & -3 & -2 & -2 & -1 & 0 & 1 & -1 & 0 & -2 & -1 & -2 & -1 & -1 & 0 & 0 & 1 & 2 & 3 & 1 & 2 & 0 & 0 & -1 & -1 & 0
& 0 & 1 & 1 & 3 & 3 \\
 -1 & -3 & -2 & -4 & -3 & -4 & -3 & -3 & -2 & -2 & -1 & 0 & 1 & -1 & 0 & -2 & -1 & -2 & -1 & -1 & 0 & 0 & 1 & 2 & 3 & 1 & 2 & 0 & 1 & 0 & 0 & 1 &
1 & 2 & 2 & 4 & 4 \\
 -1 & -3 & -2 & -5 & -4 & -6 & -5 & -4 & -3 & -2 & -1 & 0 & 1 & -1 & 0 & -3 & -2 & -4 & -3 & -2 & -1 & 0 & 1 & 2 & 3 & 1 & 2 & -1 & 0 & -2 & -2 &
-1 & -1 & 1 & 1 & 3 & 3 \\
 -1 & -3 & -2 & -5 & -4 & -6 & -5 & -4 & -3 & -2 & -1 & 0 & 1 & -1 & 0 & -3 & -2 & -4 & -3 & -2 & -1 & 0 & 1 & 2 & 3 & 1 & 2 & -1 & 0 & -2 & -1 &
0 & 0 & 2 & 2 & 4 & 4 \\
\hline
 0 & -2 & -1 & -4 & -3 & -5 & -4 & -3 & -2 & -1 & 0 & 1 & 2 & 0 & 1 & -2 & -1 & -3 & -2 & -1 & 0 & 1 & 2 & 3 & 4 & 2 & 3 & 0 & 1 & -1 & 0 & 1 & 1
& 2 & 2 & 4 & 4 \\
 0 & -2 & -1 & -4 & -3 & -5 & -4 & -3 & -2 & -1 & 0 & 1 & 2 & 0 & 1 & -2 & -1 & -3 & -2 & -1 & 0 & 1 & 2 & 3 & 4 & 2 & 3 & 0 & 1 & -1 & 0 & 1 & 2
& 3 & 3 & 5 & 5 \\
 0 & -2 & -1 & -3 & -2 & -3 & -2 & -2 & -1 & -1 & 0 & 1 & 2 & 0 & 1 & -1 & 0 & -1 & 0 & 0 & 1 & 1 & 2 & 3 & 4 & 2 & 3 & 1 & 2 & 1 & 2 & 2 & 3 & 3
& 3 & 4 & 4 \\
 0 & -2 & -1 & -3 & -2 & -3 & -2 & -2 & -1 & -1 & 0 & 1 & 2 & 0 & 1 & -1 & 0 & -1 & 0 & 0 & 1 & 1 & 2 & 3 & 4 & 2 & 3 & 1 & 2 & 1 & 2 & 2 & 3 & 3
& 4 & 5 & 5 \\
 0 & -1 & 0 & -1 & 0 & -1 & 0 & 0 & 1 & 0 & 1 & 1 & 2 & 1 & 2 & 1 & 2 & 1 & 2 & 2 & 3 & 2 & 3 & 3 & 4 & 3 & 4 & 3 & 4 & 3 & 4 & 4 & 5 & 4 & 5 & 5
& 5 \\
 0 & -1 & 0 & -1 & 0 & -1 & 0 & 0 & 1 & 0 & 1 & 1 & 2 & 1 & 2 & 1 & 2 & 1 & 2 & 2 & 3 & 2 & 3 & 3 & 4 & 3 & 4 & 3 & 4 & 3 & 4 & 4 & 5 & 4 & 5 & 5
& 6 \\
\end{array}\right)$.}\\
\end{center}
\end{appendices}

%=====================================================================
% BIBLIOGRAPHY
\newpage
\setlength{\parskip}{5mm}
\titlespacing{\chapter}{0cm}{0mm}{0mm}
\titleformat{\chapter}[display]
  {\normalfont\huge\bfseries}
  {\chaptertitlename\ \thechapter}{20pt}{\Huge}

\bibliographystyle{References/utphys}
% Add the bib file
\bibliography{References/mybibfile}

%=====================================================================
% PUBLICATIONS
%  publications if any may be listed after the literature cited.
%\addcontentsline{toc}{chapter}{List of Publications}

\end{document}